\documentclass[aps,prb,twocolumn,amsmath,amssymb,floatfix,
superscriptaddress]{revtex4-1}
\usepackage{blindtext}
\usepackage{epsfig}
\usepackage{color}
\usepackage{amsmath}
\usepackage{amsfonts}
\usepackage{amssymb}
\usepackage{graphicx}%
\usepackage{url}
\usepackage[breaklinks]{hyperref}
\usepackage[english]{babel}
\usepackage{esvect}
\usepackage[normalem]{ulem}
\usepackage[inline]{enumitem}
\usepackage{xfrac}
\usepackage{dsfont}
\usepackage{bm}
\usepackage{array}
\newcolumntype{C}[1]{>{\centering\let\newline\\\arraybackslash\hspace{0pt}}m{#1}}
\usepackage{footnote}

\hypersetup{colorlinks,citecolor=blue,filecolor=blue,linkcolor=blue,urlcolor=blue}

\numberwithin{equation}{section}

\bibpunct{[}{]}{,}{n}{}{}

\begin{document}

\title{Interacting Stochastic Topology and Mott Transition from Light Response}

\author{Philipp W. Klein}
\affiliation{CPHT, CNRS, Ecole Polytechnique, Institut Polytechnique de Paris, Route de Saclay, 91128 Palaiseau, France} 
\author{Adolfo G. Grushin}
\affiliation{Institut Ne\'el, CNRS and Universit\'e Grenoble Alpes, Grenoble, France} 
\author{Karyn Le Hur}
\affiliation{CPHT, CNRS, Ecole Polytechnique, Institut Polytechnique de Paris, Route de Saclay, 91128 Palaiseau, France} 

\date{\today}

\begin{abstract}
We develop a stochastic description of the topological properties in an interacting Chern insulator.  We confirm the Mott transition's first-order nature in the interacting Haldane model on the honeycomb geometry, from a mean-field variational approach,  supported by density matrix renormalization group results and Ginzburg-Landau arguments. From the Bloch sphere, we make predictions for circular dichroism of light related to the quantum Hall conductivity on the lattice and in the presence of interactions. This analysis shows that the topological number can be measured from the light response at the Dirac points. Electron-electron interactions can also produce a substantial number of particle-hole pairs above the band gap, which leads us to propose a ``stochastic Chern number'' as an interacting measure of the topology.  The stochastic Chern number can describe disordered situations with a fluctuating staggered potential and we build an analogy between interaction-induced particle-hole pairs and temperature effects. Our stochastic approach is physically intuitive, easy to implement and leads the way to further studies of interaction effects.
\end{abstract}

\maketitle

\section{Introduction}

Topological phases of matter have attracted a lot of interest in the last years, including generalizations of the quantum Hall effect \cite{klitzing} to the Haldane model on the honeycomb lattice \cite{haldane_1988}. This gives rise to the quantum anomalous Hall effect which has been realized in quantum materials \cite{liu_2015}, graphene \cite{McIver_2019}, photonic systems \cite{haldane_2008,Lu_2014,Rechtsman_2013,koch_2010,Le_Hur_2016,Tomoki_2019} and cold atoms in optical lattices \cite{jotzu_2014,flaschner_2016}. The
occurrence of an insulating bulk and a protected chiral edge mode can be understood through a quantized topological number, the Chern number, assuming non-interacting particles. These systems at half-filling then behave as Chern insulators (CI). While some progress has been made in the description of interacting Chern systems, in the bosonic case \cite{varney_2010,vasic_2015}, and both for spinless \cite{varney_2010,varney_2011,dai_2010,alba_2016} and spinfull \cite{hickey_2015, ruegg_2013, huber_2016, troyer_2016, vanhala_2016} fermions, and more generally in the description of interacting topological systems \cite{Stanford,Rachel_2018}, several central questions remain open. In this Letter, we introduce a many-body description for interacting CIs, taking into account interaction-induced particle-hole pairs and describing Mott physics.

To describe the effect of particle-hole pairs in the Haldane model on the honeycomb lattice with a nearest-neighbor interaction, we introduce a Hubbard-Stratonovitch transformation to re-write the quartic fermionic interaction as Gaussian variables \cite{schulz_1994}, referring to stochastic variables. One stochastic variable can be identified as a Semenoff mass \cite{Semenoff}, which will produce a sign flip of the mass term at one Dirac point only at the Mott transition. Within this approach, there is a direct relation between the creation of particle-hole pairs above the band gap due to interaction effects and fluctuations of this stochastic variable. By analogy to the studies of the two-dimensional Hubbard model \cite{plainvanilla,ReviewhighTc}, other variables dressing the kinetic term and referring to the particle-hole channel, will also play an important role. This particle-hole channel will produce a jump of the Charge Density Wave (CDW) order parameter at the Mott transition, already at a mean-field level from a variational approach, in agreement with infinite density matrix renormalization group (iDMRG) results. We are astonished that this particle-hole channel which provides a physical insight behind the Mott transition's first-order nature, has not been introduced before. The phase transition was previously shown to be first-order from exact diagonalization \cite{varney_2010,varney_2011}. We also analyse the phase transition using a Ginzburg-Landau approach \cite{ginzburg}.  

Furthermore, circular dichroism of light is related to the quantized Chern number \cite{goldman_2017,asteria_2019} for the non-interacting Haldane model. Shining light induces a population of the states in the upper band, above the band gap. The associated depletion rates depend on the orientation of the circular drive. The Chern number is encoded in the difference of rates with opposite orientation. Through a mathematical formulation on the Bloch sphere, we show that the topology of the (interacting) system is already described by the depletion rates that occur at the Dirac points of the Brillouin zone. We also consider interaction effects that contribute to the formation of particle-hole pairs and thereby stochastically change the Chern number as defined through the circular dichroism of light. We define the stochastic Chern number through an average on the ensemble of stochastic variables. We study how the stochastic topological number may be linked to disordered situations, to stochastic theories and protocols on the Bloch sphere \cite{sphere,spheremodel} and to the finite-temperature Chern number \cite{delgado_2013}.

The organization of the article is as follows. In Sec. \ref{model}, we present the interacting Haldane model with a nearest-neighbor interaction, and introduce the definitions both in the real and reciprocal spaces. In Sec. \ref{stochappro}, we derive the formalism related to the interacting stochastic topology and we obtain an effective Hamiltonian. We also present the variational mean-field approach, discuss results for the phase diagram as well as observables, and show quantitative agreement with iDMRG. In Sec. \ref{energetics}, we derive an energetics approach of the transition and build a Ginzburg-Landau theory of fluctuations to justify the nature of the Mott transition. In Sec. \ref{circularJones}, we show how the topological properties can be revealed from the Dirac points through circular dichroism of light, via a Bloch sphere quantum formalism and resorting to Stokes' theorem \cite{spheremodel}. Then, we derive an analogy with transport properties on the lattice and the quantum Hall conductivity from the Bloch sphere via the Karplus-Luttinger velocity \cite{KL}, and show
the relevance of this formalism in momentum space in the presence of interactions beyond the mean-field scheme. We show that the stochastic approach successfully reproduces the jump of the topological number at the transition directly from the light responses. In Sec. \ref{stochasticnumber}, we introduce the stochastic Chern number and show its relevance for disordered situations, e.g. in the presence of fluctuations in the underlying lattice potential. The sampling on stochastic variables here corresponds to an ensemble average. Then, we discuss applications for the light response and the Mott transition and build an analogy with temperature effects. In Sec. \ref{summary}, we summarize our findings and give further perspectives. Appendices are devoted to further details on the iDMRG approach and on temperature effects. 

\section{Model}
\label{model}

\begin{figure}[ht]\centering
\includegraphics[width=0.45\textwidth]{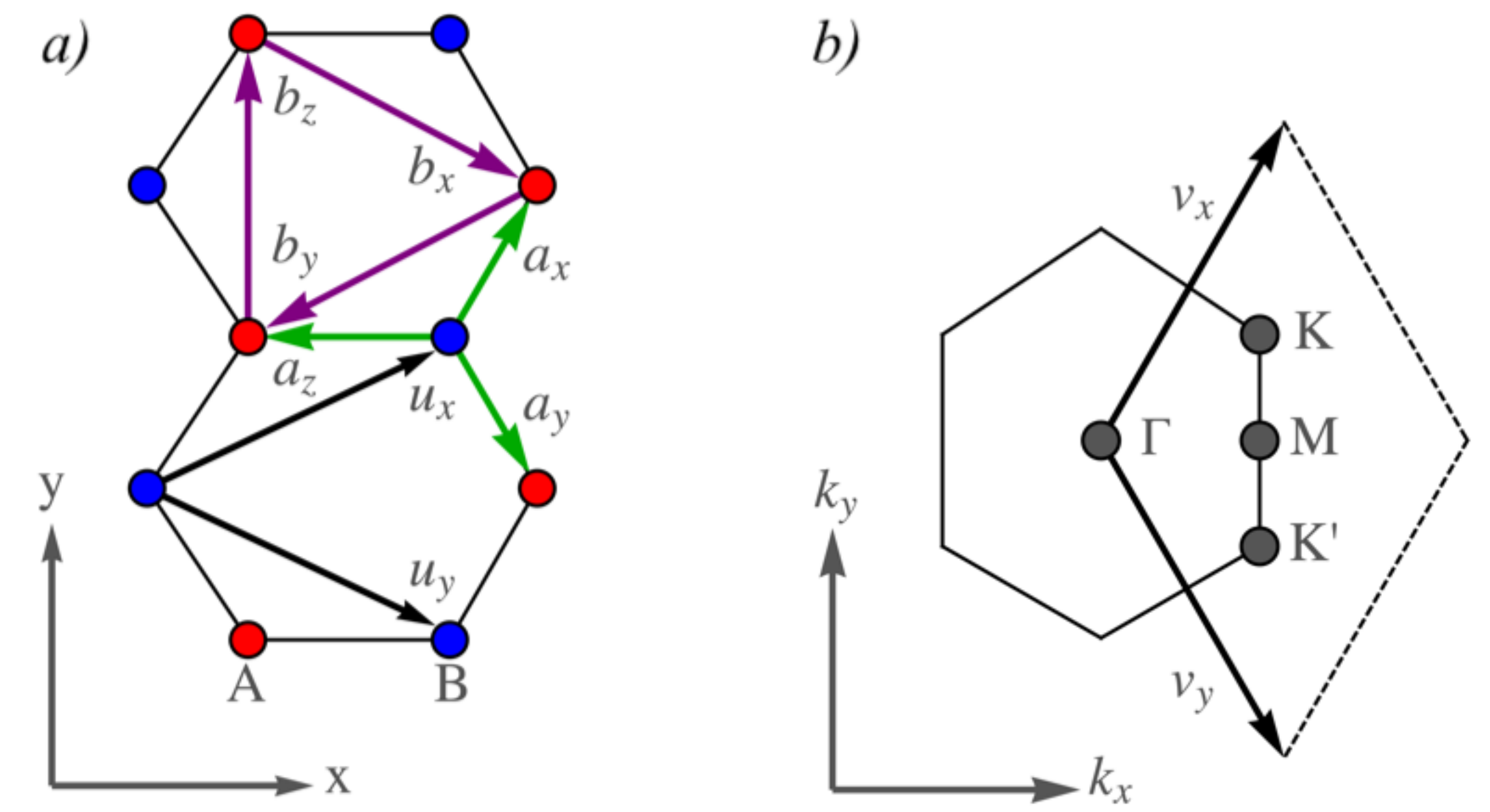}
\caption{(color online) (a) Honeycomb lattice in real space with sublattices A and B, lattice vectors, and (next-) nearest neighbor displacements. (b) Honeycomb lattice in momentum space, marking the Brillouin zone, reciprocal lattice vectors and high symmetry points.}
\label{fig_lattice_specs}
\end{figure}

Here, we introduce the model Hamiltonian $ \mathcal{H} = \mathcal{H}_0 + \mathcal{H}_V  $, where $ \mathcal{H}_0 $ is the Haldane honeycomb model for spinless fermions at half-filling \cite{haldane_1988} and $ \mathcal{H}_V $ is the nearest neighbor interaction
\begin{align}
\mathcal{H}_0 &= -\sum_{\left\langle i,j\right\rangle }t_1 c_i^{\dagger} c_j -\sum_{\left\langle\left\langle i,j \right\rangle\right\rangle} t_2 e^{\pm i \Phi} c_i^{\dagger} c_j\nonumber\\
\mathcal{H}_V &= V \sum_{\left\langle i,j\right\rangle} \left(n_i -1/2\right)\left(n_j -1/2\right).
\label{equ_model_Hamiltonian}
\end{align}
Here, $ t_1 $ represents the nearest-neighbor hopping strength which we set to unity hereafter $(t_1=1)$. Furthermore, $ t_2 e^{\pm i \Phi}$ represents the next-nearest neighbor hopping term where we fix the Peierls phase to $ \Phi=\pi/2$, for simplicity. Here, the positive (negative) sign refers to (counter-) clockwise hopping and the second nearest-neighbors are represented through the vectors $\bm{u}_i$. This set of lattice vectors in Fig. \ref{fig_lattice_specs} is given by 
\begin{equation}
{\bm u}_x = \frac{1}{2}\left(3,\sqrt{3}\right), \hspace{0.5cm} {\bm u}_y = \frac{1}{2}\left(3,-\sqrt{3}\right).
\end{equation}
The bond length has been set to unity in order to simplify the notation. The nearest neighbor displacements on the honeycomb lattice are given by the set of vectors 
\begin{equation}
{\bm a}_x=\frac{1}{2}\left(1,\sqrt{3}\right),\hspace{0.2cm}
{\bm a}_y=\frac{1}{2}\left(1,-\sqrt{3}\right),\hspace{0.2cm}
{\bm a}_z=\left(-1,0\right).
\end{equation}
We can then write the next-nearest neighbor displacements on the honeycomb lattice in terms of the $\bm{a}_p$ as $\bm{b}_i = \bm{a}_j - \bm{a}_k$ where the tuple $\left( i,j,k \right)$ is a permutation of the bond-tuple $\left( x,y,z \right)$. As in Ref. \cite{cheng_2019}, the $\bm{a}_p$ basis does not yield a Hamiltonian in Bloch form. Rather, we perform a gauge transform on the Hamiltonian to the basis, given by the lattice vectors $\bm{u}_i$, see Fig. \ref{fig_lattice_specs}.

We also introduce the definitions in the reciprocal space.
The Brillouin zone of momentum space is visualized in Fig. \ref{fig_lattice_specs}b) and shows the high symmetry M- and K-points of the Brillouin zone. Specifically, the high symmetry Dirac-points are located at 
\begin{equation}
{\bm K} = \frac{2\pi}{3}\left(1,\frac{1}{\sqrt{3}}\right), \hspace{1cm} {\bm K}^{\prime} = \frac{2\pi}{3}\left(1,-\frac{1}{\sqrt{3}}\right) \label{K-points}.
\end{equation}

\begin{figure}[t]
\includegraphics[width=8.5cm]{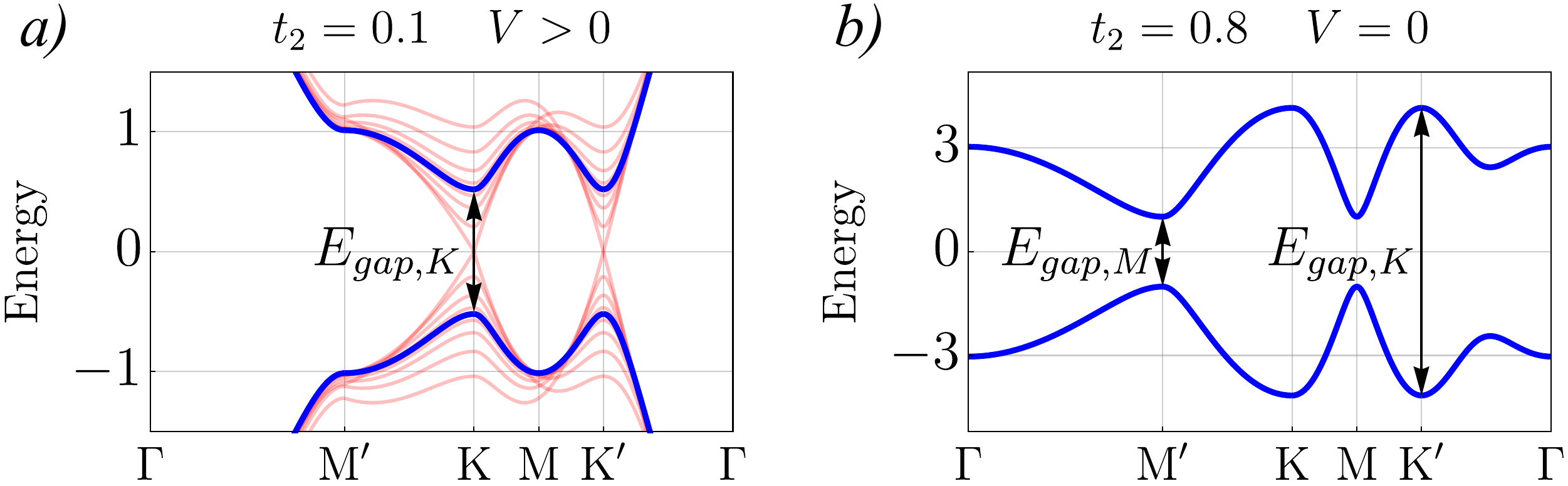}
\caption{(color online) (a) Band structure (blue pair of bands) for small $ t_2<0.2$ with $t_1=1$. The low energy physics is centered around the Dirac points, where the stochastic approach applies. If we allow for fluctuations at $ V>0 $, the sampling of $ \phi^z $ corresponds to the creation of quasi states that change the band gap at the K and K' points. (b) Haldane band structure for large $ t_2>0.2$ such that the low energy regime is located at the M-points.}
\label{fig_band_structures_all}
\vspace{-0.4cm}
\end{figure}

The energy bands in the Haldane model are represented in Fig. \ref{fig_band_structures_all}.

\section{Interacting Stochastic Topological Approach}
\label{stochappro}

Now, we introduce our stochastic approach to study interaction effects on a topological model. In this work, we study the spinless Haldane honeycomb model of Eq. (\ref{equ_model_Hamiltonian}) at half filling with nearest neighbor interactions, which is an important model giving rise to topological Bloch energy bands. We show that our approach describes correctly and quantitatively interaction effects as well as properties of the Mott transition. We highlight here that the method may be applied to other models described by a similar matrix form in the reciprocal space.

Previous studies \cite{alba_2016,dai_2010} have suggested that at a mean field level the quartic interaction term can be decoupled into a CDW order parameter which then acts as a staggered potential in sublattice space on $ \mathcal{H}_0 $. A straightforward approach, proposed in Ref. \cite{dai_2010}, would be to rewrite $ \mathcal{H}_V $ exactly as $ \mathcal{H}_V  = -\frac{V}{2}\sum_{\left\langle i,j\right\rangle} \left( n_i -n_j\right)^2$ in order to find a simple mean-field theory for the CDW order. Crucially however, this ansatz does not take into account the fact that the correlator $\langle c_i^{\dagger}c_j \rangle$, referring to the particle-hole channel \cite{plainvanilla,ReviewhighTc}, is finite in two-dimensional Hubbard models and therefore contributes to the interaction energy $\langle c_i^{\dagger}c_i c_j^{\dagger}c_j  \rangle$.  When particle-hole channels are not included, the Mott phase transition would be second-order \cite{alba_2016,dai_2010}, which is not in agreement with exact diagonalization results (at finite $t_2$) \cite{varney_2010}. 

\subsection{Stochastic Variables}

Our objective here is to introduce stochastic or Hubbard-Stratonovitch variables through the decoupling of the quartic interaction term, allowing us to derive an effective quadratic and local theory in momentum space. For this purpose, we re-write
\begin{eqnarray}
\label{eta}
\mathcal{H}_V &=& V \sum_{i,p}\left( n_i - \frac{1}{2}\right)\left( n_{i+p}-\frac{1}{2} \right) \\ \nonumber
&=& V \sum_{i,p,r} \eta_r \left( c_i^{\dagger}\sigma_{i,i+p}^r c_{i+p} \right)^2 \\ \nonumber
&-&\frac{V}{2}\sum_{i,p}\left( c_i^{\dagger}c_i + c_{i+p}^{\dagger}c_{i+p} - \frac{1}{2}\right),
\end{eqnarray}
where $ i $ denotes a unit cell, $ p $ runs over the links $ (x,y,z) $, $ r $ runs over $ (0,x,y,z) $ and $ \sigma^r $ denotes the Pauli matrices acting on sublattice space with basis $ \left( i(A),i+p(B)\right) $. The coefficients $ \eta_r $ need to fulfill the relations
\begin{equation}
\eta_0 = -\eta_z, \hspace{1cm}
\eta_x = \eta_y, \hspace{1cm}
\frac{1}{4}= \eta_0 - \eta_x.
\end{equation}
We need in principle to choose the $\eta_r$ such that the decoupling scheme incorporates particle-hole channels (i.e. $\eta_{x,y}\neq 0$) that contribute to the total energy. A priori, a generic choice of the $\eta_r$ that will ultimately minimize the total energy of the effective Hamiltonian correctly, is not obvious. Rather, a choice of coefficients $\eta_r$ needs to be justified. We can set our definitions such that $\eta^z=\eta^x$, in agreement with the variational approach presented below (The self-consistent equations (\ref{phix}) and (\ref{phiz}) will show the same normalization prefactors on the right-hand side of the equal sign). Therefore, we obtain the following choice of the $\eta_r$ variables (see also Eq. (\ref{interaction}))
\begin{equation}
- \eta_0 = \eta_x = \eta_y = \eta_z = -\frac{1}{8}.\label{equ_choice_eta_r}
\end{equation}

We can now write down the partition function and action in momentum space as
\begin{eqnarray}
\mathcal{Z} &=& \int \mathrm{D}(\Psi,\Psi^{\dagger}) e^{-\mathcal{S}}, \\ \nonumber
\mathcal{S} &=& \int_0^{\beta} \mathrm{d}\tau \sum_{\bm{k}} \Psi_{\bm{k}}^{\dagger}\left(\partial_{\tau} + \bm{h}_{0}(\bm{k})\cdot\bm{\sigma}\right)\Psi_{\bm{k}} + \mathcal{H}_V \label{equ_initial_partition_func_and_action}
\end{eqnarray}
with the spinor basis $ \Psi_{\bm{k}}^{\dagger} = \left( c_{\bm{k}A}^{\dagger},c_{\bm{k}B}^{\dagger} \right)$ such that 
\begin{eqnarray}
h_0^x({\bm{k}}) &=& -t_1\sum_p \cos(\bm{k}\cdot \bm{a}_p), \\ \nonumber
h_0^y({\bm{k}}) &=& -t_1\sum_p \sin(\bm{k}\cdot\bm{a}_p), \\ \nonumber
h_0^z({\bm{k}}) &=& -2t_2\sum_p\sin(\bm{k}\cdot \bm{b}_p).
\end{eqnarray}
Decoupling the quartic interaction term $ \mathcal{H}_V$ via a Hubbard-Stratonovich transformation  for each $ r\in\lbrace x,y,z\rbrace$ gives
\begin{widetext}
\begin{equation}
\exp\left(\frac{V}{8} \sum_{i,p} \left( c_i^{\dagger}\sigma_{i,i+p}^r c_{i+p} \right)^2\right) =
\int \mathrm{D}\phi^r \exp\left(- \sum_{i,p}2V(\phi^r_{i+p/2})^2 + V\phi^r_{i+p/2}\left(c_i^{\dagger}\sigma_{i,i+p}^r c_{i+p} \right)\right), \label{equ_Hubb_Strat_xyz}
\end{equation}
and for $ r=0 $
\begin{equation}
\exp\left(-\frac{V}{8} \sum_{i,p} \left( c_i^{\dagger}\sigma_{i,i+p}^0 c_{i+p} \right)^2\right) = \int \mathrm{D}\phi^0 \exp\left(- \sum_{i,p}2V(\phi^0_{i+p/2})^2 +i V\phi^0_{i+p/2}\left(c_i^{\dagger}\sigma_{i,i+p}^0 c_{i+p} \right)\right). \label{equ_Hubb_Strat_0}
\end{equation}
\end{widetext}
Here we introduced for each $ r $ an auxiliary field $ \phi^r_{i+p/2} $ on each link between lattice sites $ i $ (on sublattice A) and $ i+p $ (on sublattice B). The fields $ \phi^x $ and $ \phi^y $ are particle-hole channels, $ \phi^0 $ corresponds to a chemical potential and $ \phi^z $ to a staggered chemical potential in sublattice space that captures CDW order, but at the same time acts as a Semenoff mass term on the Haldane model and therefore controls the topological Chern number.

We rewrite the decoupled interaction part in Fourier space and obtain the partition function and action
\begin{widetext}
\begin{eqnarray}
\mathcal{Z} &=& \int \mathrm{D}(\Psi,\Psi^{\dagger},\phi^0,\phi^x,\phi^y,\phi^z) e^{-\mathcal{S}}, \label{equ_dec_partition_func} \\
\mathcal{S} &=& \int_0^{\beta} \mathrm{d}\tau \sum_{\bm{k}} \Psi_{\bm{k}}^{\dagger}\left(\partial_{\tau}  +\bm{h}_{0}(\bm{k})\cdot\bm{\sigma} \right)\Psi_{\bm{k}}+ \sum_{\bm{k},\bm{q},p}\Psi_{\bm{q}}^{\dagger} h_V({\bm{k},\bm{q},p})   \Psi_{\bm{k}} +  \sum_{\bm{k},r} 6V \phi^r_{\bm{k}} \phi^r_{-\bm{k}}, \label{equ_dec_action}
\end{eqnarray}
where the interaction density matrix reads
\begin{equation}
h_V({\bm{k},\bm{q},p}) = V \begin{pmatrix}
e^{-\frac{i}{2}(\bm{k}-\bm{q})\cdot \bm{a}_p}\left(i \phi^0_{\bm{k}-\bm{q}}+\phi^z_{\bm{k}-\bm{q}} \right) -\frac{1}{2} &  e^{\frac{i}{2}(\bm{k}+\bm{q})\cdot \bm{a}_p}\left( \phi^x_{\bm{k}-\bm{q}}-i\phi^y_{\bm{k}-\bm{q}} \right)\\ 
e^{-\frac{i}{2}(\bm{k}+\bm{q})\cdot \bm{a}_p}\left( \phi^x_{\bm{k}-\bm{q}}+i\phi^y_{\bm{k}-\bm{q}}\right) & e^{\frac{i}{2}(\bm{k}-\bm{q})\cdot \bm{a}_p}\left( i\phi^0_{\bm{k}-\bm{q}}-\phi^z_{\bm{k}-\bm{q}} \right)  -\frac{1}{2}
 \end{pmatrix} . \label{equ_int_matrix_initial}
\end{equation}
\end{widetext}
In principle, one could also assign an imaginary time variable $\tau$ to the stochastic variables. Then this would result in a frequency dependence of the variables $\phi^r_{\bm{k}}$. 
Below, we develop a variational approach to evaluate the stochastic variables within the ground state properties through an energy-minimization protocol. Therefore, we consider below a time-independent, static model and therefore restrict the analysis to the zero frequency contribution. For ground-state observables, this stochastic variational approach is in good agreement with the iDMRG approach. As we also show below, fluctuations of the stochastic variables around their value for the minimum of energy is well controlled. Furthermore, we restrict the discussion to leading contribution in k-space for which scattering does not change momentum. Hence, we keep only the zero momentum contribution, i.e. $ \bm{k}-\bm{q}=0 $. It's important to remind that sampling stochastic variables in time results in the sign problem for the fermionic Haldane model, which justifies our present approach. It should be emphasized that to reproduce ground-state properties, one cannot ignore the particle-hole channel $\phi^x$. 

The action $ \mathcal{S} $ now takes the form
\begin{equation}
\mathcal{S} =\sum_{\bm{k}} \Psi_{\bm{k}}^{\dagger}\left( \bm{h}_{0}(\bm{k})\cdot\bm{\sigma}+\sum_{p} h_V({\bm{k},p})    \right)\Psi_{\bm{k}} +  \sum_{\bm{k},r} 6V \phi^r_{\bm{k}} \phi^r_{-\bm{k}}, \label{equ_action_zero_contributions}
\end{equation}
where the interaction density matrix now reads
\begin{equation}
h_V({\bm{k},p})  = V \begin{pmatrix}
- \left(\phi^0 +\frac{1}{2}\right) +\phi^z &  e^{i\bm{k}\cdot \bm{a}_p}\left( \phi^x-i\phi^y \right)\\ 
e^{-i\bm{k}\cdot \bm{a}_p}\left( \phi^x+i\phi^y\right) &  -\left(\phi^0  +\frac{1}{2}\right)-\phi^z
 \end{pmatrix} .\label{equ_int_matrix_zero_contributions}
\end{equation}
Here we skipped the zero-momentum index of the fields, i.e. $ \phi^r \equiv \phi^r_0 $, and redefined the chemical potential $-i\phi^0\rightarrow \phi^0$ such that $\phi^0$ is now real for the matrix $h_V({\bm{k},p})$ to be Hermitian (where it was imaginary before the substitution, such that $ i\phi^0 $ was real).\\
We set $\mathcal{H}_{\mathrm{mf}} ({\bm{k}} ) = \bm{h}_{0}(\bm{k})\cdot\bm{\sigma}+\sum_{p} h_V({\bm{k},p}) $, and finally arrive at the effective mean field Hamiltonian $ \mathcal{H}_{\mathrm{mf}}=\sum_{\bm{k}}\Psi_{\bm{k}}^{\dagger} \mathcal{H}_{\mathrm{mf}} ({\bm{k}} )\Psi_{\bm{k}} $ where the mean field Hamiltonian density in matrix form reads
\begin{equation}
\mathcal{H}_{\mathrm{mf}} ({\bm{k}} )= 
\begin{pmatrix}
\gamma(\bm{k}) - 3V(\phi^0 +\frac{1}{2}) &  -g(\bm{k})\\ 
 -g^{\ast}(\bm{k}) & -\gamma(\bm{k})- 3V(\phi^0 +\frac{1}{2})
 \end{pmatrix} ,
\label{equ_effective_ham_mft}
\end{equation}
with the functions $ \gamma(\bm{k}) $ and $ g(\bm{k}) $ defined as 
\begin{align}
 \gamma(\bm{k}) &= 3V\phi^z -2t_2\sum_p\sin(\bm{k}\cdot \bm{b}_p),\label{equ_gamma_func}\\
 g(\bm{k}) &= \left[t_1-V(\phi^x+i\phi^y)\right]\sum_p\left( \cos(\bm{k}\cdot \bm{a}_p)-i \sin(\bm{k}\cdot \bm{a}_p)\right). \label{equ_g_func}
\end{align}
The vectors ${\bm a}_p$ and ${\bm b}_p$ are the ones from Sec. \ref{model}. 
The term $ 3V\phi^z $ assumes the role of a Semenoff mass term \cite{Semenoff} in the Haldane model, whereas the fields $ \phi^x $ and $ \phi^y $ renormalize the nearest neighbor hopping amplitude $ t_1 $.

The field $\phi^0$ can be absorbed in the chemical potential and will be fixed to $ \phi^0=-1/2 $ at half-filling. The field $ \phi^z $ changes sign in sublattice space and therefore plays the role of a staggered chemical potential. On the one hand, it measures the particle density difference between sublattices A and B, and captures CDW order. Furthermore, it acts as a Semenoff mass term \cite{Semenoff} on the Haldane model and therefore controls the Chern number of the system \cite{haldane_1988}. The variables $\phi^x$ and $\phi^y$ dress the nearest-neighbor hopping term and assuming $t_1$ is real then this favors $\phi^y=0$ while $\phi^x\neq 0$. The $\phi^z$ variable is also real in the definition of the Hubbard-Stratonovitch transformation. 

\subsection{Self consistent mean field equations from a variational approach}

Before deriving the self-consistent equations of the mean field Hamiltonian Eq. (\ref{equ_effective_ham_mft}), we provide a general remark on the derivation of self-consistent mean field equations. \\
Consider some general Hamiltonian $ \mathcal{H} = \mathcal{H}_t + \mathcal{H}_{Int} $ with a quadratic, kinetic part $ H_t $ and a quartic interaction part of the form $H_{Int}=-\sum_{i,j}U_{ij} c_i^{\dagger}c_i c_j^{\dagger} c_j \equiv -\sum_{i,j}n_i U_{ij} n_j $ with interaction matrix $ U_{ij} $. The quartic term can be decoupled by means of a Hubbard-Stratonovich transformation as \cite{altland,schulz_1994}
\begin{equation}
\exp\left( n_i U_{ij} n_j \right) = \int \mathrm{d}\phi \exp\left(  -\phi^i U_{ij} \phi^j - 2\phi^i U_{ij}n_j  \right) , \label{equ_HS-trafo_general}
\end{equation}
where we introduced some Gaussian auxiliary variable $\phi$. From the partition function and action
\begin{eqnarray}
\mathcal{Z} &=& \int \mathrm{D}(c,c^{\dagger},\phi) \exp\left(-\mathcal{S} \right), \\ \nonumber
\mathcal{S} &=& \int_0^{\beta} \mathrm{d}\tau \sum_{i,j} c_i^{\dagger} \left( \partial_{\tau} + h_t \right) c_j + \phi^i U_{ij} \phi^j + 2\phi^i U_{ij}n_j, 
\end{eqnarray}
one then usually \citep{altland,coleman_2015} computes the self consistent mean field equations via
\begin{equation}
\left\langle \frac{\delta \mathcal{S}}{\delta	\phi^i} \right\rangle \stackrel{!}{=} 0, \label{equ_saddle_point_condition} 
\end{equation}
which would yield in the above example 
\begin{equation}
0 = \left\langle U_{ij}\phi^j +2 U_{ij}n_j\right\rangle \Rightarrow \phi^j = -2\left\langle n_j \right\rangle \label{equ_saddle_point_sol_example}.
\end{equation}
Now, it's important to highlight that this result is not unique. The auxiliary field $ \phi$ can be thought of as a gauge field. Essentially, we can make a transformation as $ \phi^i \rightarrow \alpha\phi^i $ in Eq. (\ref{equ_HS-trafo_general}) with some factor $ \alpha $ to obtain
\begin{equation}
\mathcal{S} = \int_0^{\beta} \mathrm{d}\tau \sum_{i,j} c_i^{\dagger} \left( \partial_{\tau} + h_t \right) c_j + \alpha^2\phi^i U_{ij} \phi^j + 2\alpha\phi^i U_{ij}n_j.  
\end{equation}
This produces the self-consistent mean field equation
\begin{equation}
\left\langle \frac{\delta \mathcal{S}}{\delta	\phi^i} \right\rangle  \stackrel{!}{=}  0, \hspace{1cm}\Rightarrow \hspace{1cm} \phi^j = -\frac{2}{\alpha}\left\langle n_j \right\rangle.
\end{equation}
Hence, the self-consistent mean field equation depends on $ \alpha $ and is therefore not gauge independent. The problem arises, as we only minimize the action (or energy) of the decoupled, $ \phi $-dependent Hamiltonian. Instead, we need to minimize the energy of the decoupled Hamiltonian (which can be seen as a choice of a trial Hamiltonian) with respect to the original, quartic Hamiltonian. This can be done in the following way. 

Let $ \mathcal{H}_{\mathrm{mf}} $ be (a choice of) a mean field or trial Hamiltonian and $\mathcal{H}$ the original, full Hamiltonian. Then, we can rewrite formally $\mathcal{H} = \mathcal{H}_{\mathrm{mf}}  + \left( \mathcal{H} - \mathcal{H}_{\mathrm{mf}} \right)$. On the level of the free energy it follows the Bogoliubov inequality \cite{agra_2006,binney_1992} 
\begin{equation}
\mathcal{F} \leq \mathcal{F}_{\mathrm{mf}}  + \left\langle\mathcal{F} - \mathcal{F}_{\mathrm{mf}} \right\rangle.
\end{equation}
The right hand side of the inequality is a function of the mean field parameters and we need minimize it with respect to $\phi$. In our case, for the full Hamiltonian $\mathcal{H}$ in Eq. (\ref{equ_model_Hamiltonian}) and the mean field Hamiltonian in Eq. (\ref{equ_effective_ham_mft}), we obtain the following set of self-consistent mean field equations
\begin{align}
\label{meanfield}
\phi^0 =& -\frac{1}{2} \left( \left\langle c_{i}^{\dagger}c_{i} \right\rangle + \left\langle c_{i+p}^{\dagger}c_{i+p} \right\rangle \right), \\
\phi^x =& -\frac{1}{2} \left( \left\langle c_{i}^{\dagger}c_{i+p} \right\rangle + \left\langle c_{i+p}^{\dagger}c_{i} \right\rangle \right), \label{phix} \\
\phi^y =& -\frac{1}{2}i \left(- \left\langle c_{i}^{\dagger}c_{i+p} \right\rangle + \left\langle c_{i+p}^{\dagger}c_{i} \right\rangle \right),\\
\phi^z =& -\frac{1}{2} \left( \left\langle c_{i}^{\dagger}c_{i} \right\rangle - \left\langle c_{i+p}^{\dagger}c_{i+p} \right\rangle \right), \label{phiz}
\end{align}
or in short hand notation using Pauli matrices
\begin{equation}
\phi^r = -\frac{1}{2}\left\langle c_{i}^{\dagger}\sigma_{ij}^r c_{j} \right\rangle.
\end{equation}
The real space amplitudes are evaluated directly, e.g. as 
\begin{eqnarray}
 \left\langle c_{i}^{\dagger}c_{i+p} \right\rangle 
&=& \frac{2}{N_\mathrm{sites}} \sum_{\bm{k}}e^{i\bm{k}\cdot\bm{a}_p}\left\langle c_{\bm{k}A}^{\dagger}c_{\bm{k}B}\right\rangle \\ \nonumber
&=& \frac{2}{N_\mathrm{sites}} \sum_{\bm{k}}\sum_{\mu^{\prime},\nu^{\prime}}e^{i\bm{k}\cdot\bm{a}_p}\mathcal{M}_{\bm{k}A\mu^{\prime}}^{\ast}\mathcal{M}_{\bm{k}B\nu^{\prime}}\left\langle \gamma_{\bm{k}\mu^{\prime}}^{\dagger}\gamma_{\bm{k}\nu^{\prime}}\right\rangle\label{equ_munu_run}\\ \nonumber
&=& \frac{2}{N_\mathrm{sites}} \sum_{\bm{k}}\sum_{\lambda}e^{i\bm{k}\cdot\bm{a}_p}\mathcal{M}_{\bm{k}A\lambda}^{\ast}\mathcal{M}_{\bm{k}B\lambda}, \label{equ_lambda_run}
\end{eqnarray}
and $N_\mathrm{sites}$ refer to the number of sites in the system, i.e. $N_\mathrm{sites}/2$ refer to the number of unit cells. In the first line, we performed a Fourier transform of the creation and annihilation operators in real space. In the second line, we used $ \gamma_{\bm{k}}= \mathcal{M}_{\bm{k}}^{\dagger}\psi_{\bm{k}} $ where $ \mathcal{M}_{\bm{k}} $ is a unitary matrix that diagonalizes 
$\mathcal{H}_{\mathrm{mf}}$. The new spinor basis fulfills $\left\langle \gamma_{\bm{k}\mu^{\prime}}^{\dagger}\gamma_{\bm{k}\nu^{\prime}}\right\rangle =\delta_{\mu^{\prime}\nu^{\prime}}$ for occupied states. In Eq. (\ref{equ_munu_run}), $ \mu^{\prime } $ and $ \nu^{\prime} $ run over all states, whereas $ \lambda $ in Eq. (\ref{equ_lambda_run}) runs only over occupied states. The results below are obtained when solving the coupled equations above.

\subsection{Stochastic and iDMRG results}

\begin{figure}[t]
\includegraphics[width=8.4cm]{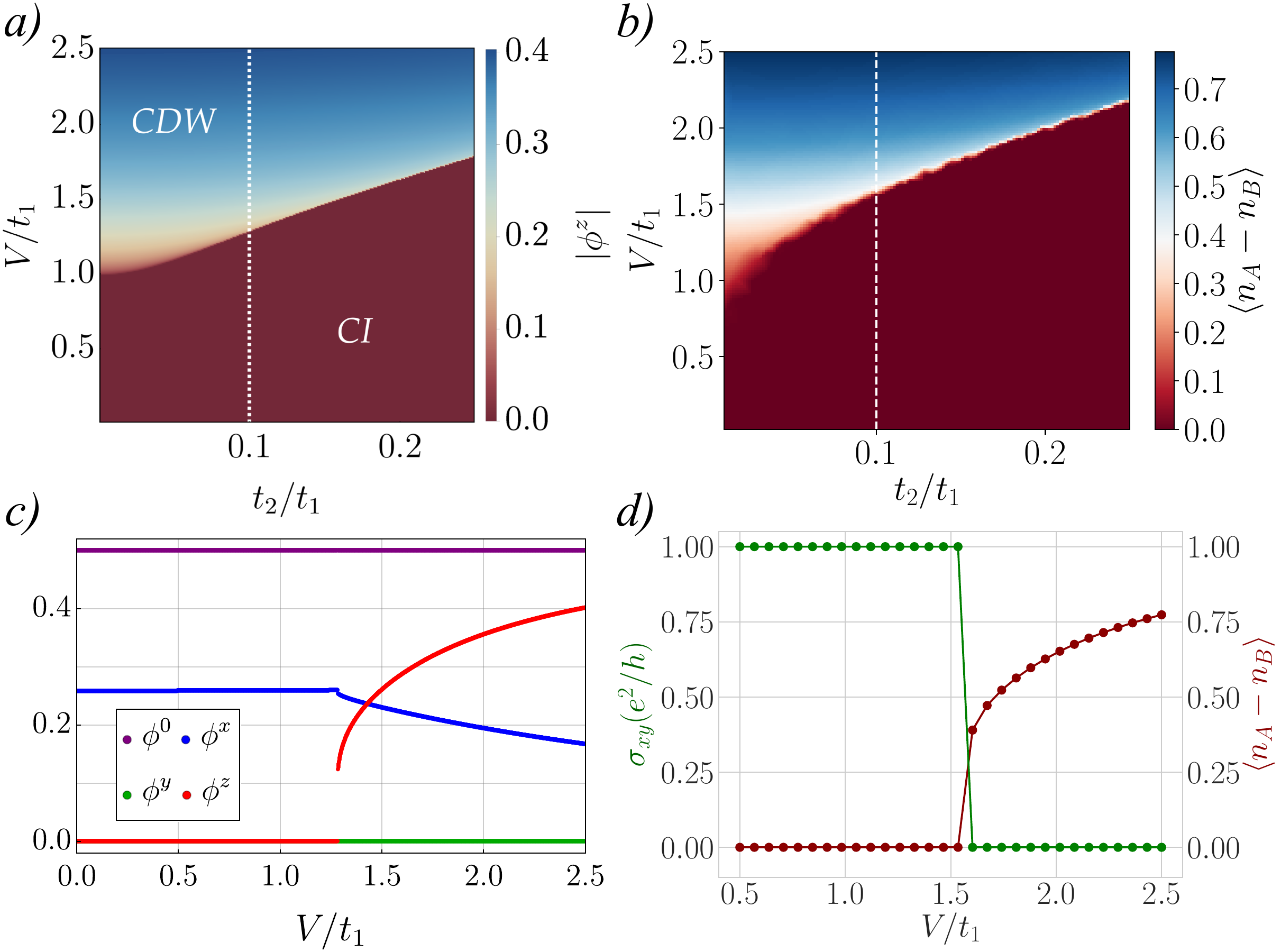}
\caption{(color online) (a) $V-t_2$ mean field phase diagram from the method. The transition marks the condensation of the CDW order parameter $\phi^z$. (b) Same phase diagram obtained with iDMRG. (c)
Absolute value of the self-consistent $\phi^r$ variable as a function of $V$ for $t_2=0.1$. (d) Hall conductivity and $\langle n_A-n_B\rangle$ from iDMRG.}
\label{fig_phase_diagrams_all}
\vspace{-0.4cm}
\end{figure}

In Fig. \ref{fig_phase_diagrams_all}a), we present the two-dimensional $t_2 - V$ phase diagram obtained with a variational approach on the mean-field theory. We confirm the presence of two phases \cite{varney_2010,vasic_2015}, a CI phase with a perfectly quantized Chern number and a Mott or CDW phase. The CDW phase is characterized by a non-zero value of $\langle n_A-n_B\rangle$ or $\phi^z$; as long as $\langle n_A-n_B\rangle$ is not equal to unity, then $\phi^x$ can remain finite above the transition as a result of quantum fluctuations. Fig. \ref{fig_phase_diagrams_all}c) shows the numerical solution of the mean field equations for $t_2=0.1$. The jump in the CDW order parameter $\phi^z$ indicates the first-order phase transition. 

Now, we give some physical insight on the occurrence of a jump in $\phi^z$, which is evaluated at the wave-vector ${\bf k}-{\bf q}={\bf 0}$. At the Mott transition, the gap closes at one Dirac point such that for the ground state we have $\langle n_A({\bf K}) \rangle = \langle n_B({\bf K})\rangle$ whereas the gap remains visible at the other Dirac point such that $\langle n_A({\bf K'})\rangle=1$. 
In real space, the system behaves (approximately) as if $\langle n_A-n_B\rangle \approx 1/2$ on a given unit cell and $|\phi^z|\approx 1/4$. It's relevant to highlight that the variable $\phi^z$ entering in the diagonal terms of the $2\times 2$ matrix describing $h_V$ is taken at the wave-vector ${\bf k}-{\bf q}=0$ instead of a Dirac point, corresponding then to an average on all the unit cells of the
lattice in real space. This argument implies a jump in the quantum Hall conductivity at the Mott transition. The closing of the gap at the ${\bf K}$ point gives a critical interaction value $V_c\approx 4\sqrt{3}t_2$ to enter into the Mott regime, suggesting then a linear relation between $V_c$ and $t_2$ as observed in the phase diagram. In the stochastic approach, the particle-hole channel allows us to determine quantitatively $\langle n_A({\bf K}) \rangle$ and $\langle n_B({\bf K}) \rangle$ and the value of $\phi^z$ at wave-vector ${\bf k}-{\bf q}={\bf 0}$ according to Eq. (\ref{equ_lambda_run}), which then results in the phase diagram of Fig. \ref{fig_phase_diagrams_all}. The linear relation between $V_c$ and $t_2$ remains visible for the range of studied parameters.

We have compared our mean field calculations with simulations using the infinite density matrix renormalization group (iDMRG) by means of the python package \textsc{TenPy}~\cite{Hauschild:2018tz}, written in the language of matrix product states. This numerical method calculates the ground-state of the model \eqref{equ_model_Hamiltonian} in the infinite cylinder geometry, as well as the expectation of the CDW order parameter, $\left\langle n_A-n_B\right\rangle$, the Hall conductivity $\sigma_{xy}$, the correlation length $\xi$ and the entanglement entropy $S$. The bond dimension $\chi$ is a measure of the maximum number of states kept by the algorithm, and sets the accuracy of the calculation. We have performed calculations up to $\chi=1200$ for cylinder circumferences of $L_y=6, 12$ sites and our results show good convergence for bond dimensions as low as $\chi \gtrsim 200$, consistent with previous iDMRG calculations~\cite{Grushin2015a}. Additional information on the method can be found in Appendix \ref{DMRGAppendix}.

The phase diagram for $\chi =200$ and $L_y=6$ is shown in Fig.~\ref{fig_phase_diagrams_all}b). In Fig.~\ref{fig_phase_diagrams_all}d) we show the CDW order parameter and the Hall conductivity along a cut at $t_2/t_1=0.1$, which show a discontinuity along the transition for all $\chi$s. These discontinuities are typical of a first-order phase transition, further supported by the saturation of the entanglement entropy at the transition as a function of correlation length. Comparing iDMRG results with the mean-field variational approach, our findings agree as long
as the smallest band gap (relevant energy scale for CDW order) is located at the K-points (relevant for topology), which is the case for $t_2 \leq 0.2$. Therefore, we focus on this parameter regime. 

\section{Energetic analysis of the phase transition}
\label{energetics}

\begin{figure}[t]
\centering
\includegraphics[width=0.35\textwidth]{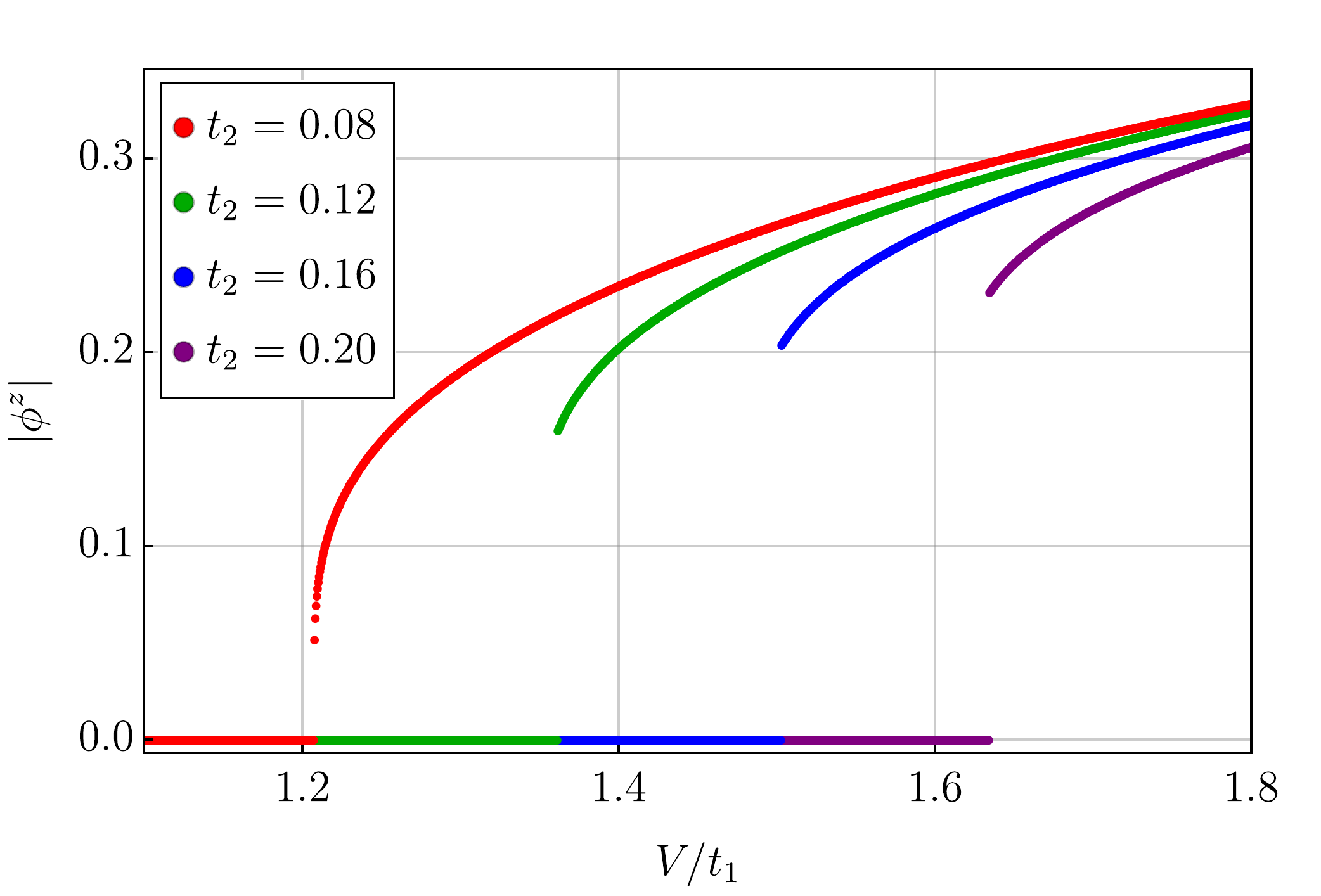}
\includegraphics[width=0.35\textwidth]{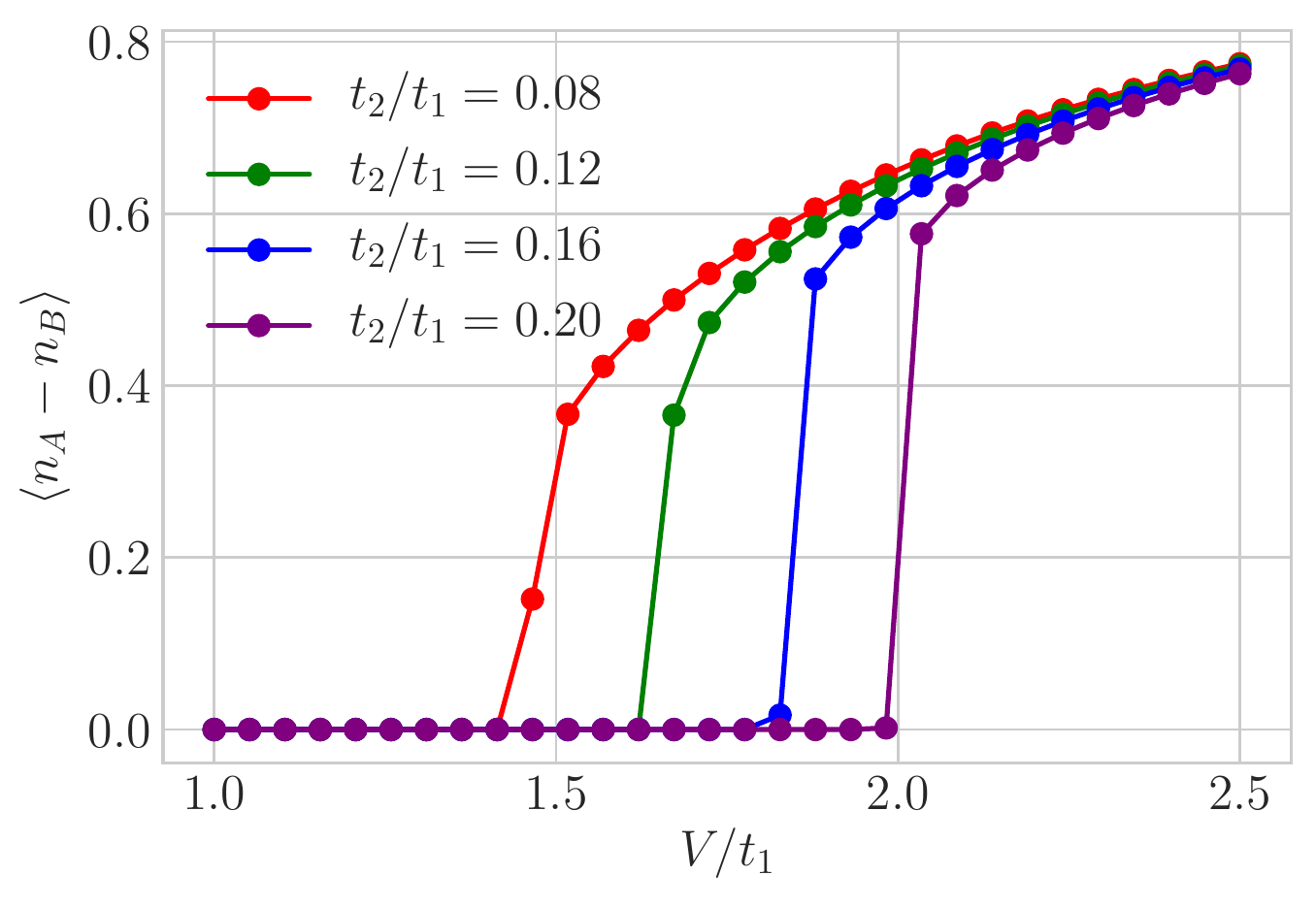}
\caption{CDW order parameter from mean field theory ($\phi^z $, top) and iDMRG ($L_y=6$, $\chi=200$, bottom) as a function of the interaction strength $V/t$ for different values of the next-nearest neighbor hopping amplitude $t_2$. In both cases the smaller $t_2$ the smaller the jump in the order parameter (in agreement with known results in the literature for $t_2=0$ \cite{Capponi_2016}). In the mean field diagram (left), we computed a solution to the self-consistent equations in small incremental steps of $\Delta V = 0.0005$ in order to show clearly the jump in the order parameter for the values of $t_2$ under investigation.}
\label{fig_CDW_order_jumps}
\end{figure}

We find at the mean field level a jump of the CDW order parameter $\phi^z$ at the phase transition for the choice of parameters' regime. In Fig. \ref{fig_CDW_order_jumps} (top), we show the CDW order parameter as a function of $V$ for different fixed values of $t_2$ ranging between $t_2=0.08$ and $t_2=0.20$. Here, the self-consistend mean equations where solved for increasing $V$ in steps of $\Delta V =0.0005$ in order to show clearly the jump in the order parameter $\phi^z$. The jump becomes smaller the smaller $t_2$ is. Therefore, at the mean field level, a clear indication of a first order phase transition can only be given when $t_2$ is sufficiently large, i.e. at the order of $t_2$. From the Ginzburg-Landau and mean field theoretical point of view a clear analysis of the nature of the mean field analysis for $t_2<0.05$ is not possible, the phase transition appears at most rather \textit{weakly} first order when $t_2$ is close to zero. This observation seems to be in accordance with the literature \cite{Capponi_2016}, where a second order phase transition is predicted for vanishing $t_2$. From our perspective, the regime of small $t_2$ seems therefore to be the middle ground between the clear indication of a first order phase transition in the range $t_2 \in (0.08, 0.20)$ and the second order phase transition for vanishing $t_2$.

In order to confirm a first order transition on the mean field level for sufficiently large $t_2$, we evaluate the total energy of system. Let $ \left\vert \Omega_{\mathrm{mf}} \right\rangle $ denote the mean field ground state which in general depends on the self-consistently obtained field $\phi^r$, i.e. $ \left\vert \Omega_{\mathrm{mf}} \right\rangle \equiv  \left\vert \Omega_{\mathrm{mf}} \right\rangle_{\bm \phi} $ . Then, we compute the energy of the system via $ \mathcal{F}(\bm{\phi})=\left\langle \Omega_{\mathrm{mf}} \right\vert \mathcal{H}\left\vert \Omega_{\mathrm{mf}} \right\rangle  \equiv \left\langle \mathcal{H}\right\rangle $ where $\mathcal{H}= \mathcal{H}_0 + \mathcal{H}_V  $ is the original Hamiltonian Eq. (\ref{equ_model_Hamiltonian}). This calculation involves exactly decomposing the quartic term $ n_i n_j = c_i^{\dagger} c_i c_j^{\dagger} c_j $ using Wick's theorem \cite{altland} as 
\begin{equation}
\langle c_i^{\dagger} c_i c_j^{\dagger} c_j \rangle = \langle c_i^{\dagger} c_i \rangle  \langle c_j^{\dagger} c_j \rangle
-\langle c_i^{\dagger} c_j^{\dagger} \rangle  \langle c_i  c_j \rangle
-\langle c_i^{\dagger} c_j \rangle  \langle c_j^{\dagger} c_i \rangle .
\end{equation}
The amplitudes such $  \langle c_i^{\dagger} c_i \rangle  $ are then evaluated similarly to the computation leading to Eq. (\ref{equ_lambda_run}).

\begin{figure}[t]
\centering
\includegraphics[width=0.48\textwidth]{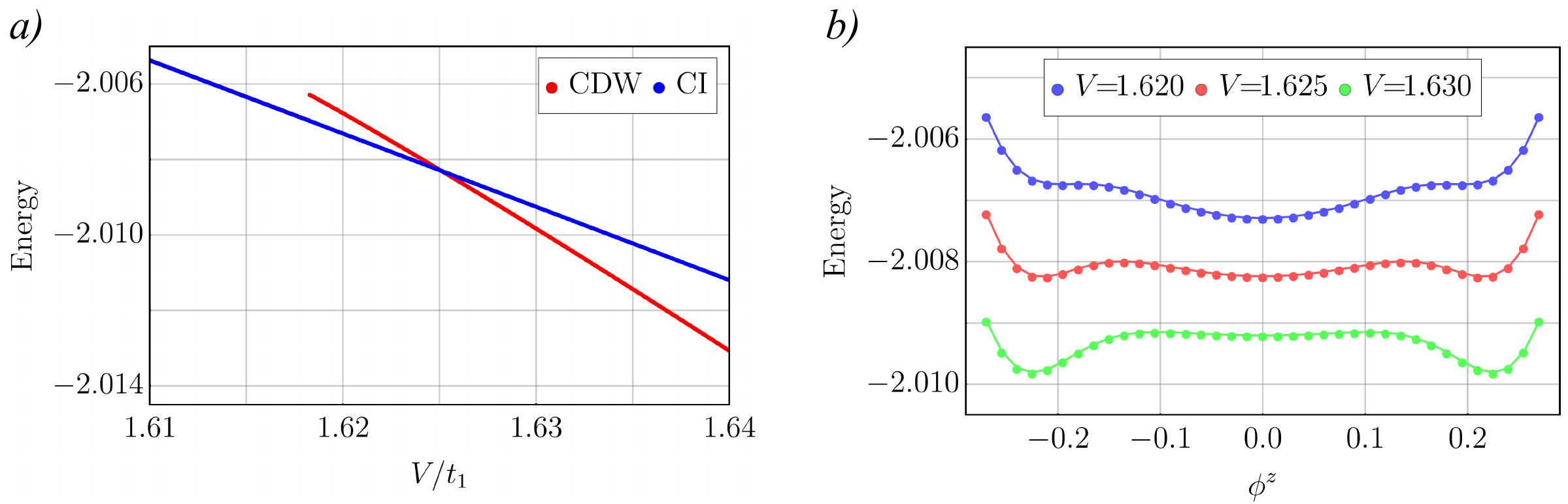}
\caption{(color online) (a) Energy of the CI and CDW Mott phases obtained from mean field theory at $t_2=0.2$. The curves cut in one point, forcing the CDW order parameter to jump as the system abruptly prefers to change the phase in order to minimize energy. (b) Energy landscape around the mean field solution at $t_2=0.2$ as a function of the CDW order parameter $\phi^z$ at the phase transition. The coexistence of local minima indicates a first order transition according to Ginzburg-Landau theory.}
\label{fig_energy_mft}
\end{figure}

Evaluating the energy in both phases around the transition shows that the energy curves cross at the transition line, see Fig. \ref{fig_energy_mft}a) for $t_2=0.2$. This indicates a first order transition as the parameter will jump at the transition to the energetically preferable solution.\\
This can be further confirmed by computing the energy explicitly for small $\phi^z$ around the saddle-point solution right before the phase transition (also for $t_2=0.2$). The curve obtained, Fig. \ref{fig_energy_mft}b) shows a typical Mexican hat form \citep{altland} with co-existing minima. We build a Ginzburg Landau theory, i.e. an expansion of the free energy curve. 
Finding appropriately relevant terms until the order $(\phi^z)^6$ is a difficult task here because $V$ is large as well as $\phi^x$, and therefore we perform this task  numerically. The free energy can be approximated by a polynomial of the form
\begin{equation}
    \mathcal{F} (\phi^z) = \mathcal{F}_0 +  \alpha ( \phi^z) ^2+ \beta ( \phi^z) ^4 + \gamma ( \phi^z) ^6, \label{equ_GL_expansion}
\end{equation}
where the coefficients fulfill in general \cite{hohenberg_2015} $ \alpha >0 $, $ \beta <0 $, and $ \gamma >0 $ to ensure the co-existence of local minima and that the free energy is bounded from below. We fit such a polynomial to the energy computed very close to the phase transition for different values of $t_2$. The results are shown in Table \ref{tab_ginzburg_landau}. In general, it is difficult to compare these coefficients for different values of $t_2$. For each value of $t_2$ we need to fix a $V$ that is close to the phase transition in order for Eq. \ref{equ_GL_expansion} to be valid. Varying $V$ in the vicinity of the phase transition slightly, i.e. moving either towards the phase transition or away from it, may change the magnitude of the coefficients in Table \ref{tab_ginzburg_landau}. However, we can make a statement on the signs of the coefficients. Since we get across all values of $t_2$ a consistent configuration of $ \alpha >0 $, $ \beta <0 $, and $ \gamma >0 $, we can confirm the first order nature of the phase transition  \cite{hohenberg_2015}.

\begin{table}
\begin{widetext}
    \centering
    \begin{tabular}{C{2.5cm} |C{1.5cm} C{1.5cm} C{1.5cm} C{1.5cm} }
        & $\mathcal{F}_0$  & $\alpha$ & $\beta$ & $\gamma$ \\
       \hline
        $t_2=0.20$ and $V=1.6250$ &-2.0913 & 0.0402 & -1.5045 & 15.6428 \\
        $t_2=0.08$ and $V=1.2074$ &-1.8445 & 0.0001 & -0.2575 & 70.7487 \\
        $t_2=0.12$ and $V=1.3610$ &-1.8974 & 0.0010 & -1.5419 & 31.4482 \\
        $t_2=0.14$ and $V=1.4250$ &-1.9232 & 0.0212 & -1.2630 & 6.7646 \\
    \end{tabular}
    \caption{Ginzburg-Landau coefficients (of the polynomial Eq. \ref{equ_GL_expansion}). The coefficients for different values of $t_2$ are in general difficult to compare since we need for each $t_2$ to fix some $V$ manually close to the phase transition, and the coefficients are subject to change in magnitude when only moving slightly towards the phase transition or away from it. Comparing signs is however possible, and the configuration at hand ($ \alpha >0 $, $ \beta <0 $, and $ \gamma >0 $) determines a first order phase transition \cite{hohenberg_2015}.}
    \label{tab_ginzburg_landau}
    \end{widetext}
\end{table}

\begin{figure}[t]
\centering
\includegraphics[width=0.4\textwidth]{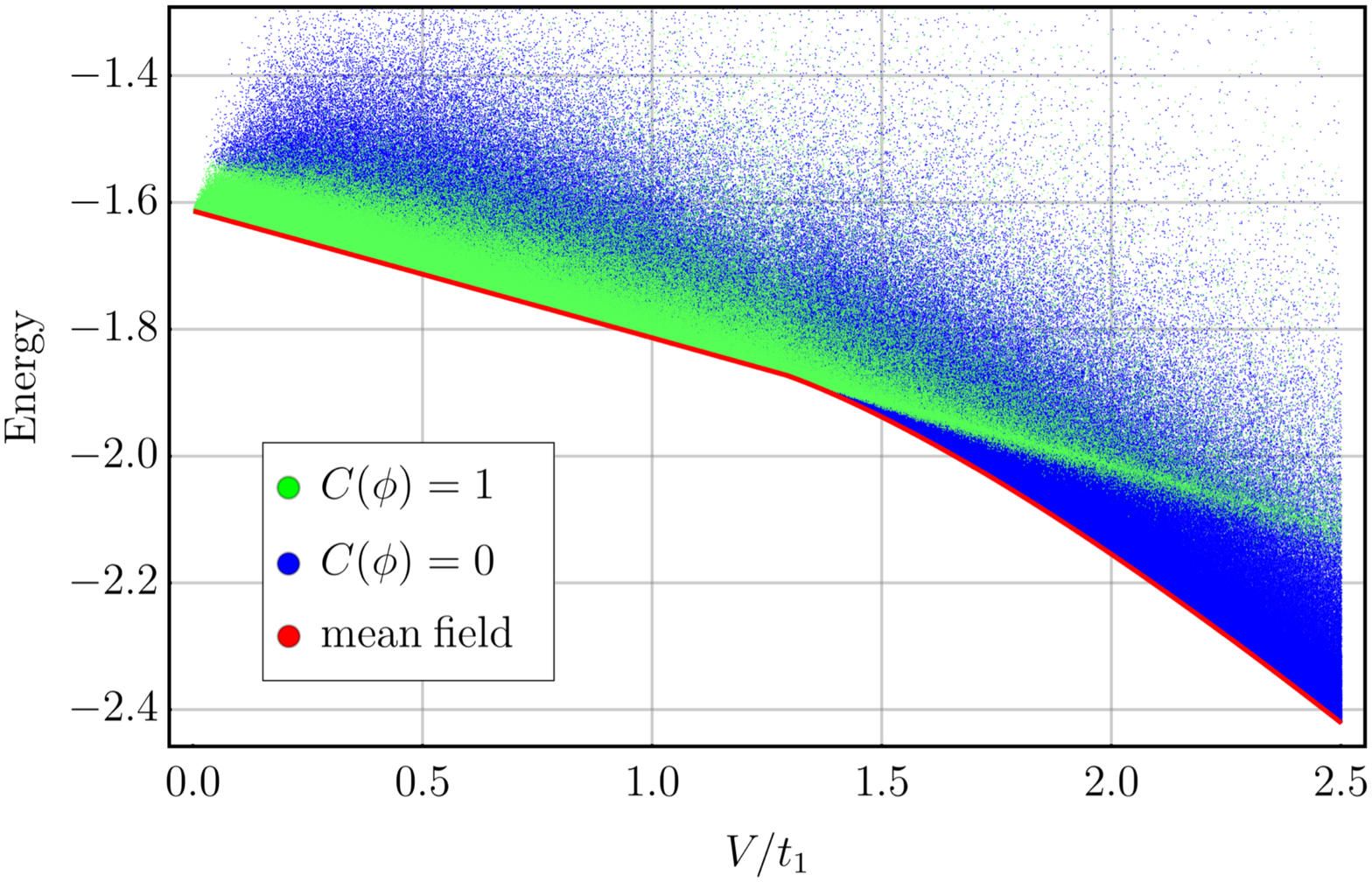}
\caption{(color online) Energy of the mean field ground state as function of $V$ (red line). Also shown is the energy distribution of quasi states obtained from sampling $(\phi^x,\phi^y,\phi^z)=\bm{\phi}$ around the saddle point. This creates quasi-excited states at energies higher than the mean field ground state. Each quasi-state can be attributed a Chern number $C(\bm{\phi})$ which will be either one (green) or zero (blue).}
\label{fig_energy_dist}
\end{figure}
Furthermore, note that a plot of the energy landscape around a mean field solution such as Fig. \ref{fig_energy_mft}b) is an important tool to check the validity of the mean field theory. If relevant mean field parameters are omitted, their weight is not correctly adjusted (the parameters $ \eta_r $ introduced above) in the self-consistent equations and the energy curve will not show a minimum.

Finally, we would like to comment on the stochastic approach to the interacting Chern insulator from the energetic point of view. Allowing the mean field parameters $\phi^r$ to fluctuate around the saddle point solution changes the energy of the quasi state under consideration. We sample the fields $\phi^{x,y,z}$ and for each configuration we can compute the energy of this quasi state with respect to the wave function $ \left\vert \Omega \right\rangle \equiv \left\vert \Omega (\phi^x,\phi^y,\phi^z) \right\rangle$. Repeating this procedure for $10^3$ sampled configurations of the $\phi^{x,y,z}$ for each respective $V$ yields Fig. \ref{fig_energy_dist}. The red line gives the energy of the mean field ground state which is the lowest energy state for each $V$. Sampling the $\phi^p$ fields will result in a quasi state at a higher energy. Each quasi state can be associated with a Chern number of either one or zero (depending on $\phi^z$). For small $V$, the mean field ground state (red line) as well as states close to it have Chern number one. Close to the phase transition, it then becomes more likely to create a state with Chern number zero when moving away from the saddle point. At the phase transition, the ground state acquires Chern number 0 and for further increasing $V$ it becomes more and more unlikely to create an excited state with non-trivial topology. This analysis also shows the occurrence of a jump in the topological Chern number at the phase transition, from the ground state, as obtained with iDMRG in Fig. \ref{fig_phase_diagrams_all}d).

\section{Circular Dichroism of Light}
\label{circularJones}

Here, we address the response of the CI to circularly polarized light with different polarizations \cite{goldman_2017}. We show that the light response can detect the Mott transition from the ground state, through mathematical arguments and through the stochastic approach. 

For clarity sake, this section
is organized as follows. First, resorting to a Bloch sphere argument where the light-matter coupling is defined through the introduction of a time-dependent vector potential in Eq. (\ref{equ_effective_ham_mft}) above, we show that topology can be described from the light-matter coupling at the K and K' points only. The depletion rates $\Gamma^{\pm}_{l \rightarrow u} (\bm{k}=\bm{K},\bm{K}')$ encode the mass term $\vert m \vert =6\sqrt{3}t_2$ which determines the size of the band gap at the K-points. The next-nearest neighbor hopping term $t_2 e^{i\Phi}$ breaks time-reversal symmetry, and leads to different signs of $m$ at the K and K' points, and therefore, to non-trivial topology \cite{haldane_1988}. 
We find a relation with the quantum Hall conductivity from the Bloch sphere and the Karplus-Luttinger velocity \cite{KL}. 
Relating to transport properties, then we show the relation between the Bloch sphere arguments and the depletion rates as derived in Ref. \cite{goldman_2017} from Fermi's golden rule:
\begin{equation}
\Gamma^{\pm}_{l \rightarrow u}\left( \omega_{\bm{k}},\bm{k} \right) = \frac{2\pi}{\hbar}\left(\frac{E}{\hbar\omega}\right)^2\left\vert \mathcal{A}^{\pm}_{l\rightarrow u} \right\vert^2  \delta\left( \epsilon_u^{\bm{k}} -\epsilon_l^{\bm{k}} -\hbar\omega \right) 
\label{equ_gamma_goldman}
\end{equation}
and $\Gamma^{\pm}_{l \rightarrow u}\left( \omega_{\bm{k}} \right)= \sum_{\bm k \in \mathrm{BZ}} \Gamma^{\pm}_{l \rightarrow u}\left( \omega_{\bm{k}},\bm{k} \right)$. Here, the transition amplitude is given by $ \mathcal{A}^{\pm}_{l\rightarrow u} = \langle u_{\bm{k}} \vert \frac{1}{i}\frac{\partial \mathcal{H}_0}{\partial k_x}\mp\frac{\partial \mathcal{H}_0}{\partial k_y}\vert l_{\bm{k}}\rangle$, $E$ is the strength of the drive or the electric field in the original basis, $\vert u_{\bm{k}} \rangle$ and $ \vert l_{\bm{k}} \rangle$ are the eigenstates corresponding to the lower and upper bands, $\epsilon_{l,u}^{\bm{k}}$ their eigenenergies, and the $\pm$ selects the polarization orientation. 
Within the Fermi golden rule, we show that it's equivalent to sum on all the momenta $\bm{k}$ in the entire Brillouin zone as in Ref. \cite{goldman_2017}, or just consider the Dirac points only.  

We also check numerically that we can evaluate the Chern number with the formula at the Dirac points only, and find for the frequency-integrated rates
\begin{equation}
\frac{1}{2}\int_0^{\infty}\mathrm{d}\omega\sum_{\bm{k}=\bm{K},\bm{K}'} \left(\Gamma^+_{l \rightarrow u}\left(\omega_{\bm{k}},\bm{k}\right)-\Gamma^-_{l \rightarrow u}\left(\omega_{\bm{k}},\bm{k}\right)\right)
= \rho C
\label{equ_depletion_K}
\end{equation}
with the constant $\rho= 16\pi^3 E^2 \sqrt{3}\left\vert t_1\right\vert^2 m^{-2}$. In the non-interacting case, $C$ is one in the topological non-trivial phase of the Haldane model and exactly zero otherwise, and is thus the (ground state) Chern number. The prefactor $1/m^2$ comes from Eq. (\ref{equ_gamma_goldman}) written in terms of $E$. 
Then, we provide general arguments related to interactions and transport in momentum space to show that Eq. (\ref{equ_gamma_goldman}) is also applicable in the presence of interactions. Through the stochastic approach, we verify that the light responses can probe the jump in the Chern number at the Mott transition for the ground state. 

\subsection{Poincar\' e-Bloch sphere formalism}

We start our analysis from the $2\times 2$ matrix form for the Hamiltonian:
\begin{equation}
\label{Haldane}
{\cal H}_0(\bm{k}) = \begin{pmatrix}
                              \gamma({\bm k}) & -g({\bm k}) \\
                              -g^*({\bm k}) & - \gamma({\bm k})
 			  \end{pmatrix} \quad,
   \end{equation}
which is the pure Haldane model with $\phi^r=0$.  Its energy eigenvalues read
   \begin{equation}
   (\epsilon_u^{\bm k} , \epsilon_l^{\bm k}) = \pm \sqrt{\gamma^2(\bm{k}) + |g({\bm k})|^2} \equiv \pm \epsilon^{\bm k},
   \end{equation}
where the positive sign refers to the upper band and the minus sign to the lower band. In analogy with a spin-1/2 Hamiltonian described by a polar angle $\theta$ and azimuthal angle $\varphi$ we can parametrize the Hamiltonian as
   \begin{equation}
   \frac{\gamma({\bm k})}{\epsilon^{\bm k}} = \cos \theta,  \hspace{1cm}
   -\frac{g^*({\bm k})}{\epsilon^{\bm k}} = e^{i\varphi}\sin\theta.
   \label{path}
   \end{equation}
From quantum mechanics and the definition of eigenstates of a spin-1/2 particle, the two eigenstates $|u\rangle$ and $|l\rangle$ referring to upper and lowest bands respectively, take the form
   \begin{eqnarray}
   \vert u\rangle &=& e^{-i\varphi/2}\cos\left(\theta/2\right) \vert a\rangle + e^{i\varphi/2}\sin\left(\theta/2 \right)\vert b\rangle,  \\ \nonumber
    \vert l\rangle &=& -e^{-i\varphi/2}\sin\left(\theta/2\right) \vert a\rangle + e^{i\varphi/2}\cos\left(\theta/2\right) \vert b\rangle.
   \end{eqnarray}
Here, the kets $|a\rangle$ and $|b\rangle$ refer to the states acting on sublattice $A$ and sublattice $B$, respectively, for a given wavevector. To define the mass inversion effects associated to the light-matter response, 
we introduce the following notations for $\epsilon_u$ and $\epsilon_l$ at the two Dirac points. Starting from the north pole and changing $\theta$ by $\pi$, then reaching the south pole, is also equivalent in Eq. (\ref{path}) to change $\epsilon^{\bm k}\rightarrow -\epsilon^{\bm k}$ and therefore to invert the definitions of lower and upper bands. Therefore, at the two Dirac points, we have $g({\bm K}/\bm{K}^{\prime})=0$ and we will use below the definitions $\epsilon_u({\bm K}) = +m/2 = - \epsilon_u({\bm K'})$ in the light response. Similarly, we have the important mass identifications $\epsilon_l({\bm K}) = -m/2 = - \epsilon_l({\bm K'})$. 
  \\
For the light-matter coupling, here we use a formulation written in terms of the vector potential. Using the spin Pauli matrix representation then this is analogous to the modification in the Dirac Hamiltonian $({\bm p}+{\bm A}e/c)\cdot \mathbf{\sigma}$ where we set $e $ and $c$ to unity. In the circular basis, the vector potential is defined as ${\bm A} = A_0 e^{-i\omega t} ({\bm e}_x \mp  i {\bm e}_y)$ where ${\bm e}_x$ and ${\bm e}_y$ are unit vectors associated with the $x$ and $y$ directions. The signs in $({\bm e}_x \mp  i {\bm e}_y)$ refer to right-handed $(+)$ and left-handed $(-)$ polarizations respectively. With such definitions, we obtain the light-matter Hamiltonian:
\begin{equation}
\label{Hmodel}
\delta{\cal H}_{\pm} = A_0 e^{\pm i\omega t} |a\rangle \langle b| + h.c.
\end{equation}
This is also equivalent to 
\begin{widetext}
\begin{equation}
\delta \mathcal{H}_{\pm} = A_0 \cos(\omega t) \left( |a\rangle \langle b| + |b\rangle \langle a| \right) 
\mp \frac{A_0}{i}\sin(\omega t) \left( |a\rangle \langle b| - |b\rangle \langle a| \right).
\end{equation}
\end{widetext}
To obtain these equations, we have considered the real parts of the vector potential components $A_x(t)$ and $A_y(t)$ to ensure that the Hamiltonian is hermitian. The speed
at the Dirac points is absorbed in $A_0$ assuming that $t_2<0.2t_1$. 
Within this formulation, changing the sense of circulation for light is equivalent to change $\omega\rightarrow -\omega$, and the subscript $\pm$ refers to a Jones polarization of light. We can then rewrite $\delta \mathcal{H}_{\pm}$ in the band basis (which diagonalizes the Hamiltonian), and we obtain two classes of terms which can be then studied by mapping the Brillouin zone onto the Bloch sphere
\begin{widetext}
\begin{equation}
\delta \mathcal{H}_{\pm} = A_0\sin\left(\theta\right) \cos\left(\omega t \pm \varphi\right)\left( \vert u\rangle \langle u\vert - \vert l\rangle \langle l\vert \right)
+ \left[A_0 (e^{\pm i \omega t} e^{i \varphi} \cos^2\left( \theta/2\right) - \sin^2\left( \theta/2\right) e^{\mp i\omega t} e^{-i\varphi})\vert u\rangle\langle l \vert +h.c.\right].
\end{equation}
\end{widetext}
The first term produces a time-dependent shift of the chemical potential in the original Haldane model, and therefore to ensure the validity of the topological
phase, we require that $A_0$ is small such that the chemical potential is in the band gap. At the poles, the first term is precisely zero. 
Now, we can apply Fermi's golden rule argument for the inter-band transitions (at the two poles). 
The transition rates $\Gamma_+({\bf k})$ and $\Gamma_-({\bf k})$ are defined such that the two terms in $\delta{\cal H}_{\pm}$ proportional to $|u\rangle \langle l|$, for a given light polarization, correspond to an absorption of a light quantum with energy $\hbar\omega$ and $-\hbar\omega$ respectively and we assume a sampling on frequencies from $-\infty$ to $+\infty$ as realized below to define the topological response.

For the $+$ polarization, then:
\begin{equation}
\label{lightresponse}
   \Gamma_+({\bm k}) = \frac{2\pi}{\hbar} |\langle u| \delta\mathcal{H}_{\pm} |l\rangle |^2\delta(\epsilon_u^{{\bm k}} - \epsilon_l ^{{\bm k}}- \hbar\omega).
\end{equation}
This results in:
\begin{equation}
\label{Gamma+}
\Gamma_+({\bm k}) = \frac{2\pi}{\hbar} A_0^2 \frac{1}{2}(1+\cos^2(\theta)) \delta(\epsilon_u^{{\bm k}} - \epsilon_l ^{{\bm k}}- \hbar\omega),
\end{equation}
corresponding to a rotating wave approximation. The time-dependent corrections average to zero
if one considers a time-Floquet average of the rates or if one goes to the rotating-wave basis. The energy conservation can be equivalently rewritten in terms of a given band $\delta(\epsilon_u^{{\bm k}} - \epsilon_l ^{{\bm k}}- \hbar\omega)=\delta(- \epsilon_l ^{{\bm k}}- \hbar\omega/2)=\delta(\epsilon_u^{{\bm k}} - \hbar\omega/2)$.

The important point within this simple argument is that if we change the polarization of light, only the sign of $\hbar\omega$ is changing in the energetic conservation law since by definition ${\cal H}_+(\omega)={\cal H}_-(-\omega)$. Therefore,
\begin{equation}
\label{Gamma-}
\Gamma_-({\bm k}) = \frac{2\pi}{\hbar} A_0^2 \frac{1}{2}(1+\cos^2(\theta)) \delta(\epsilon_u^{\bm k} - \epsilon_l^{\bm k} + \hbar\omega).
\end{equation}
One can already anticipate that the poles play a special role in the topological response as $1/2(1+\cos^2(\theta))$ is quantized to unity, for each light polarization. We show below the precise relation with the topological number and then with the quantum Hall conductivity.

\subsection{Application of Stokes Theorem}

Here, we relate the light responses with topological properties of the spin-1/2 particle applying the Stokes theorem as in Ref. \cite{spheremodel}, with the introduction of smooth fields. The Chern number is defined in terms of the Berry curvature as
\begin{equation}
\label{topo}
C = -\frac{1}{2\pi} \int_{S^2} \nabla \times {\bf A} \cdot d^2{\bf n},
\end{equation}
where $d^2 {\bf n}=d\varphi d\theta {\bf n}$ with ${\bf n}$ being the unit vector perpendicular to the surface, ${\bf A}=\langle \psi| i\mathbf{\nabla} |\psi\rangle$ with $|\psi\rangle=|l\rangle$ or $|u\rangle$ and $\mathbf{\nabla}=(\partial/\partial\varphi, \partial/\partial\theta)$.
We decompose the sphere into a north and south hemispheres such that
\begin{equation}
C = -\frac{1}{2\pi}\int_{north} \nabla \times {\bf A}_N \cdot d^2{\bf n} - \frac{1}{2\pi}\int_{south} \nabla \times {\bf A}_S \cdot d^2{\bf n}.
\end{equation}
\begin{equation}
C = -\frac{1}{2\pi} \int_0^{2\pi} d\varphi A'_{N\varphi}(\varphi,\theta_c) + \frac{1}{2\pi} \int_0^{2\pi} d\varphi A'_{S\varphi}(\varphi,\theta_c)
\end{equation}
and $\theta_c$ refers to the boundary. Here, to create a link with the circular dichroism of light we will set $\theta_c=\pi/2$. The smooth fields are defined as \cite{spheremodel}
\begin{equation}
{A}'_{N\varphi}(\varphi,\theta) = {A}_{N\varphi}(\varphi,\theta_c) - A_{\varphi}(0) 
\end{equation}
\begin{equation}
{A}'_{S\varphi}(\varphi,\theta) = {A}_{S\varphi}(\varphi,\theta_c) - A_{\varphi}(\pi),
\end{equation}
with $curl {\bf A}'=curl {\bf A}$. The smooth fields are uniquely defined on the surface of the sphere, where $A_{\varphi}(0)=\hbox{lim}_{\theta\rightarrow 0} A(\varphi,\theta)$ and $A_{\varphi}(\pi)=\hbox{lim}_{\theta\rightarrow \pi} A({\varphi,\theta})$.
The poles require special care because the $\varphi$ angle is not uniquely defined. The smooth fields satisfy ${A}'_{N\varphi}(\varphi,\theta)\rightarrow 0$ when $\theta\rightarrow 0$ and ${A}'_{S\varphi}(\varphi,\theta)\rightarrow 0$ when $\theta\rightarrow \pi$.

For the lowest band $|l\rangle$, we have
\begin{equation}
A_{\varphi}(\varphi,\theta) = -\frac{\cos\theta}{2}.
\end{equation}
At the equator, $A_{\varphi}(\varphi,\pi/2)=0={A}_{N\varphi}(\varphi,\pi/2)={A}_{S\varphi}(\varphi,\pi/2)$, and
\begin{equation}
{A}'_{N\varphi}(\varphi,\theta=\pi/2) = \frac{1}{2} = - {A}'_{S\varphi}(\varphi,\theta=\pi/2).
\end{equation}
The topological number has then two simple interpretations
\begin{equation}
\label{C}
C = A_{\varphi}(0) - A_{\varphi}(\pi) =-1
\end{equation}
and 
\begin{equation}
\label{equator}
C = {A}'_{S\varphi}(\varphi,\theta=\pi/2) - {A}'_{N\varphi}(\varphi,\theta=\pi/2) =-1.
\end{equation}
In addition, we have the identities
\begin{equation}
{A}'_{N\varphi}(\varphi,\theta) = \frac{1}{2}(-\cos\theta +1) = \sin^2\frac{\theta}{2}
\end{equation}
and 
\begin{equation}
{A}'_{S\varphi}(\varphi,\theta) = \frac{1}{2}(-\cos\theta - 1) = -\cos^2\frac{\theta}{2}.
\end{equation}
These formulae show that one can perform closed circles with $\varphi\rightarrow \varphi+2\pi$ for any angle $\theta\in ]0;\pi[$ and measure the Chern number. This is in accordance with Stokes theorem, since $\theta_c$ can be defined arbitrarily.

This equatorial plane representation seems justified since we introduce the vector potential with (real) components in the $XY$ plane $A_x=A_0\cos(\omega t)$ and $A_y = \mp A_0\sin(\omega t)$ such that the azimuthal angle can be redefined as $\varphi=\varphi_0\mp\omega t$ for the $\pm$ polarizations of light. Averaging on angles $\varphi$ is similar to perform a Floquet average on a time period.  
A closed path involving a change of the angle $\varphi\in [0;2\pi]$ in the equatorial plane corresponds to perform a similar circular closed path in the wavevector space. This situation is similar to the nuclear magnetic resonance where we obtain a signal $\frac{1}{2}\sin^2 \theta$ when performing a Floquet average of the signal. 

At the equator, for $\theta=\pi/2$, to reveal the topological properties in the light response we use the identity
\begin{equation}
\sin\theta = 2\sin(\theta/2) \cos (\theta/2) = \sqrt{2}\sin(\theta/2).
\end{equation}
Then, we re-write 
\begin{equation}
\Gamma_+({\bf k},\omega) = \frac{2\pi}{\hbar} A_0^2 (1-A'_{N\varphi}(\varphi,\theta))\delta(- \epsilon_l ^{{\bf k}} -\hbar\omega/2)
\end{equation}
and
\begin{equation}
\Gamma_+({\bf k},\omega) = \frac{2\pi}{\hbar} A_0^2 (-A'_{S\varphi}(\varphi,\theta))\delta(- \epsilon_l ^{{\bf k}} -\hbar\omega/2).
\end{equation}
This term then probes the first circle at the equator in the definition of $C$. 
\begin{widetext}
\begin{eqnarray}
\int_{-\infty}^{+\infty} d\omega\frac{1}{2\pi}\int_{2\pi}^{0} d\varphi \Gamma_+(\theta=\pi/2,\varphi,\omega) = \frac{2\pi}{\hbar} A_0^2 A_{S\varphi}'(\theta=\pi/2).
\end{eqnarray}
\end{widetext}
For the $+$ light polarization, since $\varphi=\varphi_0-\omega t$, we have defined here the angle $\varphi$ going from $2\pi$ to $0$. 
For the $-$ light polarization, 
\begin{equation}
\Gamma_-({\bf k},\omega) = \frac{2\pi}{\hbar} A_0^2 (-A'_{S\varphi}(\varphi,\theta))\delta(\epsilon_l ^{{\bf k}} -\hbar\omega/2),
\end{equation}
and under parity transformation, this is equivalent to
\begin{equation}
\Gamma_-(-{\bf k},\omega) = \frac{2\pi}{\hbar} A_0^2 A'_{N\varphi}(\varphi,\theta)\delta(\epsilon_l ^{-{\bf k}} - \hbar\omega/2).
\end{equation}
At the equator $\epsilon_l ^{-{\bf k}}=\epsilon_l ^{{\bf k}}$. 
Then, we obtain:
\begin{widetext}
\begin{equation}
\label{1}
\int_{-\infty}^{+\infty} d\omega  \frac{1}{2\pi} \int_{2\pi}^{0} d\varphi (\Gamma_+(\theta=\pi/2,\varphi,\omega) + \Gamma_-(\pi-\theta=\pi/2,-\varphi,\omega)) = \frac{2\pi}{\hbar^2}A_0^2 (A_{S\varphi}'(\theta=\pi/2) - A_{N\varphi}'(\theta=\pi/2)).
\end{equation}
If we change $+$ into $-$ and also change $\varphi\rightarrow -\varphi$, the two signals are additive. We also have
\begin{equation}
\label{2}
\int_{-\infty}^{+\infty} d\omega  \frac{1}{2\pi} \int_{2\pi}^{0} d\varphi (\Gamma_+(\pi-\theta=\pi/2,\varphi,\omega) + \Gamma_-(\theta=\pi/2,-\varphi,\omega)) = \frac{2\pi}{\hbar^2}A_0^2 (A_{S\varphi}'(\theta=\pi/2) - A_{N\varphi}'(\theta=\pi/2)).
\end{equation}
The right-hand side is invariant under a change of $\theta$ and can be equivalently re-written as
\begin{equation}
\frac{2\pi}{\hbar^2}A_0^2 (A_{S\varphi}'(\theta) - A_{N\varphi}'(\theta))=\frac{2\pi}{\hbar^2} A_0^2(A_{\varphi}(0) - A_{\varphi}(\pi)).
\end{equation}
For $\theta\rightarrow 0$, from Eqs. (\ref{1}) and (\ref{2})  then we find
\begin{equation}
\label{lightresponse}
\int_{0}^{+\infty} d\omega  \frac{1}{2\pi} \oint d\bm{l} \sum_{{\bf k}={\bf K}, {\bf K'}} \frac{(\mathbf{\Gamma}_+({\bf k},\omega) - \mathbf{\Gamma}_-(-{\bf k},\omega))}{2} = \frac{2\pi}{\hbar^2}A_0^2 C.
\end{equation}
\end{widetext}
The integration on frequencies can be equivalently re-written as an integration from $0$ to $+\infty$ and $d\bm{l}$ encodes the trajectories on circles. We introduce a vector representation $\mathbf{\Gamma}_{+}(\varphi) = \Gamma_{+}(\varphi)\bm{\nabla}\varphi$ such that 
$\mathbf{\Gamma}_{-}(-\varphi) = -\Gamma_{-}(-\varphi)\bm{\nabla}\varphi$. See also discussion below Eq. (\ref{motion}) for another definition of the light responses related to the definitions in the article. It's important to note that here $C$ measures topological properties of the lowest occupied band and therefore of the ground state. 

\subsection{Force and Relation to Transport Properties}

On the sphere, the topological number $C=A_{\varphi}(0)-A_{\varphi}(\pi)$ for a spin-1/2 can also be written in terms of the pseudo-spin response at the poles \cite{spheremodel}. In the present
calculation, since for the lower band $\langle \sigma^z\rangle = -\cos\theta$, we verify
\begin{equation}
\label{mag}
C = -\frac{1}{2}\int_0^{\pi} d\theta \frac{\partial}{\partial\theta}\langle \sigma^z\rangle = \frac{\langle \sigma^z(0)\rangle - \langle \sigma^z(\pi)\rangle}{2},
\end{equation}
which can be interpreted as a pumped charge when driving from north to south pole and changing the polar angle linearly in term $\theta=vt$, $v$ being the speed of the protocole. Then we have 
\begin{equation}
\label{jt}
C = - \frac{1}{2}\int_0^T dt \frac{\partial}{\partial t} \langle \sigma^z\rangle = - \frac{1}{2}\int_0^T dt \frac{j(t)}{e}. 
\end{equation}
Within our definitions, the final time is $T=\pi/v$ and $\sigma^z=\sigma^z({\bf k})$ is the Pauli matrix with eigenvalue $+1$ when projecting on $|a({\bf k})\rangle$ and eigenvalue $-1$ when projecting on $|b({\bf k})\rangle$. In the many-body representation, this is also equivalent to $\sigma^z({\bf k})=c^{\dagger}_{{\bf k}A}c_{{\bf k}A} - c^{\dagger}_{{\bf k}B}c_{{\bf k}B}$. The current density comes from the polarization of a dipole formed by the $A$ and $B$ sites when driving from north to south pole. Note that within this definition, the current density $j(t)$ is defined in the reciprocal space assuming the formation of Bloch energy bands.
In the protocole, it is as if one charge  $e$ moves in one direction and a charge $-e$ in the opposite direction such that $\langle \sigma^z \rangle$ changes by $+2$ in the protocole. On the lattice, this is equivalent to say that in a unit cell $i$ in real space formed with two sites $A$ and $B$ we have the relation $d \hat{n}_a/dt = - d\hat{n}_b/dt$ from the hopping term between nearest neighbors. This can be verified through the identity $d\hat{n}/dt = (i/\hbar)[{\cal H},\hat{n}]$ and since the light-matter coupling ${\cal \delta H}_{\pm}$ enters equivalently as a $t_1$ term because the associated vector potential has $x$ and $y$ components then this also ensures this correspondence. Then, by Fourier transform we check that the term $t_2$ does not affect this reasoning because it commutes with each particle density operator. Now, we also have $\hat{n}_a=1/2(1+\sigma^z)$ and $\hat{n}_b=1/2(1-\sigma^z)$ such that $d \hat{n}_a/dt=(1/2)d \sigma^z/dt$ and in the reciprocal space, then from the lattice we can simply define the current density as 
\begin{equation}
j(t)= \frac{d}{dt}(\hat{n}_a({\bf k}) - \hat{n}_b({\bf k})) =\frac{d}{dt}\sigma^z({\bf k}). 
\end{equation}

Here, we show that $C$ in Eq. (\ref{jt}) is related to the quantum Hall conductivity. For this purpose, we discuss the effect of a constant electric field such that Newton's law gives $\hbar\dot{\bf k}=e{\bf E}$. 
We introduce the vector associated to the topological number
\begin{equation}
{\bf C} = -\frac{1}{2\pi}\int d{\bf k}\times {\bf R},
\end{equation}
with ${\bf R}=\mathbb{\nabla}\times{\bf A}$ and ${\bf R}$ is a function of time. We identify a parallel and perpendicular wave-vector components $k_{\parallel}$ and $k_{\perp}$ on the unit sphere $S^2$ related to the direction of the electric field. We define the electric field ${\bf E}$ along the azimuthal angle $\varphi$. From Newton's equation, we have $k_{\parallel}=\frac{eEt}{\hbar} = \varphi$ such that $k_{\parallel}$ is a function of time and therefore in this protocol ${\bf R}$ can be viewed as a function of $k_{\parallel}(t)$ only.  Our goal is to show that the electric field along the azimuthal angle direction plays a similar role as a force driving the particle from north to south pole producing a current. The integration on $k_{\perp}=\theta$ here gives $\pi$ and from $\hbar\dot{\bf k}=e{\bf E}$ or equivalently $\hbar k_{\parallel}=eEt$, we obtain
\begin{equation}
{\bf C} = - \frac{1}{2}\int_0^T dt \frac{e{\bf E}}{\hbar}\times {\bf R}
\end{equation}
with ${\bf R}={\bf R}(k_{\parallel}(t))$. The vector ${\bf C}$ is perpendicular to the direction of the electric field so along the direction of the current density $j(t)$. If we would measure the charge $-e$ instead, then we find that ${\bf C}$ would take the opposite direction.
From Eq. (\ref{jt}), we obtain the usual form of the Karplus-Luttinger \cite{KL} anomalous velocity:
\begin{equation}
{\bf v}=+\frac{e}{\hbar}{\bf E}\times {\bf R}.
\end{equation}
The current density for a given ${\bf k}$ state reads
\begin{equation}
{\bf j}({\bf k}) = \frac{e^2}{\hbar} {\bf E}\times {\bf R},
\end{equation}
related to the quantum Hall conductivity. The importance of the term $t_2$ is hidden in the integer topological Chern number $C$. 

At this stage, from the Karplus-Luttinger velocity, we can evaluate the conductivity directly in the two-dimensional graphene plane for a general situation where ${\bf R}={\bf R}({\bf k})$ on the lattice. The quantum Hall conductivity $\sigma_{xy}$ is obtained when integrating the current density 
on all the possible ${\bf k}$ states in the Brillouin zone:
\begin{equation}
{\bf j}=\int \frac{d k_x dk_y}{(2\pi)^2} {\bf j}({\bf k})
\end{equation}
and the factor $1/(2\pi)^2$ now takes into account the symmetry between the two wave-vector components \cite{RMP}. Then, the amplitude of the current satisfies:
\begin{equation}
|{\bf j}| = \int \frac{dk_x dk_y}{(2\pi)^2} |{\bf j}({\bf k})| = \frac{e^2}{h}\int \frac{dk_x dk_y}{2\pi} |{\bf E}\times {\bf R}|.
\end{equation}
We measure the total current density which is perpendicular to the electric field along $y$ direction, so that
\begin{equation}
|j_x| = \frac{e^2}{h} |C| E_y = \sigma_{xy} E_y.
\end{equation}
Therefore, we verify that the definition of the current density in Eq. (\ref{jt}) reproduces $\sigma_{xy}=e^2/h$.

Now, writing equations of motion for the spin observables, we will then show the relation with the light response. 
Developing the tight-binding model close to the Dirac points, we have the identification
\begin{equation}
\frac{d}{dt}\sigma^z({\bf k}) = \frac{2}{\hbar}\left( (p_x +A_x)\frac{\partial {\cal H}}{\partial p_y} - (p_y+A_y)\frac{\partial {\cal H}}{\partial p_x}\right).
\end{equation}
We also have the relations $p_x = \hbar k_x$ and $p_y = \hbar k_y$, and within our notations $p_x$ and $p_y$ go to zero at the Dirac points. If we build a Fermi Golden rule argument in this current density, to second order in $A_0$, all the wave-vectors equally contribute. Therefore,
to evaluate the light response within this approach, this is equivalent either to take an average on all the wave-vectors or simply to assume the Dirac point limit with $p_x=p_y=0$. Selecting a given light polarization, we will then probe one Dirac point only through the
energy conservation. The current density operator in Eq. ({\ref{jt}) normalized to the electric charge $e$ satisfies:
\begin{equation}
\bar{j}({\bf k}) =\frac{2}{\hbar}\left(A_x\frac{\partial {\cal H}}{\partial p_y} - A_y\frac{\partial {\cal H}}{\partial p_x}\right).
\end{equation}
This justifies why we can reproduce the same response for the present Dirac Hamiltonian when considering the ${\bf K}$ and ${\bf K'}$ only. The current density here is written in terms of the Pauli matrix $\sigma^z$ and the commutation relations for the spin-1/2 then allows us to bridge with transport. 

Now, to make an identification with Eq. (\ref{equ_gamma_goldman}) we can include the forms of $A_x$ and $A_y$ for the two polarizations and therefore we obtain
\begin{equation}
\bar{j}({\bf k}) = \bar{j}({\bf k},t) = \frac{2}{\hbar}A_0\left(\cos(\omega t)\frac{\partial {\cal H}}{\partial p_y} \pm \sin(\omega t)\frac{\partial {\cal H}}{\partial p_x}\right).
\end{equation}
This is equivalent to
\begin{widetext}
\begin{equation}
\bar{j}({\bf k}) = \frac{1}{\hbar} A_0 e^{i\omega t} \left(\frac{\partial {\cal H}}{\partial p_y} \mp i\frac{\partial {\cal H}}{\partial p_x}\right)
+ \frac{1}{\hbar} A_0 e^{-i\omega t} \left(\frac{\partial {\cal H}}{\partial p_y} \pm i\frac{\partial {\cal H}}{\partial p_x}\right) = \frac{1}{\hbar} A_0 e^{i\omega t} \left(\frac{\partial {\cal H}}{\partial p_y} \mp i \frac{\partial {\cal H}}{\partial p_x}\right)
+ h.c.
\end{equation}
\end{widetext}
We verify here that $\bar{j}({\bf k})$ measured close to a Dirac point is related (proportional) to the Hamiltonian ${\cal H}_{\pm}$ in Eq. (\ref{Hmodel}), implying that from Fourier transform of the current density in real space, one can just consider the dynamics at the Dirac
points. One can then equally perform the Fermi Golden rule with the current density $\bar{j}$, which then gives 
\begin{widetext}
\begin{equation}
\label{motion}
{\Gamma}_{\pm} =\frac{2\pi}{\hbar} \sum_{{\bf k}={\bf K}, {\bf K'}} \frac{A_0^2}{\hbar^2} \left| \left\langle u\left|\left(\frac{\partial {\cal H}}{\partial p_y} \pm i \frac{\partial {\cal H}}{\partial p_x}\right)\right|l\right\rangle \right|^2 \delta(\epsilon_u^{{\bf k}} - \epsilon_l ^{{\bf k}}- \hbar\omega).
\end{equation}
\end{widetext}
One Dirac point only contributes for a given light polarization when $\omega>0$. 

From the definitions of Pauli matrices, these relations are also identical to the ones written with the light-matter Hamiltonians in Eqs. (\ref{lightresponse}). 
The light responses are the ones in Eq. (\ref{equ_gamma_goldman}) where we use the electric field strength $E$ through $A_0=E/\omega$ and the additional $(1/\hbar)^2$ factor comes from the definition of the current density. A similar formula was previously derived in Ref.  \cite{goldman_2017}, but including the summation on all the wave-vectors in the first Brillouin zone. In that paper, the relation between this formula and the quantum Hall conductivity \cite{TKNN} was shown with a different approach.  
We deduce that the Bloch sphere is useful to show why one can just perform a sum at the two Dirac points and reproduce the quantum Hall conductivity. In particular, Eq. (\ref{motion}) can also be written with an average on all the vectors {\bf k}. In Eq. (\ref{motion}), the left and right-handed light polarizations give contributions with different signs.  This justifies why $\Gamma_+$ and $\Gamma_-$ enter with a relative minus sign in Eq. (\ref{equ_depletion_K}); evaluating Eq. (\ref{motion}) directly at the Dirac points, we find a prefactor 
$\rho$ proportional to $t_1^2/m^2$ in front of the topological number. One can equivalently introduce the velocity components as $d x/dt = \dot{x} = \sigma^x$ and $dy/dt =\dot{y} = \sigma^y$. The velocity components are in agreement with the definition of the circular basis in terms of the vectors ${\bf e}_x$ and $\mp i {\bf e}_y$ such that ${\bf r} = (x,\pm i y)$ in this coordinate system and ${\bf r}=x {\bf e}_x + y {\bf e}_y$ in the plane. Therefore, $\dot{\bf r}=(\sigma^x,\pm i\sigma^y)$ such that $\Gamma_{\pm}$  involves the group velocity components.

\subsection{Topological number and Interaction Effects}

In our description of the interactions, the $V$ term is re-written self-consistently in terms of the stochastic $\phi^x$, $\phi^y$ and $\phi^z$ variables. For the ground-state properties, we can then show that a jump of the Chern number should be observed at the Mott transition both with the quantum Hall conductivity and with the light response.  More precisely, the stochastic variables $\phi^x$ and $\phi^y$ describing the particle-hole channel commute with 
$\sigma^z$ and $\phi^z=0$ for the ground state properties in the topological phase. The arguments shown above leading to Eq. (\ref{motion}) then ensure that the same topological response should be observed for ground-state properties in the topological phase through the light response or through the quantum Hall conductivity. Fluctuations in the stochastic variables then do not affect the current or light responses, but could affect the noise properties. 
In the Mott phase, the charge density-wave order becomes finite in real-space implying here a jump in $\langle\bar{j}\rangle$. In agreement with iDMRG calculations, the Chern number jumps to 0 since $\langle\sigma^z_i\rangle$ becomes polarized (fixed) in real space. 

Here, we highlight that these arguments remain in fact correct beyond the formulated mean-field arguments, within our approach. For this purpose, related to the mean-field equations (\ref{meanfield})-(\ref{phiz}), we may define operators:
\begin{align}
\hat{\phi}^0 =& -\frac{1}{2} \left(c_{i}^{\dagger}c_{i}  + c_{i+p}^{\dagger}c_{i+p}\right), \\
\hat{\phi}^x =& -\frac{1}{2} \left( c_{i}^{\dagger}c_{i+p} + c_{i+p}^{\dagger}c_{i} \right), \\
\hat{\phi}^y =& -\frac{1}{2}i \left(-c_{i}^{\dagger}c_{i+p} + c_{i+p}^{\dagger}c_{i} \right),\\
\hat{\phi}^z =& -\frac{1}{2} \left(c_{i}^{\dagger}c_{i} - c_{i+p}^{\dagger}c_{i+p}\right).
\end{align}
Each operator is defined on a link at the position $i+p/2$.
Fixing $\phi^0=\langle \hat{\phi}^0\rangle=-1/2$, which is ensured through the half-filling condition, then we obtain two equivalent manners to write the interaction terms,
\begin{align}
{\cal H}_V = V \sum_{i,r} \left((\hat{\phi}^0)^2 - (\hat{\phi}^z)^2 \right),
\end{align}
and
\begin{align}
{\cal H}_V = -V \sum_{i,r} \left( (\hat{\phi}^x)^2 +(\hat{\phi}^y)^2\right),
\end{align}
modulo a constant term. Therefore, respecting all the symmetries of the model, we can then add these two lines which then reproduce Eq. (\ref{eta})
\begin{align}
\label{interaction}
{\cal H}_V =V \sum_{i,p,r} \eta_r \left( c_i^{\dagger}\sigma_{i,i+p}^r c_{i+p} \right)^2,
\end{align}
with the same $\eta$ values as in Eq. (\ref{equ_choice_eta_r}). Now, we can perform a Hubbard-Stratonovitch transformation similar as in Eq. (\ref{equ_Hubb_Strat_xyz}) with the bosonic Hubbard-Stratonovitch quantum field variables $\phi^i$ associated
to the operators $\hat{\phi}^i$, with $i=0,x,y,z$. Then, we obtain a similar Hamiltonian as in Eq. (\ref{equ_int_matrix_zero_contributions}), which is then justified beyond the mean-field approximation. In the topological phase, $\phi^z=0$ by symmetry and therefore
the equations of motion show that the result of Eq. (\ref{motion}) is robust towards interaction effects as long as there is no gap closing.  The ${\bf k}$-space pseudo-spin magnetization, measuring here the difference of populations between $A$ and $B$ sub-lattices, is then a good and simple marker of topological properties and of the Mott transition which can be applied to other models described by similar matrix-Hamiltonians, complementing then the analysis from the Green's function formalism \cite{Zhang}. 

Here, we present numerical evaluations of Eq. (\ref{equ_depletion_K}). 
Computing the circular dichroism of light for the entire Brillouin zone gives the numerical result of Figs. \ref{fig_gammas_all}a)-\ref{fig_gammas_all}d). Although both Figs. \ref{fig_gammas_all}a) and \ref{fig_gammas_all}b) show the same ground state Chern number, the difference between the two reveals the effect of the particle-hole channel $\phi^{x}$. Increasing $V$ renormalizes $t_1$. Considering Fig. \ref{fig_gammas_all}c), the sign flip of the mass term at one K-point at the CDW transition is reflected by regions of blue curve ($\Gamma^+$) turning light red ($\Gamma^-$). 

\begin{figure}[t]
 \includegraphics[width=8.4cm]{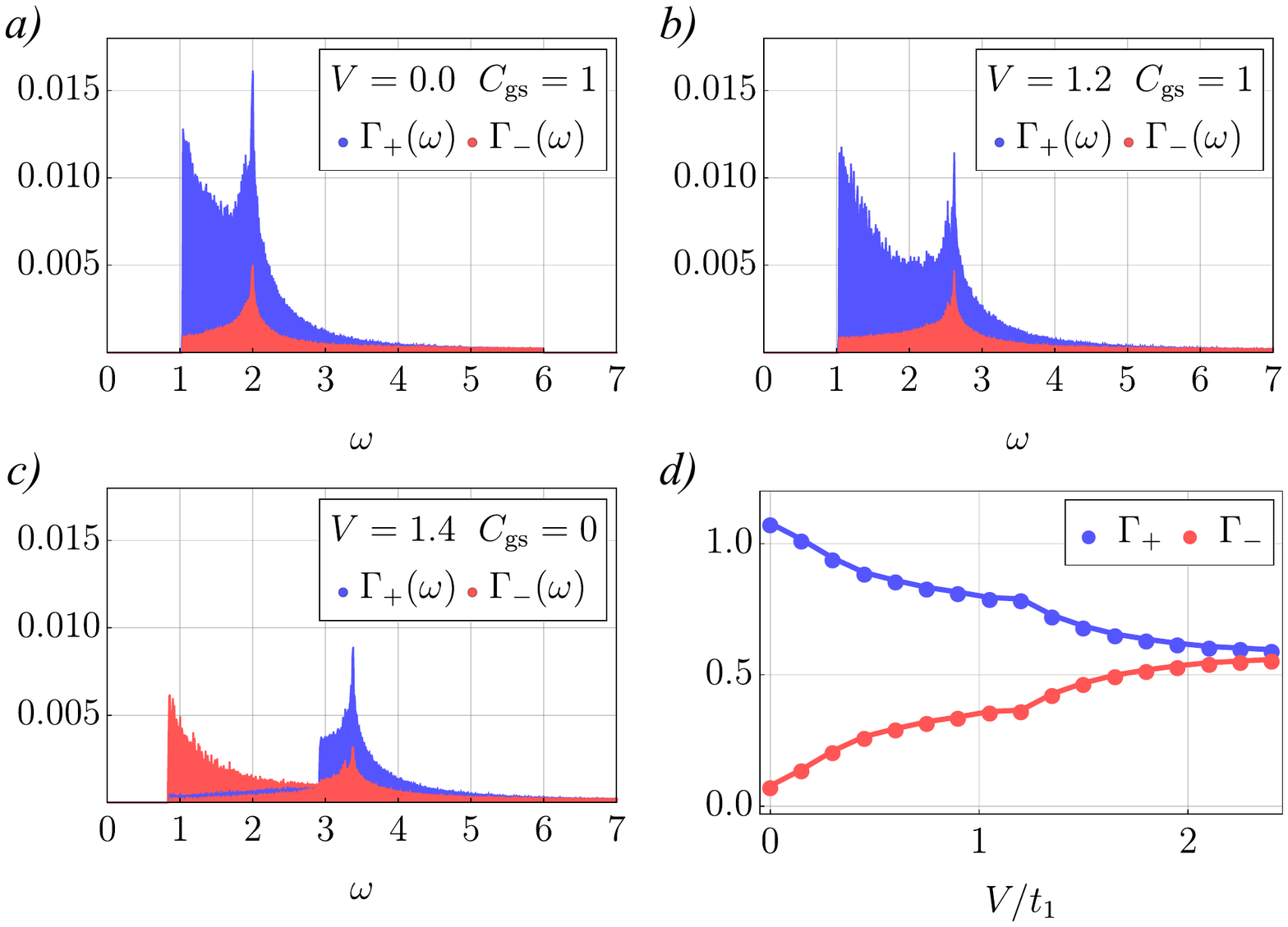}
\caption{(color online) (a-c) Ground state depletion rate  $\Gamma_{\pm} \equiv \sum_{\bm{k}\in \mathrm{BZ}}\Gamma^{\pm}_{l \rightarrow u} (\omega_{\bm k},\bm{k})$ as a function of frequency for $t_2=0.1$ and different fixed values of the interaction strength $V$. (d) Stochastic frequency-integrated depletion rate $\Gamma_{\pm}\equiv \frac{1}{2} \rho^{-1} \int\mathrm{d}\omega\sum_{\bm{k}=\bm{K},\bm{K}^{\prime}}\Gamma^{\pm}_{l \rightarrow u} (\omega_{\bm k},\bm{k})$ as a function of $V$.}
\label{fig_gammas_all}
\vspace{-0.4cm}
\end{figure}

\section{Stochastic Chern number}
\label{stochasticnumber}

Here, we introduce the stochastic Chern number which accounts for the production of particle-hole pairs in the topological phase due to deviations of the stochastic variables from their ground state values, and more precisely, when sampling on the whole distribution of stochastic variables. First, we show that this formalism 
can e.g. describe randomness in the nearest-neighbors' interaction induced by a fluctuating staggered lattice potential and include the effect of an interacting environment \cite{sphere,quantumHallKagome}. 
Then, we apply the methodology to study the light response and the Mott transition. We also show an analogy with temperature effects in the production of particle-hole pairs. This formalism then allows to classify different mechanisms
creating particle-hole pairs due to interaction effects, regarding their topological response. 

\subsection{Stochastic Topological Number and Interpretation as a Disordered Situation}

Here, we show that the sampling on the stochastic variable $\phi^z$ can be equally understood as a sampling on the interaction strength $V$. These arguments below then show that the stochastic topological number corresponds to situations with a slightly disordered interaction strength. Since we also have $n_A n_B = \hat{\phi}_0^2 - \hat{\phi}_z^2$, we deduce that fluctuations in the interaction $V$ between nearest-neighbour sites can be produced either by a fluctuating mean-density or a fluctuating staggered potential on the lattice corresponding to a Semenoff mass with zero mean and a Gaussian distribution.

We define the stochastic topological number as
\begin{align}
\label{stoch}
C = \int_{-\infty}^{+\infty} d\phi P(\phi) C(\phi),
\end{align}
with $\phi=\phi^z$ and the Gaussian distribution 
\begin{align}
\label{dist}
P(\phi)=\frac{1}{\sqrt{2\pi \xi (V)}}e^{-\frac{1}{2}(\phi-\phi_{\mathrm{mf}})^2 \xi^{-1} (V)},
\end{align}
with $\xi(V)=1/(12V)$. Since $C$ can be equally measured through the pseudo-spin magnetizations $\langle\sigma^z\rangle$ at the poles via Eq. (\ref{mag}), this is equivalent in this calculation of $C$ as if ${\phi}^x={\phi}^y=0$ since $g({\bf k})=0$ at the poles in the Hamiltonian $h_V$ of Eq. (\ref{equ_int_matrix_initial}). For a given value of $V$, we can insert the precise value of $\phi_{\mathrm{mf}}$ obtained from the variational mean-field approach with simply $\phi_{\mathrm{mf}}=0$ in the topological phase. 
From statistical physics arguments, then we have
\begin{align}
\langle \sigma^z(\phi)\rangle = \frac{1}{\cal Z}\hbox{Tr} \left(e^{-\beta h_V} \sigma^z\right),
\end{align}
assuming $\theta$ values equal to $0$ and $\pi$. Therefore, Eq. (\ref{stoch}) through Eq. (\ref{mag}) is then equivalent to define the ensembled-averaged variable
\begin{align}
\langle \sigma^z \rangle =  \int_{-\infty}^{+\infty} d\phi P(\phi) \langle \sigma^z(\phi)\rangle.
\end{align}
Now, $h_V$ is symmetric under the variables $\phi$ and $V$. 
Therefore, we equivalently have
\begin{align}
\langle \sigma^z \rangle =  \int_{-\infty}^{+\infty} d\tilde{v} P(\tilde{v}) \langle \sigma^z(\tilde{v})\rangle,
\end{align}
with
\begin{align}
P(\tilde{v}) = \frac{1}{\sqrt{2\pi \xi (V)}}e^{-\frac{1}{2}\tilde{v}^2 \xi^{-1} (V)}
\end{align}
where $\tilde{v}=(\tilde{V}-V)/V=\phi-\phi_{\mathrm{mf}}^z$ measures deviations from the mean value $V$ for the interaction strength between sub-lattices $A$ and $B$. Including fluctuations, the diagonal term in ${\cal H}_{\mathrm{mf}}$ in Eq. (\ref{equ_effective_ham_mft}) now involves $3V\tilde{v}+3V\phi_{\mathrm{mf}}^z$ showing the
relation with a fluctuating staggered potential equal to $3V\tilde{v}$ and with a Gaussian distribution $P(\tilde{v})$. From the formulation of $C$ as a current density we infer that the stochastic topological number can be measured through the quantum Hall conductivity and the circular dichroism of light corresponding e.g. to an average on different samples.

Eq. (\ref{stoch}) is therefore useful to describe lattice effects or the effect of an interacting
environment. As the term $3V\phi^z$ plays the role of a Semenoff mass term acting on the Haldane model in Eq. (\ref{equ_gamma_func}), we can also define - for a given $V$ - $\phi^z_c$ such that $3V\vert \phi^z_c\vert = 3 \sqrt{3}t_2$. Then, all states with $\vert \phi^z\vert < \vert \phi^z_c\vert$ produce a Chern number $C(\phi)=1$ to $C_{\mathrm{st}}$ while all  other $\vert \phi^z\vert$ contribute zero. When $V$ approaches the smallest band gap around the K-points in Fig. \ref{fig_band_structures_all}a), particle-hole pairs will start to form and this leads to the formation of a mixed state. The stochastic topological number is equivalent to
\begin{align}
C_{st} = 1-2\int_{|\phi^z_c|}^{+\infty} d\phi P(\phi).
\end{align}
The integral goes to zero when $V\rightarrow 0$ corresponding to a pure (ground-)state, justifying that in this case $C_{st}=C_{gs}=1$. For small interactions, $C$ can be approximated as
\begin{align}
\label{function}
C_{st} = 1 - 2\int_{|\phi^z_c|}^{+\infty} d\phi P(\phi)\delta(\phi\sim|\phi^z_c|) \approx 1-2P(\phi\sim|\phi^z_c|), 
\end{align}
keeping the dominant term in the series development of the erfc-function. Therefore, this leads to
\begin{align}
\label{Teff}
C_{st} - 1 \propto e^{-m^2/(k_B T_{eff})^2}
\end{align}
with $k_B T_{eff}\propto \sqrt{V}$. This argument then shows that deviations from unity of the topological number come from the creation of particle-hole pairs. This implies that fluctuations in the interaction strength or fluctuations in the charge environment on the
lattice is equivalent to produce a finite probability to reach the upper band. In this sense, the definition of the stochastic Chern number can describe the effect of interaction-induced particle-hole pairs in the topological phase. We also observe that $\tilde{V}$ plays a similar role
as a Landau-Zener mechanism on the sphere \cite{spheremodel}. In this sense, the stochastic Chern number may find various applications.

The parameter $k_B T_{eff}$ above leads to an analogy with temperature effects that we study below in Sec. \ref{Temperature}.

\subsection{Light-Matter Response and Mott Transition}

To evaluate the light response in a mixed state, we begin with Eq. (\ref{equ_depletion_K}), substitute $\mathcal{H}_0 \rightarrow \mathcal{H}^{\mathrm{mf}}$ and here sample all the stochastic variables with a distribution $P(\phi^r)$ according to Eq. (\ref{dist}).  Importantly, $\phi^z$ acts as a Semenoff mass term on the Haldane model modifying the band gap at the Dirac points. Sampling $\phi^z$ generates excited states with smaller energy band gaps, see the light red bands in Fig.~\ref{fig_band_structures_all}a). Here, we sample the fields $(\phi^x,\phi^y,\phi^z)=\bm{\phi}$ according to $P(\phi^r)$ while keeping the chemical potential constant at half-filling, i.e. $\phi^0=-1/2$. In Fig. \ref{fig_gammas_all}d), we show
the evolution of the ensemble-averaged rates $\Gamma_+$ and $\Gamma_-$ as a function of $V$, when sampling on the variables $\bm{\phi}$. These variables are now hidden in the eigenenergies in Eq. (\ref{equ_gamma_goldman}). 

For each configuration we can also associate a $\bm{\phi}$-dependent Chern number $C(\bm{\phi})$ via Eq. (\ref{equ_depletion_K}) that will be either one or zero. Then, for completeness, we evaluate
\begin{equation}
C_{\mathrm{st}} \equiv \int_{-\infty}^{+\infty}\mathrm{d}\bm{\phi} P (\bm{\phi})C(\bm{\phi}),
\end{equation}
which can take non-integer values when it refers to a mixed state. Computing $C_{\mathrm{st}}$ for $10^5$ random configurations, as a function of $V$, then we obtain the result in red in Fig. \ref{fig_Chern_all}a), which can be compared to the ground state Chern number $C_{\mathrm{gs}}$ in blue obtained when $\phi^z = \phi^z_{\mathrm{mf}}$. The quantity $C_{\mathrm{gs}}$ determines the quantum Hall conductivity, in agreement with iDMRG (see Fig.~\ref{fig_phase_diagrams_all}d)) and with the Bloch sphere arguments.
Hence, we can also write $C_{\mathrm{st}}$ via Eq. (\ref{function}), which results in the grey curve in Fig. \ref{fig_Chern_all}a). This highlights the correspondence between the ensemble-averaged values of $\Gamma_+ - \Gamma_-$ in Fig. \ref{fig_gammas_all}d), and $C_{\mathrm{st}}$ as a function of $V$. It's interesting to observe that $C_{\mathrm{st}}$ still reveals the first-order Mott transition through a small jump in Fig. \ref{fig_Chern_all}a). 

\begin{figure}[t]
\includegraphics[width=8cm]{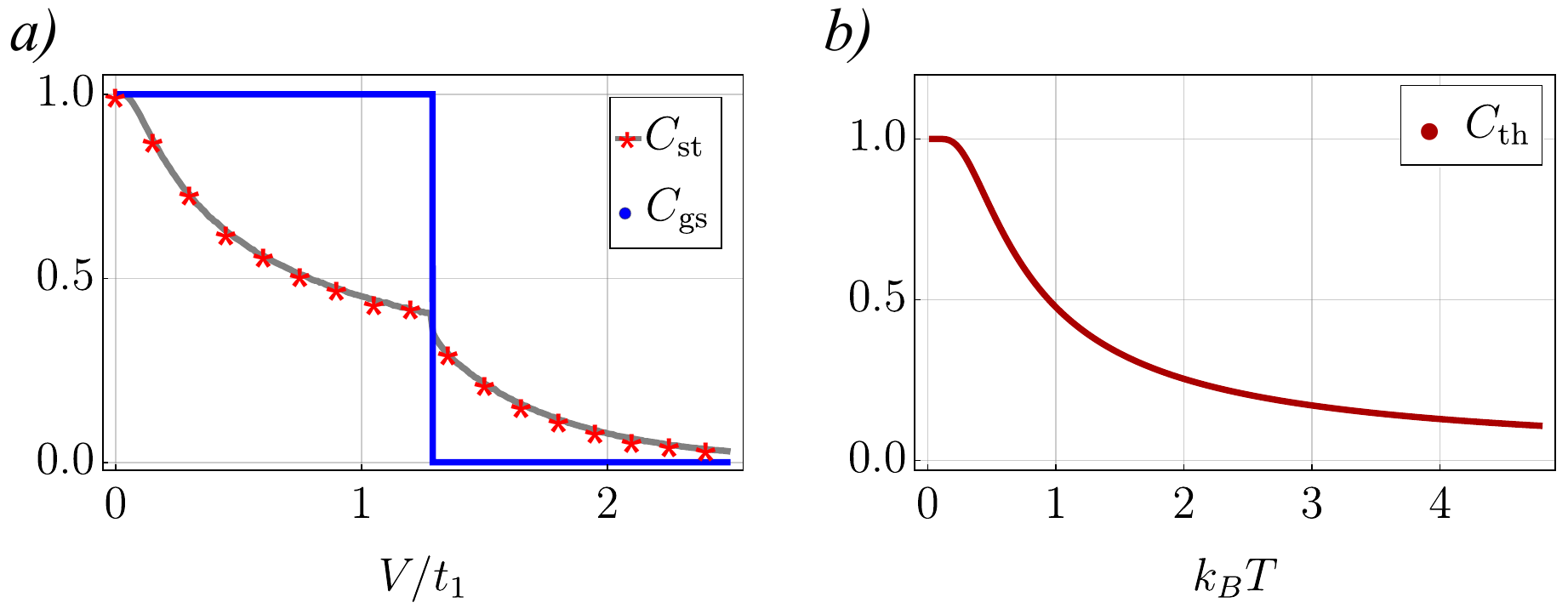}\label{fig_Chern_numbers}
\caption{(color online) (a) Evaluation of the ground state Chern number $C_{\mathrm{gs}}$ and stochastic Chern number $C_{\mathrm{st}}$ as a function of $V$. The grey curve comes from the analytical formula in Eq. (\ref{function}). (b) $C_{\mathrm{th}}$ from Eq. (\ref{equ_depletion_K_finite_T}) at $V=0$ as a function of $k_B T$.}
\label{fig_Chern_all}
\vspace{-0.4cm}
\end{figure}

\subsection{Analogy with Temperature Effects}
\label{Temperature}

Here, we formulate an analogy with the finite-temperature version of the Hall conductivity \cite{delgado_2013} and introduce a finite-temperature version of Eq. (\ref{equ_depletion_K}) 
\begin{equation}
\frac{1}{2}\int_0^{\infty}\mathrm{d}\omega\sum_{\alpha,\bm{k}} \left(p^{\bm k}\Gamma^+_{\alpha}\left(\omega_{\bm{k}},\bm{k}\right)-p^{\bm k}\Gamma^-_{\alpha}\left(\omega_{\bm{k}},\bm{k}\right)\right)
= \rho C_{\mathrm{th}}
\label{equ_depletion_K_finite_T}
\end{equation}
where $p^{\bm k} = (1 + \exp (\epsilon^{\bm k}_{u,l}/k_B T))^{-1}$ is the Fermi distribution, $k_B$ is the Boltzmann constant, and the variable $\alpha$ here refers to $\lbrace l \rightarrow u, u\rightarrow l \rbrace $ such that $p^{\bm k}$ effectively mixes the states of the lower and upper band. We then allow in Eq. (\ref{equ_depletion_K_finite_T}) for heating of the bulk to contribute to $C_{\mathrm{th}}$ \cite{delgado_2013}. From Eq. (\ref{equ_depletion_K_finite_T}), we find that at low temperatures $(k_B T\ll m)$, the finite-temperature Chern number $C_{\mathrm{th}}$ decreases smoothly as $1-e^{-m/k_B T}$ in Fig. \ref{fig_Chern_all}. For further information, see Appendix \ref{TemperatureAppendix}. In the presence of interactions, we observe an analogy with heating in the sense that the probability to create a particle-hole pair in the topological phase will be dominated by values of $|\phi^z|\sim|\phi^z_c|$, producing a reduction of $C_{\mathrm{st}}$ evolving as $P(|\phi^z|\sim |\phi^z_c|) \propto e^{-m^2/(k_B T_{eff})^2}$ from Eq. (\ref{equ_effective_ham_mft}), with an effective temperature  such that $k_B T_{eff}\propto \sqrt{V}$ in Fig. \ref{fig_Chern_all}. We also observe a similar behavior of $C$ in the presence of band-crossing effects on the Bloch sphere \cite{spheremodel}, which then suggests various possible further applications of this formalism.

\section{Conclusion}
\label{summary}

We have introduced a stochastic theory to describe interaction effects in the fermionic Haldane model with nearest-neighbor interactions. We have shown that the method accurately gives quantitative results on the Mott transition and on its nature already from a variational mean-field approach, in agreement with iDMRG. Then, we have studied the effect of light-matter coupling and we have shown that the Mott transition can be probed through circular dichroism of light. We have also derived general relations linking the Bloch sphere, the force due to a constant electric field and transport properties on the lattice, from the momentum or reciprocal space, and in the presence of interactions through the stochastic approach formalism. 
Then, we have introduced a stochastic Chern number which corresponds to sample the ground state Chern number on the whole ensemble of stochastic variables. Physically, this situation can describe disorder effects in the interaction strength resulting e.g. from fluctuations in the lattice potential and producing a mixed state.  We have shown that such fluctuations result in the production of particle-hole pairs, which then engender a non-quantized topological response. We have built analogies with temperature effects and protocols on the Bloch sphere. The method can be generalized in various directions, such as bilayer systems \cite{cheng_2019} and the Kane-Mele \cite{KM,KM1,KM2,KM3} model and lead to further insight on probing particle-hole pairs' responses in interacting topological systems. 
\\

We acknowledge discussions with N. Regnault, S. Biermann, M. O. Goerbig, G. Uhrig, C. Weitenberg, N. Goldman, W. Wu, A. Rosch, J. Hutchinson, J. Legendre, C. Repellin, L. Herviou, H.-Y. Sit and J. Benabou. This research was funded by the Deutsche Forschungsgemeinschaft (DFG, German Research Foundation) via Research Unit FOR 2414 under project number 277974659 (PWK and KLH). Furthermore, KLH acknowledges funding from the French ANR BOCA and from CIFAR in Canada. AGG is supported by the ANR JCJC ANR-18-CE30-0001-01. 
\\

\begin{appendix}

\section{DMRG computational details}
\label{DMRGAppendix}

In this section we give further details of our infinite density matrix renormalization group results (iDMRG). All our results are computed using the open source package \textsc{TenPy}~\cite{Hauschild:2018tz}, which implements the iDMRG algorithm~\cite{White:1993fb} in the language of matrix-product states. The bond dimension $\chi$ sets the maximum linear dimension of the matrices that are stored to represent a given ground-state, and determines the accuracy of the method. For a given $\chi$ the variational algorithm determines the ground-state of the system in the infinite cylinder geometry, which uses translational invariance to reach system sizes of $[L_y,\infty]$. Our conclusion and the following discussion are based on simulations with cylinder circumferences $L_y=6,12$ and $\chi \in \left[50,1200\right]$ which we find sufficient to reach convergence.

\begin{widetext}
\onecolumngrid
\begin{figure}[t]
	\centering
\includegraphics[width=0.6\textwidth]{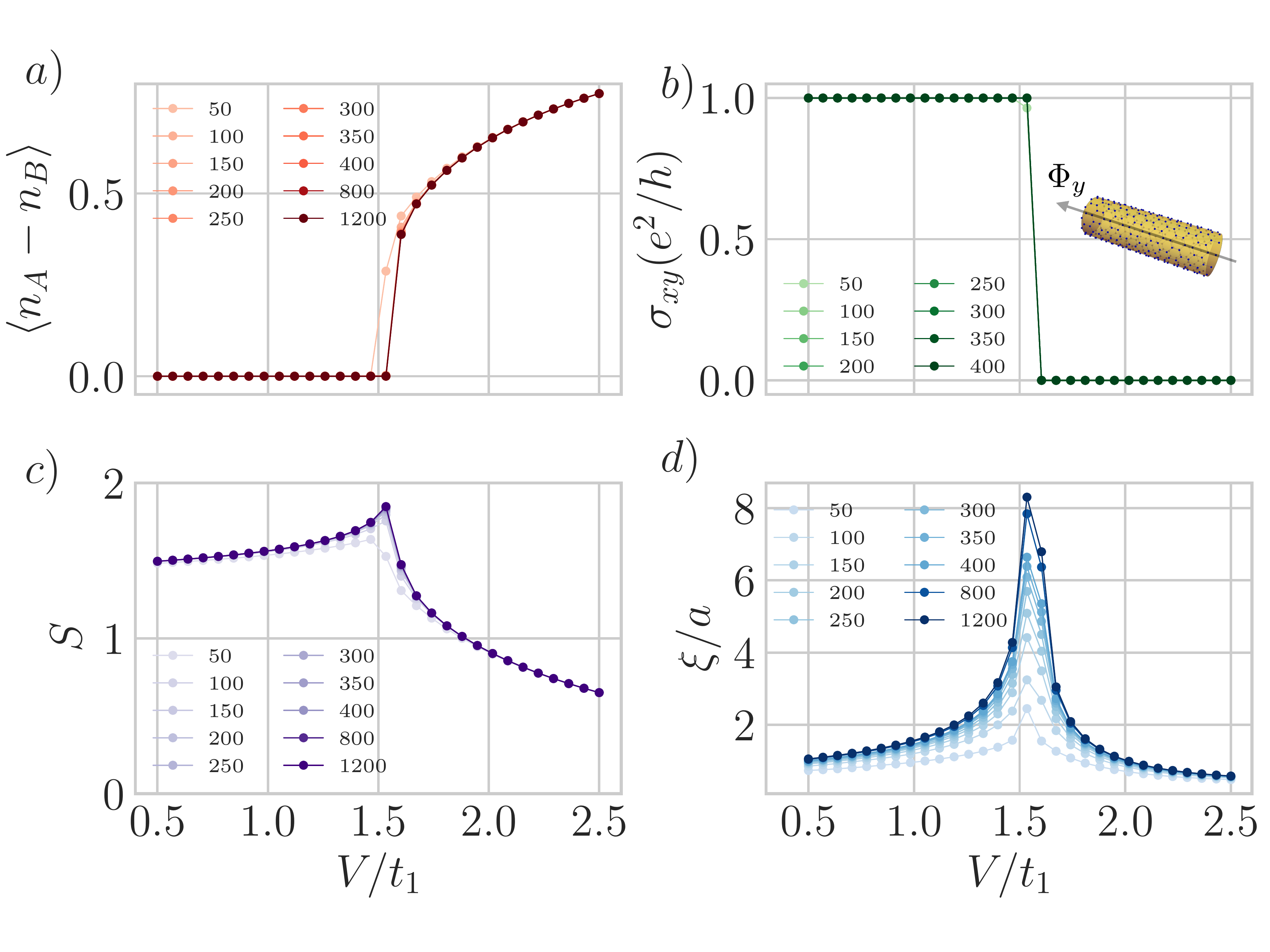}
\centering
\caption{iDMRG convergence. In a) the figure shows the charge density-wave order parameter. b) shows the Hall conductivity calculated by adiabatically threading a flux $\Phi_y/\Phi_0 \in [0,2\pi]$ in units of the flux quantum $\Phi_0$ through the cylinder geometry (inset). Within the Chern insulator or CDW phase one or zero charges are pumped leading to $\sigma_{xy}=e^2/h$ or  $\sigma_{xy}=0$ respectively. c) shows the entanglement entropy $S$ and the figure d) shows the correlation length $\xi/a$  in units of the lattice constant $a$.
d) the charge density wave order parameter. Except for $\sigma_{xy}$ which converges at $\chi=100$, these quantities are calculated for bond dimensions $\chi \in [50,1200]$ and convergence is achieved for $\chi\sim 200$.}
\label{fig_convergence}
\end{figure}
\end{widetext}

\begin{figure}[b]
	\centering
\includegraphics[width=0.3\textwidth]{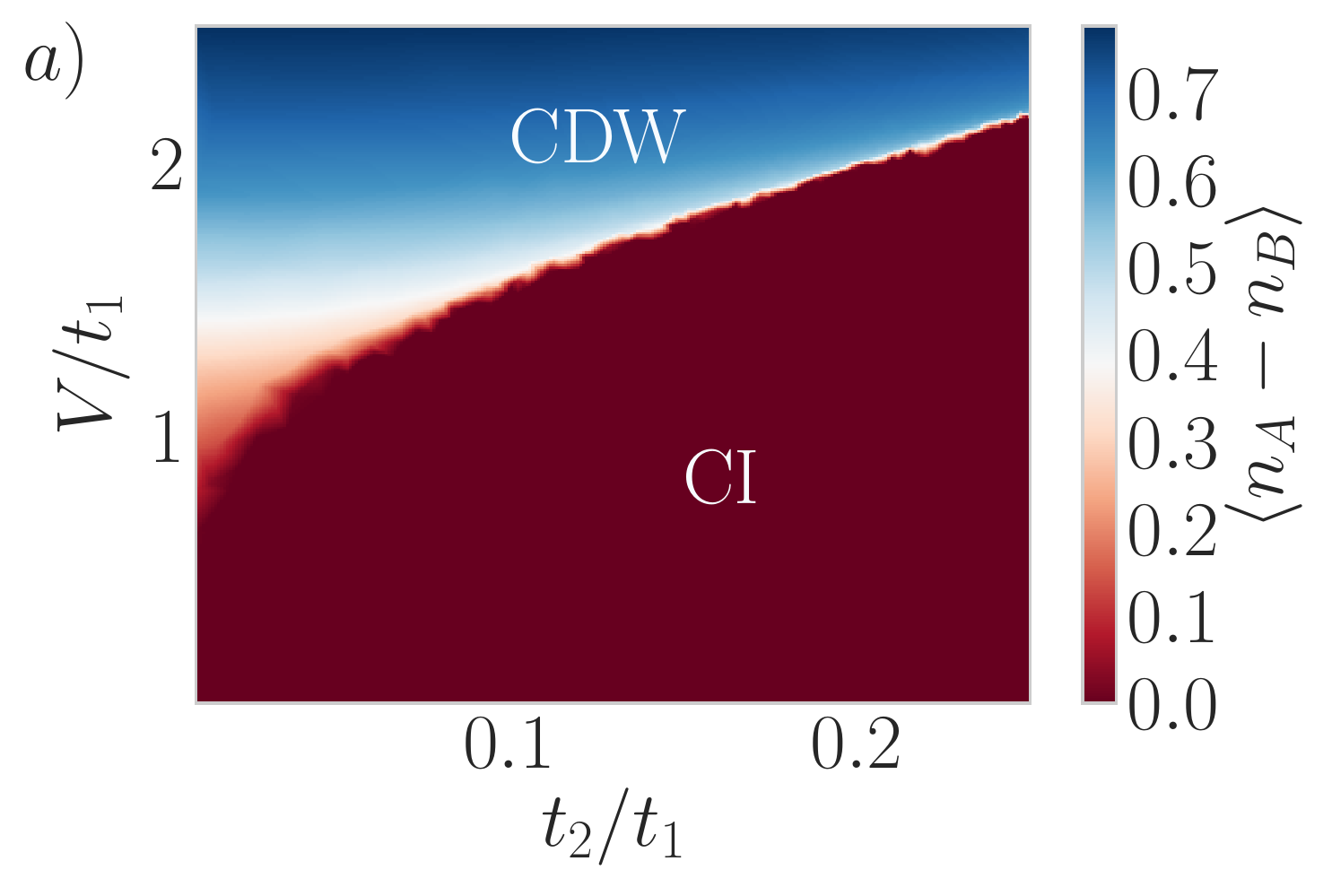}
\includegraphics[width=0.3\textwidth]{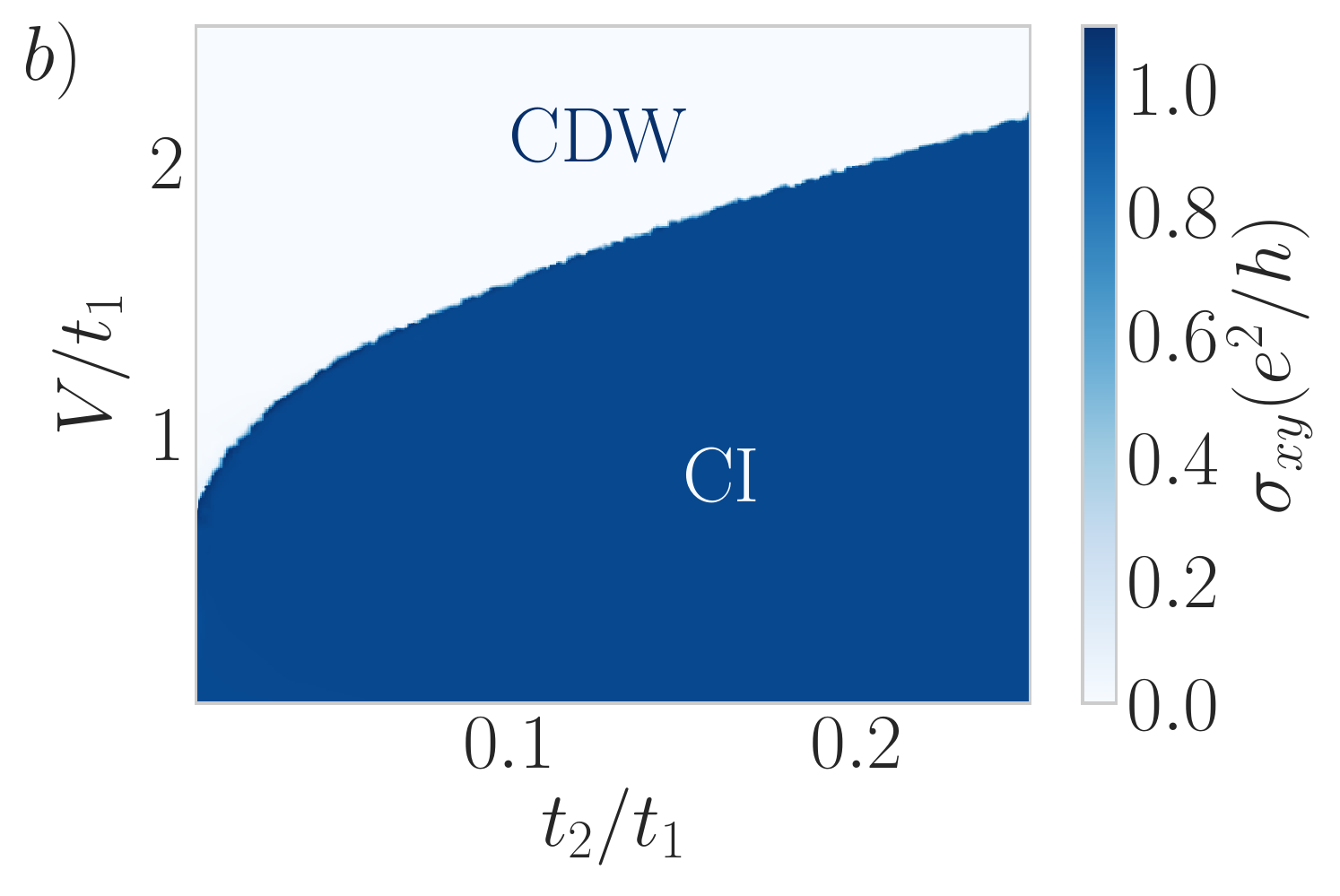}
\centering
\caption{Phase diagrams. a) CDW order parameter and b) $\sigma_{xy}$ calculated in iDMRG for $L_y=6$, $\chi=200$. The phase boundary for both order parameters fall on top of each other indicating that there is no coexistence of a finite CDW order with a non-zero Hall conductivity. }
\label{fig_pd}
\end{figure}

\begin{figure}[t]
	\centering
\includegraphics[width=0.3\textwidth]{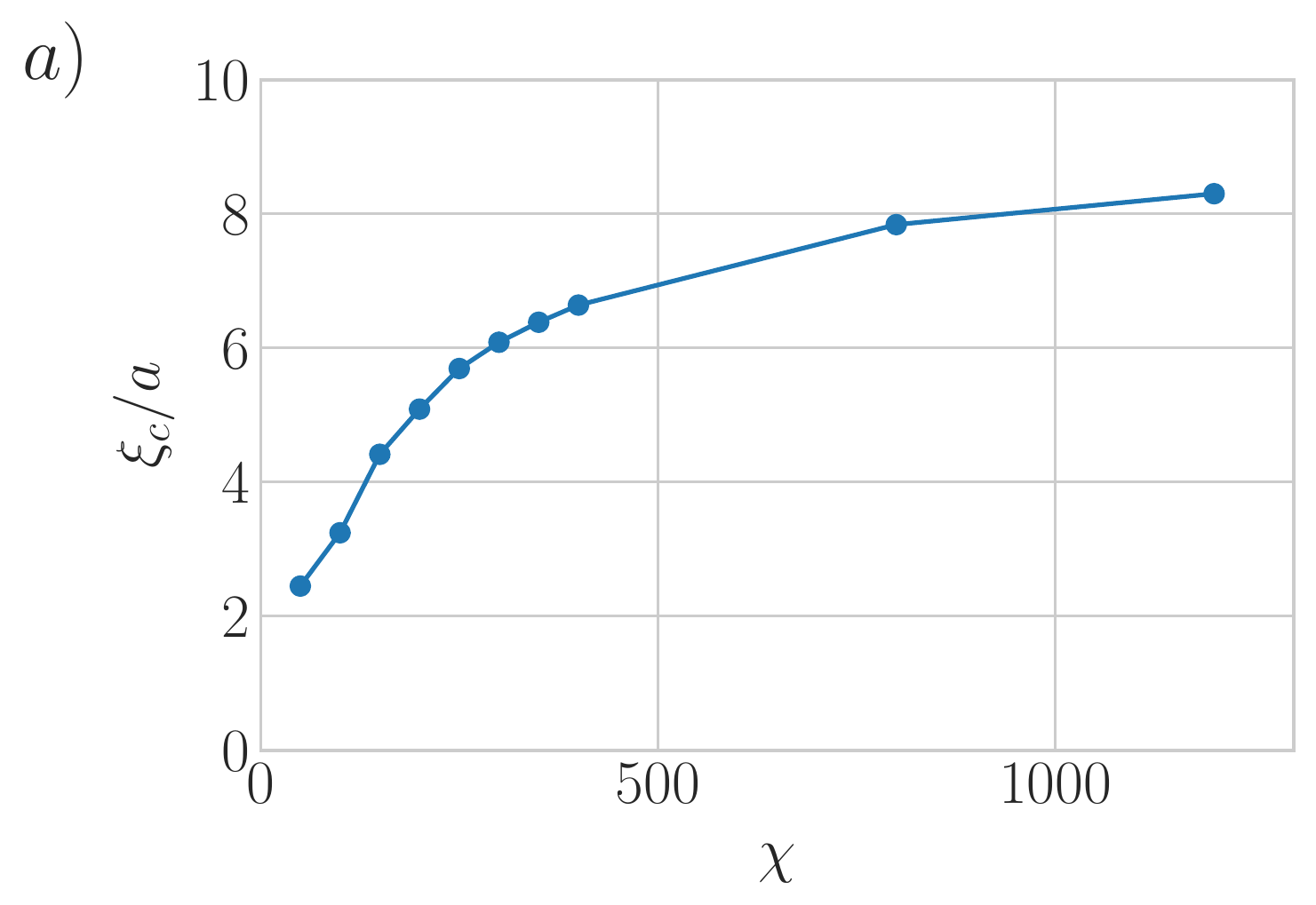}
\includegraphics[width=0.3\textwidth]{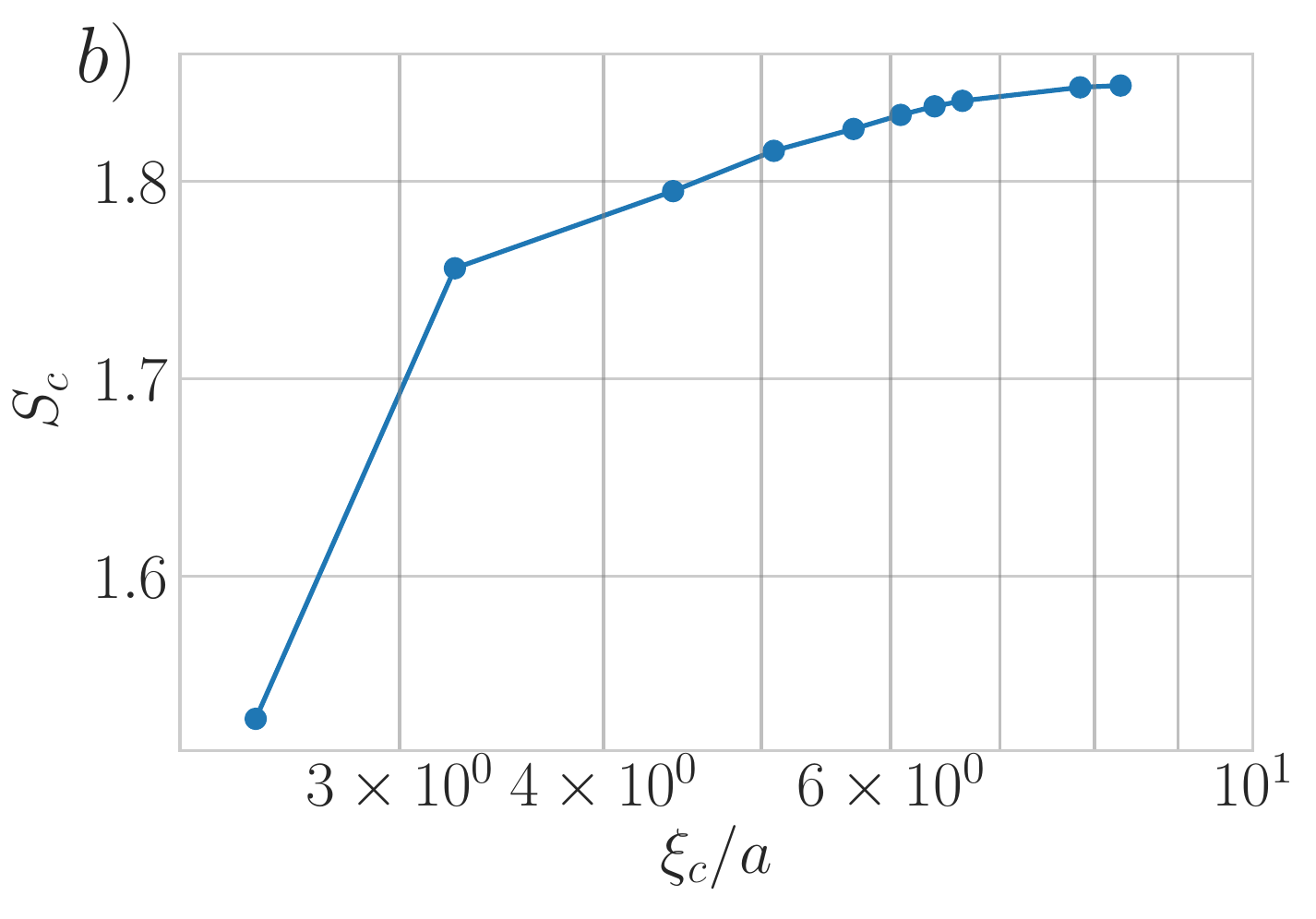}
\centering
\caption{Order of the phase transition in iDMRG. a) shows the correlation length at the transition $\xi_c$ in units of the lattice constant $a$. A second order phase transition would show a divergent correlation length at the transition. Instead the correlation length seems to saturate with increasing $\chi$. b) Entanglement entropy at the transition $S_c$ plotted as a function of the correlation length at the transition $\xi_c$ in units of the lattice constant $a$. The entanglement saturates as a function of correlation length, deviating from the critical scaling law $S_c = c/6 \ln(\xi_c/a)+s_0$.}
\label{fig_trans}
\end{figure}

\subsection{Finite bond dimension effects and order of the phase transition}

For the interacting Haldane hamiltonian, we provide representative results in Fig.~\ref{fig_convergence}. They are obtained for a cut in the phase diagram for $L_y=6$ lattice sites and $t_2/t_1=0.1$ as a function of $V/t_1$. We show the CDW order parameter, the Hall conductivity $\sigma_{xy}$, the entanglement entropy $S$, and correlation length $\xi/a$ in units of the lattice constant $a$ of the ground-state wave-function with different $\chi\in [50,1200]$. We compute the CDW order parameter through the ground-state expectation values of the difference of the two sub-lattice densities $\left\langle n_A-n_B\right\rangle$. The entanglement entropy $S$ is computed by cutting the system in real space and computing the Schmidt decomposition of the groundstate, and relating the corresponding Schmidt eigenvalues to the entanglement entropy~\cite{Hauschild:2018tz}. The correlation length $\xi$ is defined as the largest decay of correlation functions, which is set by the second largest eigenvalue of the transfer matrix~\cite{Hauschild:2018tz}. Finally, to obtain the Hall conductivity $\sigma_{xy}$ we use the method of adiabatic flux insertion implemented numerically as detailed in Refs.~\cite{Zaletel_2014,Grushin2015a}. It consists of threading a magnetic flux through the axis of the cylinder by twisting the boundary conditions by a phase $\Phi_y$ (see inset in Fig.~\ref{fig_convergence}b)). As the flux is threaded from $0$ to $2\pi$ in units of the flux quantum $\Phi_0$, we monitor the pumped charged through a given real-space cut of the cylinder. One charge pumped in such a cycle is equivalent to $\sigma_{xy}=e^2/h$ while no charge pumped amounts to $\sigma_{xy}=0$~\cite{Grushin2015a}.

In Fig.~\ref{fig_convergence} we observe that all quantities within the CI and CDW phases do not change significantly above $\chi\sim 200$, a size for the bond dimension that is consistent with previous work~\cite{Grushin2015a}. The topological origin of the Hall conductivity allows it to converge faster even close to the phase transition, and we find it fully converged at $\chi\sim100$. The inverse gap in the CDW and Chern insulator phases determines the largest length scale up to which correlations spread. Thus the convergence is better as the parameters are chosen deeper in these phases, where the gap is expected to be larger, and the correlation length shorter, consistent with Fig.~\ref{fig_convergence}a).

In Fig.~\ref{fig_pd} we show the CDW and $\sigma_{xy}$ phase diagrams for $\chi=200$. The two phase boundaries coincide for this range of parameters; we find no intermediate phase. Although extracting the nature of the phase transition is a subtle issue in iDMRG data there are hints that this transition is first order. First, both the CDW order parameter and $\sigma_{xy}$ show a clear discontinuity, which can be interpreted as a first order phase transition, consistent with mean field theory. Interestingly, the correlation length $\xi$ does not show such an evident discontinuity. We remind the reader that a second order phase transition presents a divergent correlation length $\xi$ at the transition as a function of increasing bond dimension $\chi$~\cite{Kjall2013}. Furthermore, the entanglement entropy of a critical system is expected to scale as $S = c/6 \ln(\xi/a)+s_0$ where $c$ is the central charge and $s_0$ is a constant~\cite{Kjall2013}. A first order phase transition on the other hand presents either a discontinuity in $\chi$~\cite{Grushin2015a,Motruk2015}, or a saturating behaviour of the maximal or critical correlation length, $\xi_c$, as a function of $\chi$ at the phase transition~\cite{Repellin2017}. Our iDMRG results shown in Fig.~\ref{fig_convergence} point to the latter scenario since the correlation length at the critical point seems to saturate as a function of bond dimension. This is supported by a more careful analysis, shown in Fig.~\ref{fig_trans}, where we plot the maximum correlation length $\xi_c$ as a function of $\chi$ (a), and the maximum entanglement entropy $S_c$ as a function of $\xi_c$ in semi-logarithmic scale (b). Both $\xi_c$ and $S_c$ show an increasing yet saturating behaviour, supporting the absence of a critical state with divergent correlation length. Such an increasing but non-divergent correlation length at the transition is sometimes referred to as a weak first order transition. 

\subsection{Finite size effects}

\begin{figure}[t]
	\centering
\includegraphics[width=0.35\textwidth]{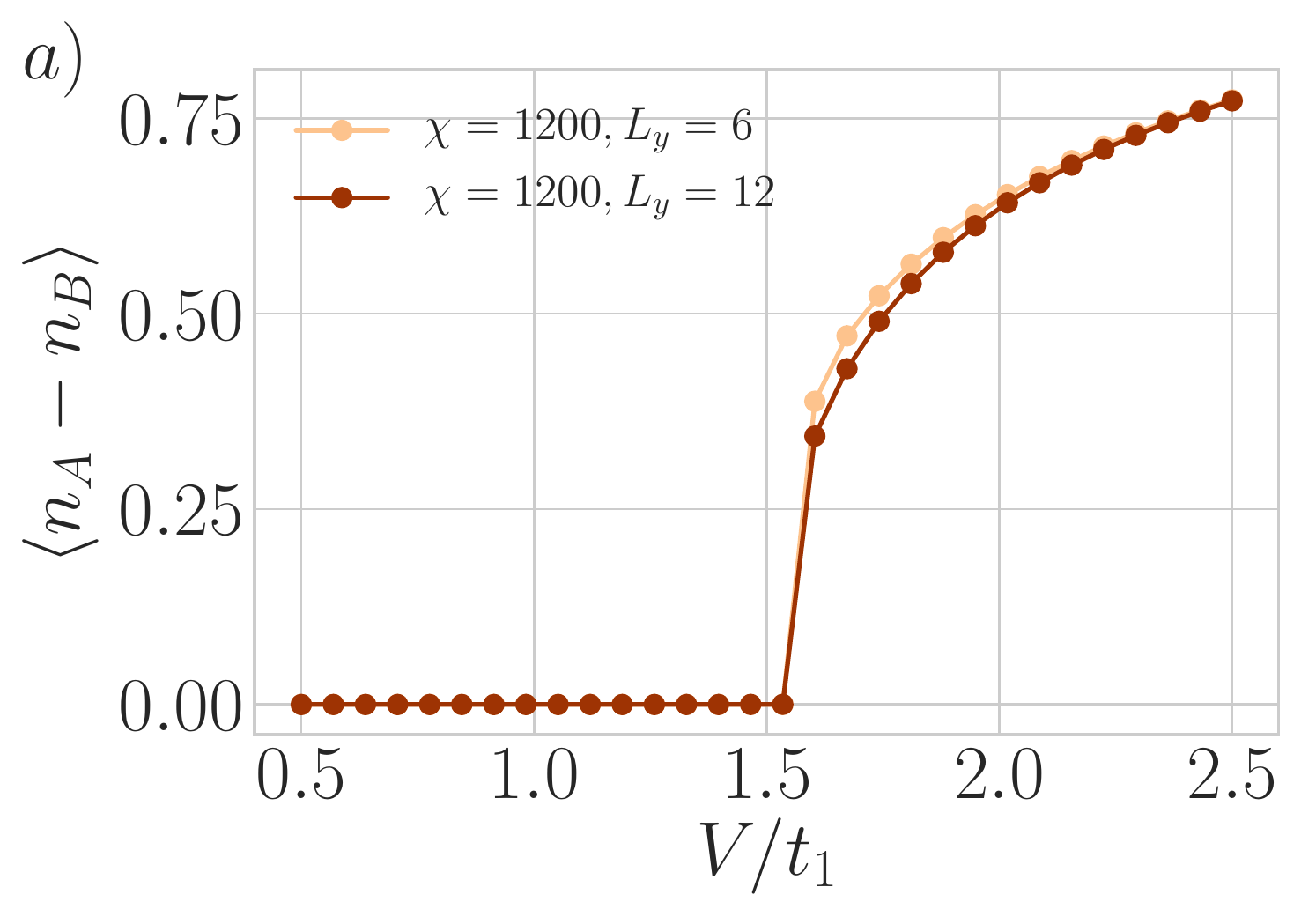}
\includegraphics[width=0.35\textwidth]{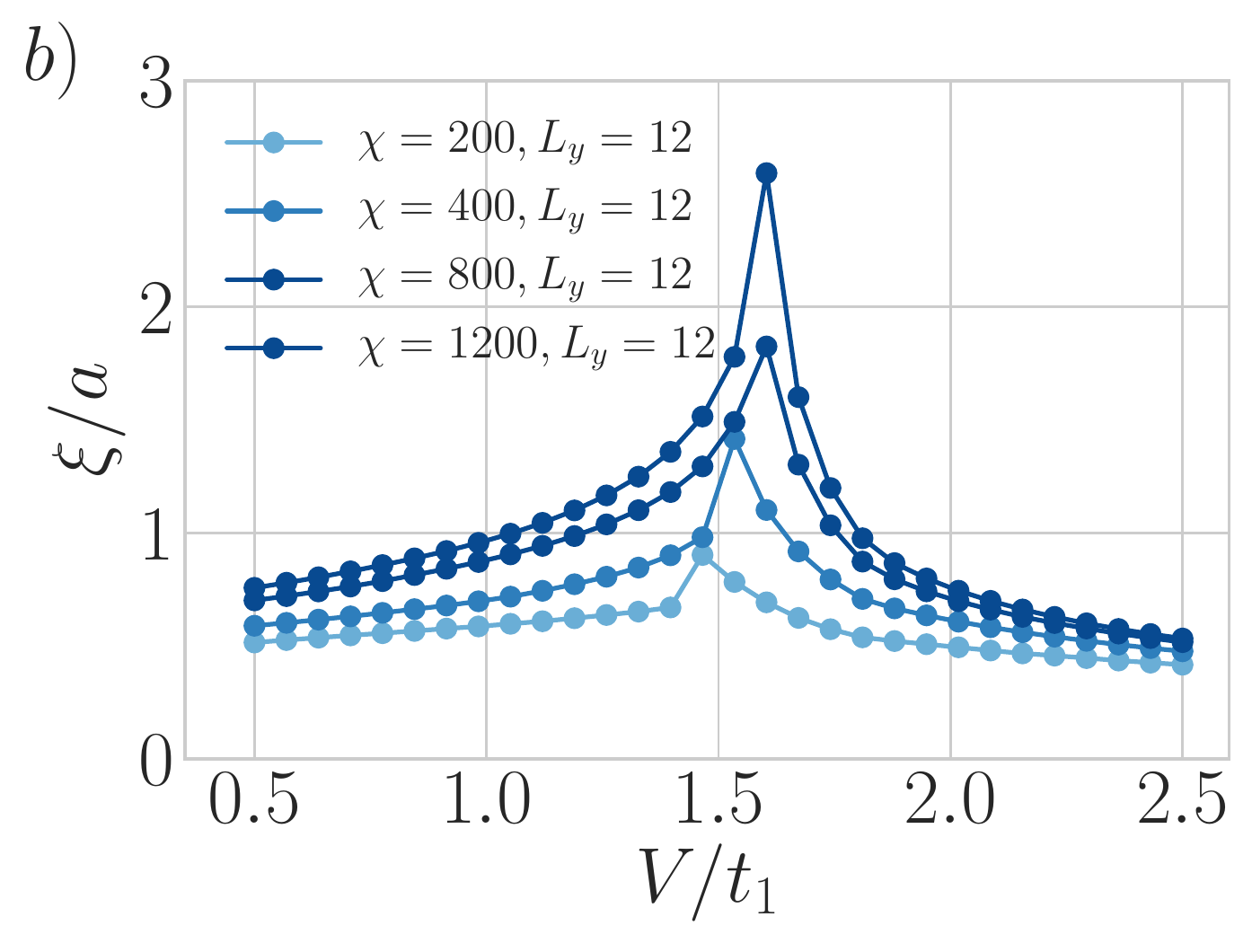}
\centering
\caption{Fine size effects. We show the CDW order parameter (left) and correlation length in units of the lattice constant ($\xi/a$, right) calculated in iDMRG for $L_y=6$ and $L_y=12$, and $\chi=1200$. The phase boundary for the CDW order parameter does not change, and the correlation length drops when increasing the system size. These indicate that large system sizes lead to qualitatively the same results.}
\label{fig_sizes}
\end{figure}

To address the robustness of our results we have performed the calculations for $L_y=12$ with $\chi$ up to $\chi=1200$ (see Fig.~\ref{fig_sizes}) for $t_2/t_1=0.1$. The results are shown in Fig.~\ref{fig_sizes}. Fig.~\ref{fig_sizes}a) shows that the CDW transition is still abrupt and occurs at the same value of $V/t_1$ for larger system sizes. For $L_y=12$ the correlation length $\xi$ close to the transition presents the same peaked structure but is smaller in magnitude that that of $L_y=6$ (see Fig.~\ref{fig_convergence}d)). The decrease in magnitude of $\xi$ can be interpreted as a sign of an increase in the gap size of both the CDW and Chern insulator phases.

\section{Finite temperature version of the topological number}
\label{TemperatureAppendix}

\begin{figure}[t]
\centering
\includegraphics[width=0.35\textwidth]{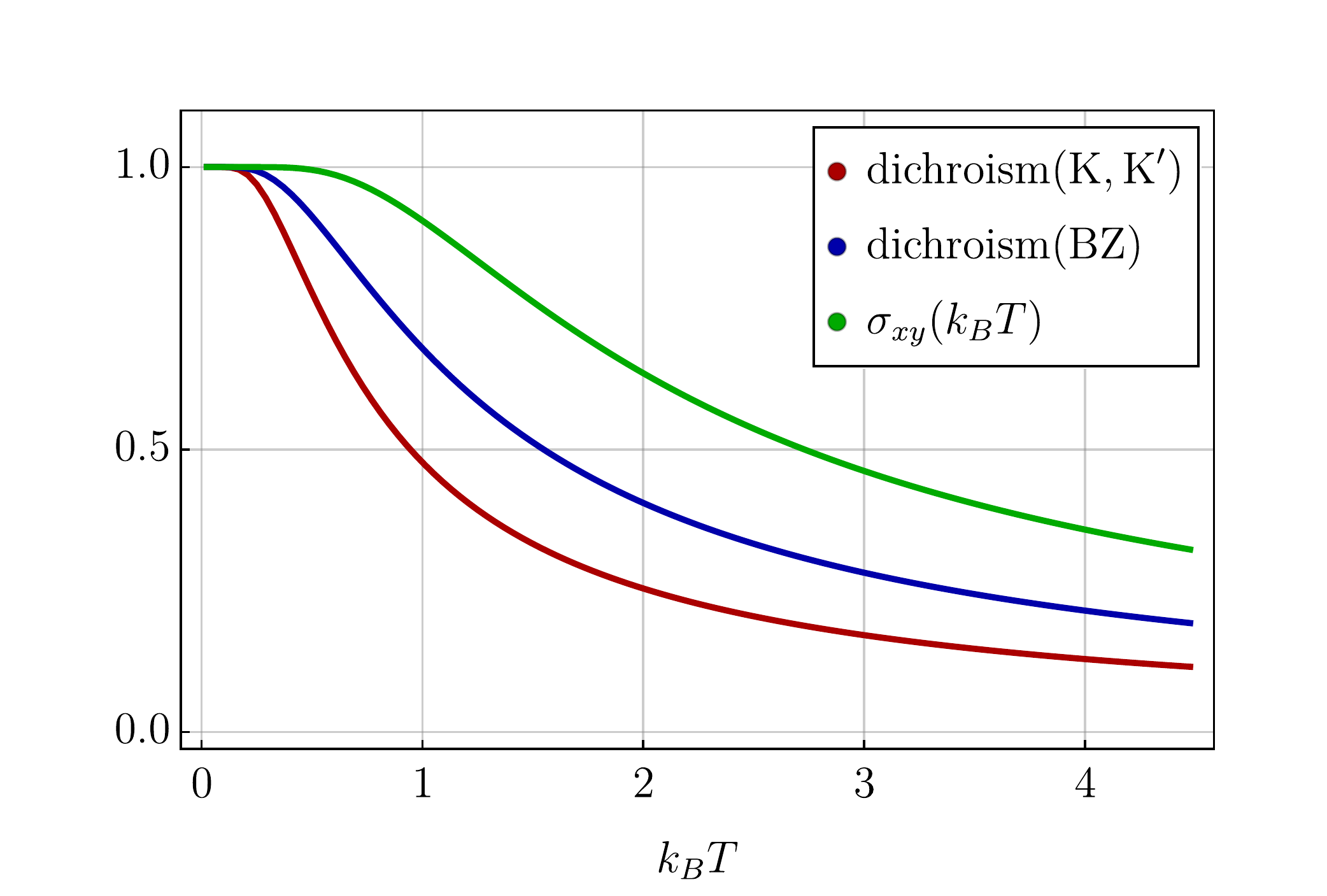}
\caption{(color online) Finite temperature version of the topological index. Green: definition from Ref. \cite{delgado_2013} as in Eq. (\ref{equ_finiteT_Chern_delgado}). Blue: based on Eq. (\ref{equ_finiteT_Chern_dichroism}). Red: Circular dichroism of light as presented in Ref. \cite{goldman_2017}, with a momentum-integral over the entire Brillouin zone.}
\label{fig_finiteT_Cherns}
\end{figure}

In Ref. \cite{delgado_2013}, a finite temperature version of the Hall conductivity was introduced as 
\begin{equation}
\sigma_{xy}(k_BT) h/e^2 = \sum_{\alpha\in\lbrace u,l\rbrace} \int_{\bm{k}\in \mathrm{BZ}} \mathrm{d}\bm{k} \hspace{0.1cm}p^{\bm k}(k_B T ) F^{\alpha}_{x,y}(\bm{k}) \label{equ_finiteT_Chern_delgado}
\end{equation}
with the Fermi distribution Fermi distribution $p^{\bm k}(\epsilon_{u,l}^{\bm k})=(\exp(\epsilon_{u,l}^{\bm k}/k_B T) +1)^{-1}$. Here, the factor  $p^{\bm k}$ effectively mixes upper and lower band states. Hence, Eq. (\ref{equ_finiteT_Chern_delgado}) will show a topological number one for zero temperature and a many-body topological index that approaches zero in the limit of large temperatures, see Fig. \ref{fig_finiteT_Cherns}. Furthermore note, that Eq. (\ref{equ_finiteT_Chern_delgado}) takes heating effects from the bulk into account such that even in a zero-temperature trivial phase, Eq. (\ref{equ_finiteT_Chern_delgado}) will show a many-body Hall conductivity larger than zero for $T>0$ \cite{delgado_2013}.

For the circular dichroism of light, we can produce a similar finite temperature version as in Ref. \cite{delgado_2013} such that the finite temperature version of Eq. (\ref{lightresponse}) reads
\begin{equation}
\int_0^{+\infty} \mathrm{d}\omega\sum_{{\bm k}={\bm K}, {\bm K'}} \sum_{\alpha= l\rightarrow u, u\rightarrow l} \frac{ p^{\bm k} \Gamma^+_{\alpha}({\bm k}) - p^{\bm k} \Gamma^-_{\alpha}({\bm k})}{2}. \label{equ_finiteT_Chern_dichroism}
\end{equation}
Note that Eq. (\ref{equ_finiteT_Chern_delgado}) and Eq. (\ref{equ_finiteT_Chern_dichroism}) give in principle the same result, however the finite temperature topological index scales slightly differently with $k_BT$. Also a finite temperature version of the circular dichroism that integrates over the entire Brillouin zone shows a different scaling behavior, see Fig. \ref{fig_finiteT_Cherns}.\\
Taking into account that at very low temperatures
\begin{eqnarray}
\left(1+\exp\left(-\epsilon^{\bm k}/k_B T\right)\right)^{-1}\approx 1 \hspace{1cm}
\end{eqnarray}
and
\begin{eqnarray}
\left(1+\exp\left(\epsilon^{\bm k}/k_B T\right)\right)^{-1}\approx \exp\left(-\epsilon^{\bm k}/k_B T\right),
\end{eqnarray}
then we obtain that in the formula above, $C$ would acquire ${\cal O}(e^{-|m|/k_B T})$ corrections in agreement with the Arrhenius law for the distribution of particle-hole pairs in the upper band at finite temperature. We then check that combined interaction and drive effects studied in the preceding section have a similar effect as heating in the sense that they produce an effective temperature effect scaling as $\sqrt{V}$.

\end{appendix}

\bibliographystyle{apsrev4-1}
\bibliography{bibi}

\begin{thebibliography}{57}%
\makeatletter
\providecommand \@ifxundefined [1]{%
 \@ifx{#1\undefined}
}%
\providecommand \@ifnum [1]{%
 \ifnum #1\expandafter \@firstoftwo
 \else \expandafter \@secondoftwo
 \fi
}%
\providecommand \@ifx [1]{%
 \ifx #1\expandafter \@firstoftwo
 \else \expandafter \@secondoftwo
 \fi
}%
\providecommand \natexlab [1]{#1}%
\providecommand \enquote  [1]{``#1''}%
\providecommand \bibnamefont  [1]{#1}%
\providecommand \bibfnamefont [1]{#1}%
\providecommand \citenamefont [1]{#1}%
\providecommand \href@noop [0]{\@secondoftwo}%
\providecommand \href [0]{\begingroup \@sanitize@url \@href}%
\providecommand \@href[1]{\@@startlink{#1}\@@href}%
\providecommand \@@href[1]{\endgroup#1\@@endlink}%
\providecommand \@sanitize@url [0]{\catcode `\\12\catcode `\$12\catcode
  `\&12\catcode `\#12\catcode `\^12\catcode `\_12\catcode `\%12\relax}%
\providecommand \@@startlink[1]{}%
\providecommand \@@endlink[0]{}%
\providecommand \url  [0]{\begingroup\@sanitize@url \@url }%
\providecommand \@url [1]{\endgroup\@href {#1}{\urlprefix }}%
\providecommand \urlprefix  [0]{URL }%
\providecommand \Eprint [0]{\href }%
\providecommand \doibase [0]{http://dx.doi.org/}%
\providecommand \selectlanguage [0]{\@gobble}%
\providecommand \bibinfo  [0]{\@secondoftwo}%
\providecommand \bibfield  [0]{\@secondoftwo}%
\providecommand \translation [1]{[#1]}%
\providecommand \BibitemOpen [0]{}%
\providecommand \bibitemStop [0]{}%
\providecommand \bibitemNoStop [0]{.\EOS\space}%
\providecommand \EOS [0]{\spacefactor3000\relax}%
\providecommand \BibitemShut  [1]{\csname bibitem#1\endcsname}%
\let\auto@bib@innerbib\@empty
\bibitem [{\citenamefont {Klitzing}\ \emph {et~al.}(1980)\citenamefont
  {Klitzing}, \citenamefont {Dorda},\ and\ \citenamefont {Pepper}}]{klitzing}%
  \BibitemOpen
  \bibfield  {author} {\bibinfo {author} {\bibfnamefont {K.~v.}\ \bibnamefont
  {Klitzing}}, \bibinfo {author} {\bibfnamefont {G.}~\bibnamefont {Dorda}}, \
  and\ \bibinfo {author} {\bibfnamefont {M.}~\bibnamefont {Pepper}},\ }\href
  {\doibase 10.1103/PhysRevLett.45.494} {\bibfield  {journal} {\bibinfo
  {journal} {Phys. Rev. Lett.}\ }\textbf {\bibinfo {volume} {45}},\ \bibinfo
  {pages} {494} (\bibinfo {year} {1980})}\BibitemShut {NoStop}%
\bibitem [{\citenamefont {Haldane}(1988)}]{haldane_1988}%
  \BibitemOpen
  \bibfield  {author} {\bibinfo {author} {\bibfnamefont {F.~D.~M.}\
  \bibnamefont {Haldane}},\ }\href {\doibase 10.1103/PhysRevLett.61.2015}
  {\bibfield  {journal} {\bibinfo  {journal} {Phys. Rev. Lett.}\ }\textbf
  {\bibinfo {volume} {61}},\ \bibinfo {pages} {2015} (\bibinfo {year}
  {1988})}\BibitemShut {NoStop}%
\bibitem [{\citenamefont {Liu}\ \emph {et~al.}(2016)\citenamefont {Liu},
  \citenamefont {Zhang},\ and\ \citenamefont {Qi}}]{liu_2015}%
  \BibitemOpen
  \bibfield  {author} {\bibinfo {author} {\bibfnamefont {C.-X.}\ \bibnamefont
  {Liu}}, \bibinfo {author} {\bibfnamefont {S.-C.}\ \bibnamefont {Zhang}}, \
  and\ \bibinfo {author} {\bibfnamefont {X.-L.}\ \bibnamefont {Qi}},\ }\href
  {\doibase https://doi.org/10.1146/annurev-conmatphys-031115-011417}
  {\bibfield  {journal} {\bibinfo  {journal} {Annual Review of Condensed Matter
  Physics}\ }\textbf {\bibinfo {volume} {7}},\ \bibinfo {pages} {301} (\bibinfo
  {year} {2016})}\BibitemShut {NoStop}%
\bibitem [{\citenamefont {McIver}\ \emph {et~al.}(2019)\citenamefont {McIver},
  \citenamefont {Schulte}, \citenamefont {Stein}, \citenamefont {Matsuyama},
  \citenamefont {Jotzu}, \citenamefont {Meier},\ and\ \citenamefont
  {Cavalleri}}]{McIver_2019}%
  \BibitemOpen
  \bibfield  {author} {\bibinfo {author} {\bibfnamefont {J.~W.}\ \bibnamefont
  {McIver}}, \bibinfo {author} {\bibfnamefont {B.}~\bibnamefont {Schulte}},
  \bibinfo {author} {\bibfnamefont {F.-U.}\ \bibnamefont {Stein}}, \bibinfo
  {author} {\bibfnamefont {T.}~\bibnamefont {Matsuyama}}, \bibinfo {author}
  {\bibfnamefont {G.}~\bibnamefont {Jotzu}}, \bibinfo {author} {\bibfnamefont
  {G.}~\bibnamefont {Meier}}, \ and\ \bibinfo {author} {\bibfnamefont
  {A.}~\bibnamefont {Cavalleri}},\ }\href {\doibase 10.1038/s41567-019-0698-y}
  {\bibfield  {journal} {\bibinfo  {journal} {Nature Physics}\ } (\bibinfo
  {year} {2019}),\ 10.1038/s41567-019-0698-y}\BibitemShut {NoStop}%
\bibitem [{\citenamefont {Haldane}\ and\ \citenamefont
  {Raghu}(2008)}]{haldane_2008}%
  \BibitemOpen
  \bibfield  {author} {\bibinfo {author} {\bibfnamefont {F.~D.~M.}\
  \bibnamefont {Haldane}}\ and\ \bibinfo {author} {\bibfnamefont
  {S.}~\bibnamefont {Raghu}},\ }\href {\doibase 10.1103/PhysRevLett.100.013904}
  {\bibfield  {journal} {\bibinfo  {journal} {Phys. Rev. Lett.}\ }\textbf
  {\bibinfo {volume} {100}},\ \bibinfo {pages} {013904} (\bibinfo {year}
  {2008})}\BibitemShut {NoStop}%
\bibitem [{\citenamefont {Lu}\ \emph {et~al.}(2014)\citenamefont {Lu},
  \citenamefont {Joannopoulos},\ and\ \citenamefont {Soljacic}}]{Lu_2014}%
  \BibitemOpen
  \bibfield  {author} {\bibinfo {author} {\bibfnamefont {L.}~\bibnamefont
  {Lu}}, \bibinfo {author} {\bibfnamefont {J.~D.}\ \bibnamefont
  {Joannopoulos}}, \ and\ \bibinfo {author} {\bibfnamefont {M.}~\bibnamefont
  {Soljacic}},\ }\href {\doibase 10.1038/nphoton.2014.248} {\bibfield
  {journal} {\bibinfo  {journal} {Nature Photonics}\ }\textbf {\bibinfo
  {volume} {8}},\ \bibinfo {pages} {821} (\bibinfo {year} {2014})}\BibitemShut
  {NoStop}%
\bibitem [{\citenamefont {Rechtsman}\ \emph {et~al.}(2013)\citenamefont
  {Rechtsman}, \citenamefont {Zeuner}, \citenamefont {Plotnik}, \citenamefont
  {Lumer}, \citenamefont {Podolsky}, \citenamefont {Dreisow}, \citenamefont
  {Nolte}, \citenamefont {Segev},\ and\ \citenamefont
  {Szameit}}]{Rechtsman_2013}%
  \BibitemOpen
  \bibfield  {author} {\bibinfo {author} {\bibfnamefont {M.~C.}\ \bibnamefont
  {Rechtsman}}, \bibinfo {author} {\bibfnamefont {J.~M.}\ \bibnamefont
  {Zeuner}}, \bibinfo {author} {\bibfnamefont {Y.}~\bibnamefont {Plotnik}},
  \bibinfo {author} {\bibfnamefont {Y.}~\bibnamefont {Lumer}}, \bibinfo
  {author} {\bibfnamefont {D.}~\bibnamefont {Podolsky}}, \bibinfo {author}
  {\bibfnamefont {F.}~\bibnamefont {Dreisow}}, \bibinfo {author} {\bibfnamefont
  {S.}~\bibnamefont {Nolte}}, \bibinfo {author} {\bibfnamefont
  {M.}~\bibnamefont {Segev}}, \ and\ \bibinfo {author} {\bibfnamefont
  {A.}~\bibnamefont {Szameit}},\ }\href {\doibase 10.1038/nature12066}
  {\bibfield  {journal} {\bibinfo  {journal} {Nature}\ }\textbf {\bibinfo
  {volume} {496}},\ \bibinfo {pages} {196} (\bibinfo {year}
  {2013})}\BibitemShut {NoStop}%
\bibitem [{\citenamefont {Koch}\ \emph {et~al.}(2010)\citenamefont {Koch},
  \citenamefont {Houck}, \citenamefont {Le~Hur},\ and\ \citenamefont
  {Girvin}}]{koch_2010}%
  \BibitemOpen
  \bibfield  {author} {\bibinfo {author} {\bibfnamefont {J.}~\bibnamefont
  {Koch}}, \bibinfo {author} {\bibfnamefont {A.~A.}\ \bibnamefont {Houck}},
  \bibinfo {author} {\bibfnamefont {K.}~\bibnamefont {Le~Hur}}, \ and\ \bibinfo
  {author} {\bibfnamefont {S.~M.}\ \bibnamefont {Girvin}},\ }\href {\doibase
  10.1103/PhysRevA.82.043811} {\bibfield  {journal} {\bibinfo  {journal} {Phys.
  Rev. A}\ }\textbf {\bibinfo {volume} {82}},\ \bibinfo {pages} {043811}
  (\bibinfo {year} {2010})}\BibitemShut {NoStop}%
\bibitem [{\citenamefont {Le~Hur}\ \emph {et~al.}(2016)\citenamefont {Le~Hur},
  \citenamefont {Henriet}, \citenamefont {Petrescu}, \citenamefont {Plekhanov},
  \citenamefont {Roux},\ and\ \citenamefont {Schiro}}]{Le_Hur_2016}%
  \BibitemOpen
  \bibfield  {author} {\bibinfo {author} {\bibfnamefont {K.}~\bibnamefont
  {Le~Hur}}, \bibinfo {author} {\bibfnamefont {L.}~\bibnamefont {Henriet}},
  \bibinfo {author} {\bibfnamefont {A.}~\bibnamefont {Petrescu}}, \bibinfo
  {author} {\bibfnamefont {K.}~\bibnamefont {Plekhanov}}, \bibinfo {author}
  {\bibfnamefont {G.}~\bibnamefont {Roux}}, \ and\ \bibinfo {author}
  {\bibfnamefont {M.}~\bibnamefont {Schiro}},\ }\href {\doibase
  10.1016/j.crhy.2016.05.003} {\bibfield  {journal} {\bibinfo  {journal}
  {Comptes Rendus Physique}\ }\textbf {\bibinfo {volume} {17}},\ \bibinfo
  {pages} {808} (\bibinfo {year} {2016})}\BibitemShut {NoStop}%
\bibitem [{\citenamefont {Ozawa}\ \emph {et~al.}(2019)\citenamefont {Ozawa},
  \citenamefont {Price}, \citenamefont {Amo}, \citenamefont {Goldman},
  \citenamefont {Hafezi}, \citenamefont {Lu}, \citenamefont {Rechtsman},
  \citenamefont {Schuster}, \citenamefont {Simon}, \citenamefont {Zilberberg},\
  and\ \citenamefont {Carusotto}}]{Tomoki_2019}%
  \BibitemOpen
  \bibfield  {author} {\bibinfo {author} {\bibfnamefont {T.}~\bibnamefont
  {Ozawa}}, \bibinfo {author} {\bibfnamefont {H.~M.}\ \bibnamefont {Price}},
  \bibinfo {author} {\bibfnamefont {A.}~\bibnamefont {Amo}}, \bibinfo {author}
  {\bibfnamefont {N.}~\bibnamefont {Goldman}}, \bibinfo {author} {\bibfnamefont
  {M.}~\bibnamefont {Hafezi}}, \bibinfo {author} {\bibfnamefont
  {L.}~\bibnamefont {Lu}}, \bibinfo {author} {\bibfnamefont {M.~C.}\
  \bibnamefont {Rechtsman}}, \bibinfo {author} {\bibfnamefont {D.}~\bibnamefont
  {Schuster}}, \bibinfo {author} {\bibfnamefont {J.}~\bibnamefont {Simon}},
  \bibinfo {author} {\bibfnamefont {O.}~\bibnamefont {Zilberberg}}, \ and\
  \bibinfo {author} {\bibfnamefont {I.}~\bibnamefont {Carusotto}},\ }\href
  {\doibase 10.1103/RevModPhys.91.015006} {\bibfield  {journal} {\bibinfo
  {journal} {Rev. Mod. Phys.}\ }\textbf {\bibinfo {volume} {91}},\ \bibinfo
  {pages} {015006} (\bibinfo {year} {2019})}\BibitemShut {NoStop}%
\bibitem [{\citenamefont {Jotzu}\ \emph {et~al.}(2014)\citenamefont {Jotzu},
  \citenamefont {Messer}, \citenamefont {Desbuquois}, \citenamefont {Lebrat},
  \citenamefont {Uehlinger}, \citenamefont {Greif},\ and\ \citenamefont
  {Esslinger}}]{jotzu_2014}%
  \BibitemOpen
  \bibfield  {author} {\bibinfo {author} {\bibfnamefont {G.}~\bibnamefont
  {Jotzu}}, \bibinfo {author} {\bibfnamefont {M.}~\bibnamefont {Messer}},
  \bibinfo {author} {\bibfnamefont {R.}~\bibnamefont {Desbuquois}}, \bibinfo
  {author} {\bibfnamefont {M.}~\bibnamefont {Lebrat}}, \bibinfo {author}
  {\bibfnamefont {T.}~\bibnamefont {Uehlinger}}, \bibinfo {author}
  {\bibfnamefont {D.}~\bibnamefont {Greif}}, \ and\ \bibinfo {author}
  {\bibfnamefont {T.}~\bibnamefont {Esslinger}},\ }\href {\doibase
  10.1038/nature13915} {\bibfield  {journal} {\bibinfo  {journal} {Nature}\
  }\textbf {\bibinfo {volume} {515}},\ \bibinfo {pages} {237} (\bibinfo {year}
  {2014})}\BibitemShut {NoStop}%
\bibitem [{\citenamefont {Flaschner}\ \emph {et~al.}(2016)\citenamefont
  {Flaschner}, \citenamefont {Rem}, \citenamefont {Tarnowski}, \citenamefont
  {Vogel}, \citenamefont {Luhmann}, \citenamefont {Sengstock},\ and\
  \citenamefont {Weitenberg}}]{flaschner_2016}%
  \BibitemOpen
  \bibfield  {author} {\bibinfo {author} {\bibfnamefont {N.}~\bibnamefont
  {Flaschner}}, \bibinfo {author} {\bibfnamefont {B.~S.}\ \bibnamefont {Rem}},
  \bibinfo {author} {\bibfnamefont {M.}~\bibnamefont {Tarnowski}}, \bibinfo
  {author} {\bibfnamefont {D.}~\bibnamefont {Vogel}}, \bibinfo {author}
  {\bibfnamefont {D.-S.}\ \bibnamefont {Luhmann}}, \bibinfo {author}
  {\bibfnamefont {K.}~\bibnamefont {Sengstock}}, \ and\ \bibinfo {author}
  {\bibfnamefont {C.}~\bibnamefont {Weitenberg}},\ }\href {\doibase
  10.1126/science.aad4568} {\bibfield  {journal} {\bibinfo  {journal}
  {Science}\ }\textbf {\bibinfo {volume} {352}},\ \bibinfo {pages} {1091}
  (\bibinfo {year} {2016})}\BibitemShut {NoStop}%
\bibitem [{\citenamefont {Varney}\ \emph {et~al.}(2010)\citenamefont {Varney},
  \citenamefont {Sun}, \citenamefont {Rigol},\ and\ \citenamefont
  {Galitski}}]{varney_2010}%
  \BibitemOpen
  \bibfield  {author} {\bibinfo {author} {\bibfnamefont {C.~N.}\ \bibnamefont
  {Varney}}, \bibinfo {author} {\bibfnamefont {K.}~\bibnamefont {Sun}},
  \bibinfo {author} {\bibfnamefont {M.}~\bibnamefont {Rigol}}, \ and\ \bibinfo
  {author} {\bibfnamefont {V.}~\bibnamefont {Galitski}},\ }\href {\doibase
  10.1103/PhysRevB.82.115125} {\bibfield  {journal} {\bibinfo  {journal} {Phys.
  Rev. B}\ }\textbf {\bibinfo {volume} {82}},\ \bibinfo {pages} {115125}
  (\bibinfo {year} {2010})}\BibitemShut {NoStop}%
\bibitem [{\citenamefont {Vasi\ifmmode~\acute{c}\else \'{c}\fi{}}\ \emph
  {et~al.}(2015)\citenamefont {Vasi\ifmmode~\acute{c}\else \'{c}\fi{}},
  \citenamefont {Petrescu}, \citenamefont {Le~Hur},\ and\ \citenamefont
  {Hofstetter}}]{vasic_2015}%
  \BibitemOpen
  \bibfield  {author} {\bibinfo {author} {\bibfnamefont {I.}~\bibnamefont
  {Vasi\ifmmode~\acute{c}\else \'{c}\fi{}}}, \bibinfo {author} {\bibfnamefont
  {A.}~\bibnamefont {Petrescu}}, \bibinfo {author} {\bibfnamefont
  {K.}~\bibnamefont {Le~Hur}}, \ and\ \bibinfo {author} {\bibfnamefont
  {W.}~\bibnamefont {Hofstetter}},\ }\href {\doibase
  10.1103/PhysRevB.91.094502} {\bibfield  {journal} {\bibinfo  {journal} {Phys.
  Rev. B}\ }\textbf {\bibinfo {volume} {91}},\ \bibinfo {pages} {094502}
  (\bibinfo {year} {2015})}\BibitemShut {NoStop}%
\bibitem [{\citenamefont {Varney}\ \emph {et~al.}(2011)\citenamefont {Varney},
  \citenamefont {Sun}, \citenamefont {Rigol},\ and\ \citenamefont
  {Galitski}}]{varney_2011}%
  \BibitemOpen
  \bibfield  {author} {\bibinfo {author} {\bibfnamefont {C.~N.}\ \bibnamefont
  {Varney}}, \bibinfo {author} {\bibfnamefont {K.}~\bibnamefont {Sun}},
  \bibinfo {author} {\bibfnamefont {M.}~\bibnamefont {Rigol}}, \ and\ \bibinfo
  {author} {\bibfnamefont {V.}~\bibnamefont {Galitski}},\ }\href {\doibase
  10.1103/PhysRevB.84.241105} {\bibfield  {journal} {\bibinfo  {journal} {Phys.
  Rev. B}\ }\textbf {\bibinfo {volume} {84}},\ \bibinfo {pages} {241105}
  (\bibinfo {year} {2011})}\BibitemShut {NoStop}%
\bibitem [{\citenamefont {Wang}\ \emph {et~al.}(2010)\citenamefont {Wang},
  \citenamefont {Shi}, \citenamefont {Zhang}, \citenamefont {Wang},
  \citenamefont {Dai},\ and\ \citenamefont {Xie}}]{dai_2010}%
  \BibitemOpen
  \bibfield  {author} {\bibinfo {author} {\bibfnamefont {L.}~\bibnamefont
  {Wang}}, \bibinfo {author} {\bibfnamefont {H.}~\bibnamefont {Shi}}, \bibinfo
  {author} {\bibfnamefont {S.}~\bibnamefont {Zhang}}, \bibinfo {author}
  {\bibfnamefont {X.}~\bibnamefont {Wang}}, \bibinfo {author} {\bibfnamefont
  {X.}~\bibnamefont {Dai}}, \ and\ \bibinfo {author} {\bibfnamefont {X.~C.}\
  \bibnamefont {Xie}},\ }\href@noop {} {\enquote {\bibinfo {title}
  {Charge-density-wave and topological transitions in interacting haldane
  model},}\ } (\bibinfo {year} {2010}),\ \Eprint
  {http://arxiv.org/abs/1012.5163} {arXiv:1012.5163 [cond-mat.str-el]}
  \BibitemShut {NoStop}%
\bibitem [{\citenamefont {Alba}\ \emph {et~al.}(2016)\citenamefont {Alba},
  \citenamefont {Pachos},\ and\ \citenamefont
  {Garc{\'{\i}}a-Ripoll}}]{alba_2016}%
  \BibitemOpen
  \bibfield  {author} {\bibinfo {author} {\bibfnamefont {E.}~\bibnamefont
  {Alba}}, \bibinfo {author} {\bibfnamefont {J.~K.}\ \bibnamefont {Pachos}}, \
  and\ \bibinfo {author} {\bibfnamefont {J.~J.}\ \bibnamefont
  {Garc{\'{\i}}a-Ripoll}},\ }\href {\doibase 10.1088/1367-2630/18/3/033022}
  {\bibfield  {journal} {\bibinfo  {journal} {New Journal of Physics}\ }\textbf
  {\bibinfo {volume} {18}},\ \bibinfo {pages} {033022} (\bibinfo {year}
  {2016})}\BibitemShut {NoStop}%
\bibitem [{\citenamefont {Hickey}\ \emph {et~al.}(2015)\citenamefont {Hickey},
  \citenamefont {Rath},\ and\ \citenamefont {Paramekanti}}]{hickey_2015}%
  \BibitemOpen
  \bibfield  {author} {\bibinfo {author} {\bibfnamefont {C.}~\bibnamefont
  {Hickey}}, \bibinfo {author} {\bibfnamefont {P.}~\bibnamefont {Rath}}, \ and\
  \bibinfo {author} {\bibfnamefont {A.}~\bibnamefont {Paramekanti}},\ }\href
  {\doibase 10.1103/PhysRevB.91.134414} {\bibfield  {journal} {\bibinfo
  {journal} {Phys. Rev. B}\ }\textbf {\bibinfo {volume} {91}},\ \bibinfo
  {pages} {134414} (\bibinfo {year} {2015})}\BibitemShut {NoStop}%
\bibitem [{\citenamefont {Maciejko}\ and\ \citenamefont
  {R\"uegg}(2013)}]{ruegg_2013}%
  \BibitemOpen
  \bibfield  {author} {\bibinfo {author} {\bibfnamefont {J.}~\bibnamefont
  {Maciejko}}\ and\ \bibinfo {author} {\bibfnamefont {A.}~\bibnamefont
  {R\"uegg}},\ }\href {\doibase 10.1103/PhysRevB.88.241101} {\bibfield
  {journal} {\bibinfo  {journal} {Phys. Rev. B}\ }\textbf {\bibinfo {volume}
  {88}},\ \bibinfo {pages} {241101} (\bibinfo {year} {2013})}\BibitemShut
  {NoStop}%
\bibitem [{\citenamefont {Prychynenko}\ and\ \citenamefont
  {Huber}(2016)}]{huber_2016}%
  \BibitemOpen
  \bibfield  {author} {\bibinfo {author} {\bibfnamefont {D.}~\bibnamefont
  {Prychynenko}}\ and\ \bibinfo {author} {\bibfnamefont {S.~D.}\ \bibnamefont
  {Huber}},\ }\href {\doibase 10.1016/j.physb.2015.10.027} {\bibfield
  {journal} {\bibinfo  {journal} {Physica B: Condensed Matter}\ }\textbf
  {\bibinfo {volume} {481}},\ \bibinfo {pages} {53} (\bibinfo {year}
  {2016})}\BibitemShut {NoStop}%
\bibitem [{\citenamefont {Imri\ifmmode~\check{s}\else \v{s}\fi{}ka}\ \emph
  {et~al.}(2016)\citenamefont {Imri\ifmmode~\check{s}\else \v{s}\fi{}ka},
  \citenamefont {Wang},\ and\ \citenamefont {Troyer}}]{troyer_2016}%
  \BibitemOpen
  \bibfield  {author} {\bibinfo {author} {\bibfnamefont {J.}~\bibnamefont
  {Imri\ifmmode~\check{s}\else \v{s}\fi{}ka}}, \bibinfo {author} {\bibfnamefont
  {L.}~\bibnamefont {Wang}}, \ and\ \bibinfo {author} {\bibfnamefont
  {M.}~\bibnamefont {Troyer}},\ }\href {\doibase 10.1103/PhysRevB.94.035109}
  {\bibfield  {journal} {\bibinfo  {journal} {Phys. Rev. B}\ }\textbf {\bibinfo
  {volume} {94}},\ \bibinfo {pages} {035109} (\bibinfo {year}
  {2016})}\BibitemShut {NoStop}%
\bibitem [{\citenamefont {Vanhala}\ \emph {et~al.}(2016)\citenamefont
  {Vanhala}, \citenamefont {Siro}, \citenamefont {Liang}, \citenamefont
  {Troyer}, \citenamefont {Harju},\ and\ \citenamefont
  {T\"orm\"a}}]{vanhala_2016}%
  \BibitemOpen
  \bibfield  {author} {\bibinfo {author} {\bibfnamefont {T.~I.}\ \bibnamefont
  {Vanhala}}, \bibinfo {author} {\bibfnamefont {T.}~\bibnamefont {Siro}},
  \bibinfo {author} {\bibfnamefont {L.}~\bibnamefont {Liang}}, \bibinfo
  {author} {\bibfnamefont {M.}~\bibnamefont {Troyer}}, \bibinfo {author}
  {\bibfnamefont {A.}~\bibnamefont {Harju}}, \ and\ \bibinfo {author}
  {\bibfnamefont {P.}~\bibnamefont {T\"orm\"a}},\ }\href {\doibase
  10.1103/PhysRevLett.116.225305} {\bibfield  {journal} {\bibinfo  {journal}
  {Phys. Rev. Lett.}\ }\textbf {\bibinfo {volume} {116}},\ \bibinfo {pages}
  {225305} (\bibinfo {year} {2016})}\BibitemShut {NoStop}%
\bibitem [{\citenamefont {Qi}\ and\ \citenamefont {Zhang}(2011)}]{Stanford}%
  \BibitemOpen
  \bibfield  {author} {\bibinfo {author} {\bibfnamefont {X.-L.}\ \bibnamefont
  {Qi}}\ and\ \bibinfo {author} {\bibfnamefont {S.-C.}\ \bibnamefont {Zhang}},\
  }\href {\doibase https://doi.org/10.1103/RevModPhys.83.1057} {\bibfield
  {journal} {\bibinfo  {journal} {Rev. Mod. Phys.}\ }\textbf {\bibinfo {volume}
  {83}},\ \bibinfo {pages} {1057} (\bibinfo {year} {2011})}\BibitemShut
  {NoStop}%
\bibitem [{\citenamefont {Rachel}(2018)}]{Rachel_2018}%
  \BibitemOpen
  \bibfield  {author} {\bibinfo {author} {\bibfnamefont {S.}~\bibnamefont
  {Rachel}},\ }\href {\doibase 10.1088/1361-6633/aad6a6} {\bibfield  {journal}
  {\bibinfo  {journal} {Reports on Progress in Physics}\ }\textbf {\bibinfo
  {volume} {81}},\ \bibinfo {pages} {116501} (\bibinfo {year}
  {2018})}\BibitemShut {NoStop}%
\bibitem [{\citenamefont {Schulz}(1994)}]{schulz_1994}%
  \BibitemOpen
  \bibfield  {author} {\bibinfo {author} {\bibfnamefont {H.~J.}\ \bibnamefont
  {Schulz}},\ }\href@noop {} {\enquote {\bibinfo {title} {Functional integrals
  for correlated electrons},}\ } (\bibinfo {year} {1994}),\ \Eprint
  {http://arxiv.org/abs/cond-mat/9402103} {arXiv:cond-mat/9402103 [cond-mat]}
  \BibitemShut {NoStop}%
\bibitem [{\citenamefont {Semenoff}(1984)}]{Semenoff}%
  \BibitemOpen
  \bibfield  {author} {\bibinfo {author} {\bibfnamefont {G.~W.}\ \bibnamefont
  {Semenoff}},\ }\href {\doibase 10.1103/PhysRevLett.53.2449} {\bibfield
  {journal} {\bibinfo  {journal} {Phys. Rev. Lett.}\ }\textbf {\bibinfo
  {volume} {53}},\ \bibinfo {pages} {2449} (\bibinfo {year}
  {1984})}\BibitemShut {NoStop}%
\bibitem [{\citenamefont {Anderson}\ \emph {et~al.}(2004)\citenamefont
  {Anderson}, \citenamefont {Lee}, \citenamefont {Randeria}, \citenamefont
  {Rice}, \citenamefont {Trivedi},\ and\ \citenamefont {Zhang}}]{plainvanilla}%
  \BibitemOpen
  \bibfield  {author} {\bibinfo {author} {\bibfnamefont {P.~W.}\ \bibnamefont
  {Anderson}}, \bibinfo {author} {\bibfnamefont {P.~A.}\ \bibnamefont {Lee}},
  \bibinfo {author} {\bibfnamefont {M.}~\bibnamefont {Randeria}}, \bibinfo
  {author} {\bibfnamefont {T.~M.}\ \bibnamefont {Rice}}, \bibinfo {author}
  {\bibfnamefont {N.}~\bibnamefont {Trivedi}}, \ and\ \bibinfo {author}
  {\bibfnamefont {F.~C.}\ \bibnamefont {Zhang}},\ }\href {\doibase
  10.1088/0953-8984/16/24/R02} {\bibfield  {journal} {\bibinfo  {journal} {J
  Phys. Condens. Matter}\ }\textbf {\bibinfo {volume} {16}},\ \bibinfo {pages}
  {R755} (\bibinfo {year} {2004})}\BibitemShut {NoStop}%
\bibitem [{\citenamefont {Le~Hur}\ and\ \citenamefont
  {Rice}(2009)}]{ReviewhighTc}%
  \BibitemOpen
  \bibfield  {author} {\bibinfo {author} {\bibfnamefont {K.}~\bibnamefont
  {Le~Hur}}\ and\ \bibinfo {author} {\bibfnamefont {T.~M.}\ \bibnamefont
  {Rice}},\ }\href
  {https://www.sciencedirect.com/science/article/abs/pii/S0003491609000451?via%3Dihubhttps://doi.org/10.1016/j.aop.2009.02.004}
  {\bibfield  {journal} {\bibinfo  {journal} {Annals of Physics}\ }\textbf
  {\bibinfo {volume} {324}},\ \bibinfo {pages} {1452} (\bibinfo {year}
  {2009})}\BibitemShut {NoStop}%
\bibitem [{\citenamefont {Ginzburg}\ and\ \citenamefont
  {Landau}(1950)}]{ginzburg}%
  \BibitemOpen
  \bibfield  {author} {\bibinfo {author} {\bibfnamefont {V.~L.}\ \bibnamefont
  {Ginzburg}}\ and\ \bibinfo {author} {\bibfnamefont {L.~D.}\ \bibnamefont
  {Landau}},\ }\href@noop {} {\bibfield  {journal} {\bibinfo  {journal} {Zh.
  Eksp. Teor. Fiz.}\ }\textbf {\bibinfo {volume} {20}},\ \bibinfo {pages}
  {1064} (\bibinfo {year} {1950})}\BibitemShut {NoStop}%
\bibitem [{\citenamefont {Tran}\ \emph {et~al.}(2017)\citenamefont {Tran},
  \citenamefont {Dauphin}, \citenamefont {Grushin}, \citenamefont {Zoller},\
  and\ \citenamefont {Goldman}}]{goldman_2017}%
  \BibitemOpen
  \bibfield  {author} {\bibinfo {author} {\bibfnamefont {D.~T.}\ \bibnamefont
  {Tran}}, \bibinfo {author} {\bibfnamefont {A.}~\bibnamefont {Dauphin}},
  \bibinfo {author} {\bibfnamefont {A.~G.}\ \bibnamefont {Grushin}}, \bibinfo
  {author} {\bibfnamefont {P.}~\bibnamefont {Zoller}}, \ and\ \bibinfo {author}
  {\bibfnamefont {N.}~\bibnamefont {Goldman}},\ }\href {\doibase
  10.1126/sciadv.1701207} {\bibfield  {journal} {\bibinfo  {journal} {Science
  Advances}\ }\textbf {\bibinfo {volume} {3}},\ \bibinfo {pages} {e1701207}
  (\bibinfo {year} {2017})}\BibitemShut {NoStop}%
\bibitem [{\citenamefont {Asteria}\ \emph {et~al.}(2019)\citenamefont
  {Asteria}, \citenamefont {Tran}, \citenamefont {Ozawa}, \citenamefont
  {Tarnowski}, \citenamefont {Rem}, \citenamefont {Fl\"aschner}, \citenamefont
  {Sengstock}, \citenamefont {Goldman},\ and\ \citenamefont
  {Weitenberg}}]{asteria_2019}%
  \BibitemOpen
  \bibfield  {author} {\bibinfo {author} {\bibfnamefont {L.}~\bibnamefont
  {Asteria}}, \bibinfo {author} {\bibfnamefont {D.~T.}\ \bibnamefont {Tran}},
  \bibinfo {author} {\bibfnamefont {T.}~\bibnamefont {Ozawa}}, \bibinfo
  {author} {\bibfnamefont {M.}~\bibnamefont {Tarnowski}}, \bibinfo {author}
  {\bibfnamefont {B.~S.}\ \bibnamefont {Rem}}, \bibinfo {author} {\bibfnamefont
  {N.}~\bibnamefont {Fl\"aschner}}, \bibinfo {author} {\bibfnamefont
  {K.}~\bibnamefont {Sengstock}}, \bibinfo {author} {\bibfnamefont
  {N.}~\bibnamefont {Goldman}}, \ and\ \bibinfo {author} {\bibfnamefont
  {C.}~\bibnamefont {Weitenberg}},\ }\href {\doibase 10.1038/s41567-019-0417-8}
  {\bibfield  {journal} {\bibinfo  {journal} {Nature Physics}\ }\textbf
  {\bibinfo {volume} {15}},\ \bibinfo {pages} {449} (\bibinfo {year}
  {2019})}\BibitemShut {NoStop}%
\bibitem [{\citenamefont {Henriet}\ \emph {et~al.}(2017)\citenamefont
  {Henriet}, \citenamefont {Sclocchi}, \citenamefont {Orth},\ and\
  \citenamefont {Le~Hur}}]{sphere}%
  \BibitemOpen
  \bibfield  {author} {\bibinfo {author} {\bibfnamefont {L.}~\bibnamefont
  {Henriet}}, \bibinfo {author} {\bibfnamefont {A.}~\bibnamefont {Sclocchi}},
  \bibinfo {author} {\bibfnamefont {P.~P.}\ \bibnamefont {Orth}}, \ and\
  \bibinfo {author} {\bibfnamefont {K.}~\bibnamefont {Le~Hur}},\ }\href
  {\doibase https://doi.org/10.1103/PhysRevB.95.054307} {\bibfield  {journal}
  {\bibinfo  {journal} {Phys. Rev. B}\ }\textbf {\bibinfo {volume} {95}},\
  \bibinfo {pages} {054307} (\bibinfo {year} {2017})}\BibitemShut {NoStop}%
\bibitem [{\citenamefont {Hutchinson}\ and\ \citenamefont
  {Le~Hur}(2020)}]{spheremodel}%
  \BibitemOpen
  \bibfield  {author} {\bibinfo {author} {\bibfnamefont {J.}~\bibnamefont
  {Hutchinson}}\ and\ \bibinfo {author} {\bibfnamefont {K.}~\bibnamefont
  {Le~Hur}},\ }\href@noop {} {\bibfield  {journal} {\bibinfo  {journal}
  {arXiv:2002.11823}\ } (\bibinfo {year} {2020})}\BibitemShut {NoStop}%
\bibitem [{\citenamefont {Rivas}\ \emph {et~al.}(2013)\citenamefont {Rivas},
  \citenamefont {Viyuela},\ and\ \citenamefont
  {Martin-Delgado}}]{delgado_2013}%
  \BibitemOpen
  \bibfield  {author} {\bibinfo {author} {\bibfnamefont {A.}~\bibnamefont
  {Rivas}}, \bibinfo {author} {\bibfnamefont {O.}~\bibnamefont {Viyuela}}, \
  and\ \bibinfo {author} {\bibfnamefont {M.~A.}\ \bibnamefont
  {Martin-Delgado}},\ }\href {\doibase 10.1103/PhysRevB.88.155141} {\bibfield
  {journal} {\bibinfo  {journal} {Phys. Rev. B}\ }\textbf {\bibinfo {volume}
  {88}},\ \bibinfo {pages} {155141} (\bibinfo {year} {2013})}\BibitemShut
  {NoStop}%
\bibitem [{\citenamefont {Karplus}\ and\ \citenamefont {Luttinger}(1954)}]{KL}%
  \BibitemOpen
  \bibfield  {author} {\bibinfo {author} {\bibfnamefont {R.}~\bibnamefont
  {Karplus}}\ and\ \bibinfo {author} {\bibfnamefont {J.~M.}\ \bibnamefont
  {Luttinger}},\ }\href {\doibase https://doi.org/10.1103/PhysRev.95.1154}
  {\bibfield  {journal} {\bibinfo  {journal} {Phys. Rev.}\ }\textbf {\bibinfo
  {volume} {95}},\ \bibinfo {pages} {1154} (\bibinfo {year}
  {1954})}\BibitemShut {NoStop}%
\bibitem [{\citenamefont {Cheng}\ \emph {et~al.}(2019)\citenamefont {Cheng},
  \citenamefont {Klein}, \citenamefont {Plekhanov}, \citenamefont {Sengstock},
  \citenamefont {Aidelsburger}, \citenamefont {Weitenberg},\ and\ \citenamefont
  {Le~Hur}}]{cheng_2019}%
  \BibitemOpen
  \bibfield  {author} {\bibinfo {author} {\bibfnamefont {P.}~\bibnamefont
  {Cheng}}, \bibinfo {author} {\bibfnamefont {P.~W.}\ \bibnamefont {Klein}},
  \bibinfo {author} {\bibfnamefont {K.}~\bibnamefont {Plekhanov}}, \bibinfo
  {author} {\bibfnamefont {K.}~\bibnamefont {Sengstock}}, \bibinfo {author}
  {\bibfnamefont {M.}~\bibnamefont {Aidelsburger}}, \bibinfo {author}
  {\bibfnamefont {C.}~\bibnamefont {Weitenberg}}, \ and\ \bibinfo {author}
  {\bibfnamefont {K.}~\bibnamefont {Le~Hur}},\ }\href {\doibase
  10.1103/PhysRevB.100.081107} {\bibfield  {journal} {\bibinfo  {journal}
  {Phys. Rev. B}\ }\textbf {\bibinfo {volume} {100}},\ \bibinfo {pages}
  {081107} (\bibinfo {year} {2019})}\BibitemShut {NoStop}%
\bibitem [{\citenamefont {Altland}\ and\ \citenamefont
  {Simons}(2010)}]{altland}%
  \BibitemOpen
  \bibfield  {author} {\bibinfo {author} {\bibfnamefont {A.}~\bibnamefont
  {Altland}}\ and\ \bibinfo {author} {\bibfnamefont {B.~D.}\ \bibnamefont
  {Simons}},\ }\href {\doibase 10.1017/CBO9780511789984} {\emph {\bibinfo
  {title} {Condensed Matter Field Theory}}},\ \bibinfo {edition} {2nd}\ ed.\
  (\bibinfo  {publisher} {Cambridge University Press},\ \bibinfo {year}
  {2010})\BibitemShut {NoStop}%
\bibitem [{\citenamefont {Coleman}(2015)}]{coleman_2015}%
  \BibitemOpen
  \bibfield  {author} {\bibinfo {author} {\bibfnamefont {P.}~\bibnamefont
  {Coleman}},\ }\href {\doibase 10.1017/CBO9781139020916} {\emph {\bibinfo
  {title} {Introduction to Many-Body Physics}}}\ (\bibinfo  {publisher}
  {Cambridge University Press},\ \bibinfo {year} {2015})\BibitemShut {NoStop}%
\bibitem [{\citenamefont {Agra}\ \emph {et~al.}(2006)\citenamefont {Agra},
  \citenamefont {Wijland},\ and\ \citenamefont {Trizac}}]{agra_2006}%
  \BibitemOpen
  \bibfield  {author} {\bibinfo {author} {\bibfnamefont {R.}~\bibnamefont
  {Agra}}, \bibinfo {author} {\bibfnamefont {F.~v.}\ \bibnamefont {Wijland}}, \
  and\ \bibinfo {author} {\bibfnamefont {E.}~\bibnamefont {Trizac}},\ }\href
  {\doibase 10.1088/0143-0807/27/2/022} {\bibfield  {journal} {\bibinfo
  {journal} {European Journal of Physics}\ }\textbf {\bibinfo {volume} {27}},\
  \bibinfo {pages} {407} (\bibinfo {year} {2006})}\BibitemShut {NoStop}%
\bibitem [{\citenamefont {Binney}\ \emph {et~al.}(1992)\citenamefont {Binney},
  \citenamefont {Dowrick}, \citenamefont {Fisher},\ and\ \citenamefont
  {Newman}}]{binney_1992}%
  \BibitemOpen
  \bibfield  {author} {\bibinfo {author} {\bibfnamefont {J.~J.}\ \bibnamefont
  {Binney}}, \bibinfo {author} {\bibfnamefont {N.~J.}\ \bibnamefont {Dowrick}},
  \bibinfo {author} {\bibfnamefont {A.~J.}\ \bibnamefont {Fisher}}, \ and\
  \bibinfo {author} {\bibfnamefont {M.}~\bibnamefont {Newman}},\ }\href@noop {}
  {\emph {\bibinfo {title} {The Theory of Critical Phenomena: An Introduction
  to the Renormalization Group}}}\ (\bibinfo  {publisher} {Oxford University
  Press, Inc.},\ \bibinfo {address} {New York, NY, USA},\ \bibinfo {year}
  {1992})\BibitemShut {NoStop}%
\bibitem [{\citenamefont {Hauschild}\ and\ \citenamefont
  {Pollmann}(2018)}]{Hauschild:2018tz}%
  \BibitemOpen
  \bibfield  {author} {\bibinfo {author} {\bibfnamefont {J.}~\bibnamefont
  {Hauschild}}\ and\ \bibinfo {author} {\bibfnamefont {F.}~\bibnamefont
  {Pollmann}},\ }\href {\doibase 10.21468/SciPostPhysLectNotes.5} {\bibfield
  {journal} {\bibinfo  {journal} {SciPost Phys. Lect. Notes}\ ,\ \bibinfo
  {pages} {5}} (\bibinfo {year} {2018})}\BibitemShut {NoStop}%
\bibitem [{\citenamefont {Grushin}\ \emph {et~al.}(2015)\citenamefont
  {Grushin}, \citenamefont {Motruk}, \citenamefont {Zaletel},\ and\
  \citenamefont {Pollmann}}]{Grushin2015a}%
  \BibitemOpen
  \bibfield  {author} {\bibinfo {author} {\bibfnamefont {A.~G.}\ \bibnamefont
  {Grushin}}, \bibinfo {author} {\bibfnamefont {J.}~\bibnamefont {Motruk}},
  \bibinfo {author} {\bibfnamefont {M.~P.}\ \bibnamefont {Zaletel}}, \ and\
  \bibinfo {author} {\bibfnamefont {F.}~\bibnamefont {Pollmann}},\ }\href
  {\doibase https://doi.org/10.1103/PhysRevB.91.035136} {\bibfield  {journal}
  {\bibinfo  {journal} {Phys. Rev. B}\ }\textbf {\bibinfo {volume} {91}},\
  \bibinfo {pages} {035136} (\bibinfo {year} {2015})}\BibitemShut {NoStop}%
\bibitem [{\citenamefont {Capponi}(2016)}]{Capponi_2016}%
  \BibitemOpen
  \bibfield  {author} {\bibinfo {author} {\bibfnamefont {S.}~\bibnamefont
  {Capponi}},\ }\href {\doibase 10.1088/1361-648x/29/4/043002} {\bibfield
  {journal} {\bibinfo  {journal} {Journal of Physics: Condensed Matter}\
  }\textbf {\bibinfo {volume} {29}},\ \bibinfo {pages} {043002} (\bibinfo
  {year} {2016})}\BibitemShut {NoStop}%
\bibitem [{\citenamefont {Hohenberg}\ and\ \citenamefont
  {Krekhov}(2015)}]{hohenberg_2015}%
  \BibitemOpen
  \bibfield  {author} {\bibinfo {author} {\bibfnamefont {P.}~\bibnamefont
  {Hohenberg}}\ and\ \bibinfo {author} {\bibfnamefont {A.}~\bibnamefont
  {Krekhov}},\ }\href {\doibase 10.1016/j.physrep.2015.01.001} {\bibfield
  {journal} {\bibinfo  {journal} {Physics Reports}\ }\textbf {\bibinfo {volume}
  {572}},\ \bibinfo {pages} {1} (\bibinfo {year} {2015})}\BibitemShut {NoStop}%
\bibitem [{\citenamefont {Xiao}\ \emph {et~al.}(2010)\citenamefont {Xiao},
  \citenamefont {Chang},\ and\ \citenamefont {Niu}}]{RMP}%
  \BibitemOpen
  \bibfield  {author} {\bibinfo {author} {\bibfnamefont {D.}~\bibnamefont
  {Xiao}}, \bibinfo {author} {\bibfnamefont {M.-C.}\ \bibnamefont {Chang}}, \
  and\ \bibinfo {author} {\bibfnamefont {Q.}~\bibnamefont {Niu}},\ }\href
  {\doibase https://doi.org/10.1103/PhysRevLett.49.405} {\bibfield  {journal}
  {\bibinfo  {journal} {Rev. Mod. Phys.}\ }\textbf {\bibinfo {volume} {82}},\
  \bibinfo {pages} {1959} (\bibinfo {year} {2010})}\BibitemShut {NoStop}%
\bibitem [{\citenamefont {Thouless}\ \emph {et~al.}(1982)\citenamefont
  {Thouless}, \citenamefont {Kohmoto}, \citenamefont {Nightingale},\ and\
  \citenamefont {den Nijs}}]{TKNN}%
  \BibitemOpen
  \bibfield  {author} {\bibinfo {author} {\bibfnamefont {D.}~\bibnamefont
  {Thouless}}, \bibinfo {author} {\bibfnamefont {M.}~\bibnamefont {Kohmoto}},
  \bibinfo {author} {\bibfnamefont {M.~P.}\ \bibnamefont {Nightingale}}, \ and\
  \bibinfo {author} {\bibfnamefont {M.}~\bibnamefont {den Nijs}},\ }\href
  {\doibase https://doi.org/10.1103/PhysRevLett.49.405} {\bibfield  {journal}
  {\bibinfo  {journal} {Phys. Rev. Lett.}\ }\textbf {\bibinfo {volume} {49}},\
  \bibinfo {pages} {405} (\bibinfo {year} {1982})}\BibitemShut {NoStop}%
\bibitem [{\citenamefont {Wang}\ and\ \citenamefont {Zhang}(2012)}]{Zhang}%
  \BibitemOpen
  \bibfield  {author} {\bibinfo {author} {\bibfnamefont {Z.}~\bibnamefont
  {Wang}}\ and\ \bibinfo {author} {\bibfnamefont {S.-C.}\ \bibnamefont
  {Zhang}},\ }\href {\doibase https://doi.org/10.1103/PhysRevX.2.031008}
  {\bibfield  {journal} {\bibinfo  {journal} {Phys. Rev. X}\ }\textbf {\bibinfo
  {volume} {2}},\ \bibinfo {pages} {031008} (\bibinfo {year}
  {2012})}\BibitemShut {NoStop}%
\bibitem [{\citenamefont {Legendre}\ and\ \citenamefont
  {Le~Hur}(2020)}]{quantumHallKagome}%
  \BibitemOpen
  \bibfield  {author} {\bibinfo {author} {\bibfnamefont {J.}~\bibnamefont
  {Legendre}}\ and\ \bibinfo {author} {\bibfnamefont {K.}~\bibnamefont
  {Le~Hur}},\ }\href {\doibase
  https://doi.org/10.1103/PhysRevResearch.2.022043} {\bibfield  {journal}
  {\bibinfo  {journal} {Phys. Rev. Research}\ }\textbf {\bibinfo {volume}
  {2}},\ \bibinfo {pages} {022043(R)} (\bibinfo {year} {2020})}\BibitemShut
  {NoStop}%
\bibitem [{\citenamefont {Kane}\ and\ \citenamefont {Mele}(2005)}]{KM}%
  \BibitemOpen
  \bibfield  {author} {\bibinfo {author} {\bibfnamefont {C.~L.}\ \bibnamefont
  {Kane}}\ and\ \bibinfo {author} {\bibfnamefont {E.}~\bibnamefont {Mele}},\
  }\href {\doibase 10.1103/PhysRevLett.95.226801} {\bibfield  {journal}
  {\bibinfo  {journal} {Phys. Rev. Lett.}\ }\textbf {\bibinfo {volume} {95}},\
  \bibinfo {pages} {226801} (\bibinfo {year} {2005})}\BibitemShut {NoStop}%
\bibitem [{\citenamefont {Rachel}\ and\ \citenamefont {Le~Hur}(2010)}]{KM1}%
  \BibitemOpen
  \bibfield  {author} {\bibinfo {author} {\bibfnamefont {S.}~\bibnamefont
  {Rachel}}\ and\ \bibinfo {author} {\bibfnamefont {K.}~\bibnamefont
  {Le~Hur}},\ }\href {\doibase https://doi.org/10.1103/PhysRevB.82.075106}
  {\bibfield  {journal} {\bibinfo  {journal} {Phys. Rev. B}\ }\textbf {\bibinfo
  {volume} {82}},\ \bibinfo {pages} {075106} (\bibinfo {year}
  {2010})}\BibitemShut {NoStop}%
\bibitem [{\citenamefont {Wu}\ \emph {et~al.}(2010)\citenamefont {Wu},
  \citenamefont {Rachel}, \citenamefont {Liu},\ and\ \citenamefont
  {Le~Hur}}]{KM2}%
  \BibitemOpen
  \bibfield  {author} {\bibinfo {author} {\bibfnamefont {W.}~\bibnamefont
  {Wu}}, \bibinfo {author} {\bibfnamefont {S.}~\bibnamefont {Rachel}}, \bibinfo
  {author} {\bibfnamefont {W.-M.}\ \bibnamefont {Liu}}, \ and\ \bibinfo
  {author} {\bibfnamefont {K.}~\bibnamefont {Le~Hur}},\ }\href {\doibase
  https://doi.org/10.1103/PhysRevB.85.205102} {\bibfield  {journal} {\bibinfo
  {journal} {Phys. Rev. B}\ }\textbf {\bibinfo {volume} {85}},\ \bibinfo
  {pages} {205102} (\bibinfo {year} {2010})}\BibitemShut {NoStop}%
\bibitem [{\citenamefont {Plekhanov}\ \emph {et~al.}(2018)\citenamefont
  {Plekhanov}, \citenamefont {Vasic}, \citenamefont {Petrescu}, \citenamefont
  {Nirvan}, \citenamefont {Roux}, \citenamefont {Hofsetter},\ and\
  \citenamefont {Le~Hur}}]{KM3}%
  \BibitemOpen
  \bibfield  {author} {\bibinfo {author} {\bibfnamefont {K.}~\bibnamefont
  {Plekhanov}}, \bibinfo {author} {\bibfnamefont {I.}~\bibnamefont {Vasic}},
  \bibinfo {author} {\bibfnamefont {A.}~\bibnamefont {Petrescu}}, \bibinfo
  {author} {\bibfnamefont {R.}~\bibnamefont {Nirvan}}, \bibinfo {author}
  {\bibfnamefont {G.}~\bibnamefont {Roux}}, \bibinfo {author} {\bibfnamefont
  {W.}~\bibnamefont {Hofsetter}}, \ and\ \bibinfo {author} {\bibfnamefont
  {K.}~\bibnamefont {Le~Hur}},\ }\href {\doibase
  https://doi.org/10.1103/PhysRevLett.120.157201} {\bibfield  {journal}
  {\bibinfo  {journal} {Phys. Rev. Lett.}\ }\textbf {\bibinfo {volume} {120}},\
  \bibinfo {pages} {157201} (\bibinfo {year} {2018})}\BibitemShut {NoStop}%
\bibitem [{\citenamefont {White}(1993)}]{White:1993fb}%
  \BibitemOpen
  \bibfield  {author} {\bibinfo {author} {\bibfnamefont {S.~R.}\ \bibnamefont
  {White}},\ }\href {\doibase 10.1103/PhysRevB.48.10345} {\bibfield  {journal}
  {\bibinfo  {journal} {Phys. Rev. B}\ }\textbf {\bibinfo {volume} {48}},\
  \bibinfo {pages} {10345} (\bibinfo {year} {1993})}\BibitemShut {NoStop}%
\bibitem [{\citenamefont {Zaletel}\ \emph {et~al.}(2014)\citenamefont
  {Zaletel}, \citenamefont {Mong},\ and\ \citenamefont
  {Pollmann}}]{Zaletel_2014}%
  \BibitemOpen
  \bibfield  {author} {\bibinfo {author} {\bibfnamefont {M.~P.}\ \bibnamefont
  {Zaletel}}, \bibinfo {author} {\bibfnamefont {R.~S.~K.}\ \bibnamefont
  {Mong}}, \ and\ \bibinfo {author} {\bibfnamefont {F.}~\bibnamefont
  {Pollmann}},\ }\href {\doibase 10.1088/1742-5468/2014/10/p10007} {\bibfield
  {journal} {\bibinfo  {journal} {Journal of Statistical Mechanics: Theory and
  Experiment}\ }\textbf {\bibinfo {volume} {2014}},\ \bibinfo {pages} {P10007}
  (\bibinfo {year} {2014})}\BibitemShut {NoStop}%
\bibitem [{\citenamefont {Kj\"all}\ \emph {et~al.}(2013)\citenamefont
  {Kj\"all}, \citenamefont {Zaletel}, \citenamefont {Mong}, \citenamefont
  {Bardarson},\ and\ \citenamefont {Pollmann}}]{Kjall2013}%
  \BibitemOpen
  \bibfield  {author} {\bibinfo {author} {\bibfnamefont {J.~A.}\ \bibnamefont
  {Kj\"all}}, \bibinfo {author} {\bibfnamefont {M.~P.}\ \bibnamefont
  {Zaletel}}, \bibinfo {author} {\bibfnamefont {R.~S.~K.}\ \bibnamefont
  {Mong}}, \bibinfo {author} {\bibfnamefont {J.~H.}\ \bibnamefont {Bardarson}},
  \ and\ \bibinfo {author} {\bibfnamefont {F.}~\bibnamefont {Pollmann}},\
  }\href {\doibase 10.1103/PhysRevB.87.235106} {\bibfield  {journal} {\bibinfo
  {journal} {Phys. Rev. B}\ }\textbf {\bibinfo {volume} {87}},\ \bibinfo
  {pages} {235106} (\bibinfo {year} {2013})}\BibitemShut {NoStop}%
\bibitem [{\citenamefont {Motruk}\ \emph {et~al.}(2015)\citenamefont {Motruk},
  \citenamefont {Grushin}, \citenamefont {de~Juan},\ and\ \citenamefont
  {Pollmann}}]{Motruk2015}%
  \BibitemOpen
  \bibfield  {author} {\bibinfo {author} {\bibfnamefont {J.}~\bibnamefont
  {Motruk}}, \bibinfo {author} {\bibfnamefont {A.~G.}\ \bibnamefont {Grushin}},
  \bibinfo {author} {\bibfnamefont {F.}~\bibnamefont {de~Juan}}, \ and\
  \bibinfo {author} {\bibfnamefont {F.}~\bibnamefont {Pollmann}},\ }\href
  {\doibase 10.1103/PhysRevB.92.085147} {\bibfield  {journal} {\bibinfo
  {journal} {Phys. Rev. B}\ }\textbf {\bibinfo {volume} {92}},\ \bibinfo
  {pages} {085147} (\bibinfo {year} {2015})}\BibitemShut {NoStop}%
\bibitem [{\citenamefont {Repellin}\ \emph {et~al.}(2017)\citenamefont
  {Repellin}, \citenamefont {He},\ and\ \citenamefont
  {Pollmann}}]{Repellin2017}%
  \BibitemOpen
  \bibfield  {author} {\bibinfo {author} {\bibfnamefont {C.}~\bibnamefont
  {Repellin}}, \bibinfo {author} {\bibfnamefont {Y.-C.}\ \bibnamefont {He}}, \
  and\ \bibinfo {author} {\bibfnamefont {F.}~\bibnamefont {Pollmann}},\ }\href
  {\doibase 10.1103/PhysRevB.96.205124} {\bibfield  {journal} {\bibinfo
  {journal} {Phys. Rev. B}\ }\textbf {\bibinfo {volume} {96}},\ \bibinfo
  {pages} {205124} (\bibinfo {year} {2017})}\BibitemShut {NoStop}%
\end{thebibliography}%

\end{document}